\documentclass[a4paper,11pt]{article}
\pdfoutput=1 

\usepackage{jcappub} 

\usepackage[table,svgnames,dvipsnames]{xcolor}
\usepackage[normalem]{ulem}
\usepackage{aas_macros}
\usepackage{graphicx}
\usepackage{graphics}
\usepackage{amsmath}
\usepackage{bm}
\usepackage{xcolor}
\usepackage{makecell}
\usepackage{hyperref}
\usepackage{verbatim}
\usepackage{comment}
\usepackage{lineno}
\usepackage{soul}
\usepackage{multicol,multirow}
\usepackage{fontawesome}
\usepackage{orcidlink}
\usepackage{xfrac}
\usepackage{hanging}
\usepackage{todonotes}
\usepackage{cancel}
\usepackage{amssymb,amsfonts}
\usepackage{physics}
\usepackage{lineno}

\usepackage[draft,author=]{fixme}
\fxsetup{layout=footnote}

\newcommand{\fNL}{f_{\rm NL}}

\newcommand{\hMpc}{\ h^{-1}\text{Mpc}}

\newcommand{\ihMpc}{\ h\,\text{Mpc}^{-1}}

\DeclareMathOperator{\cov}{cov}

\newcommand{\be}{\begin{equation}}
\newcommand{\ee}{\end{equation}}

\renewcommand{\vec}{\bm}

\newcommand{\kmax}{{k_{\rm max}}}

\newcommand{\Om}{\Omega_\mathrm{m}}

\newcommand{\aiso}{\alpha_{\rm iso}}
\newcommand{\alap}{\alpha_{\rm AP}}

\newcommand{\abacus}{{\tt Abacus}}
\newcommand{\abacussecond}{{\tt Abacus-2}}
\newcommand{\abacusfirst}{{\tt Abacus-1}}
\newcommand{\ezmock}{{\tt EZmock}}
\newcommand{\ezmocks}{{\tt EZmocks}}

\newcommand{\altmtl}{{`altmtl'}}

\newcommand{\bgs}{{\tt BGS}}
\newcommand{\elgo}{{\tt ELG1}}
\newcommand{\elgt}{{\tt ELG2}}

\newcommand{\lrgo}{{\tt LRG1}}
\newcommand{\lrgt}{{\tt LRG2}}
\newcommand{\lrgth}{{\tt LRG3}}

\newcommand{\qso}{{\tt QSO}}
\newcommand{\desidrone}{{DESI DR1}}
\newcommand{\desione}{{DESI DR1}}
\newcommand{\planck}{{\it Planck} 2018-$\Lambda$CDM}

\newcommand{\velocileptors}{\texttt{velocileptors}}
\newcommand{\folps}{\texttt{Folps$\nu$}}
\newcommand{\pybird}{\texttt{PyBird}}
\newcommand{\eftgsm}{\texttt{EFT-GSM}}

\newcommand{\hinvmpc}{\ h^{-1}\,{\rm Mpc}}

\newcommand{\rd}{r_{\mathrm{d}}}
\newcommand{\DM}{D_{\mathrm{M}}}
\newcommand{\Dh}{D_{\mathrm{H}}}
\newcommand{\DV}{D_{\mathrm{V}}}

\renewcommand*\vec[1]{\ensuremath{\boldsymbol{#1}}}
\newcommand*\tens[1]{\ensuremath{\mathsf{#1}}}

\usepackage[nameinlink,noabbrev]{cleveref}
\crefname{equation}{Eq.}{Eqs.}
\crefname{section}{Section}{Sections}
\crefname{figure}{Figure}{Figures}
\crefname{table}{Table}{Tables}
\crefname{appendix}{Appendix}{Appendices}
\Crefname{figure}{Figure}{Figures}
\Crefname{equation}{Equation}{Equations}
\Crefname{section}{Section}{Sections}
\Crefname{table}{Table}{Tables}

\usepackage{xspace}

\title{\boldmath DESI 2024 V: Full-Shape Galaxy Clustering from Galaxies and Quasars}

\author{{DESI Collaboration}:}
\emailAdd{spokespersons@desi.lbl.gov}
\affiliation{Affiliations are in Appendix \ref{sec:affiliations}}

\author[1]{{A.~G.~Adame},}
\author[2]{{J.~Aguilar},}
\author[3]{{S.~Ahlen}\orcidlink{0000-0001-6098-7247},}
\author[4]{{S.~Alam}\orcidlink{0000-0002-3757-6359},}
\author[5,6]{{D.~M.~Alexander}\orcidlink{0000-0002-5896-6313},}
\author[2]{{M.~Alvarez},}
\author[7]{{O.~Alves},}
\author[2]{{A.~Anand}\orcidlink{0000-0003-2923-1585},}
\author[8,7]{{U.~Andrade}\orcidlink{0000-0002-4118-8236},}
\author[9]{{E.~Armengaud}\orcidlink{0000-0001-7600-5148},}
% ...

\date{\today}

\abstract{
We present the measurements and cosmological implications of the galaxy two-point clustering using over 4.7 million unique galaxy and quasar redshifts in the range $0.1<z<2.1$ divided into six redshift bins over a $\sim 7,500$ square degree footprint, from the first year of observations with the Dark Energy Spectroscopic Instrument (DESI Data Release 1). By fitting the full power spectrum, we extend previous DESI DR1 baryon acoustic oscillation (BAO) measurements to include redshift-space distortions and signals from the matter-radiation equality scale. 
For the first time, this Full-Shape analysis is blinded at the catalogue-level to avoid confirmation bias and the systematic errors are accounted for at the two-point clustering level, which automatically propagates them into any cosmological parameter. 
When analysing the data in terms of compressed model-agnostic variables, we obtain a combined precision of 4.7\% on the amplitude of the redshift space distortion (RSD) signal reaching a similar precision with just one year of DESI data than with twenty years of observation from the previous generation survey.
We also analyse the data to directly constrain the cosmological parameters within the $\Lambda$CDM model using perturbation theory and combine this information with the reconstructed DESI DR1 galaxy BAO. Using a Big Bang Nucleosynthesis Gaussian prior on the baryon density parameter, $\omega_b$, and a weak Gaussian prior on the spectral index, $n_s$, we constrain the matter density is $\Omega_m = 0.296\pm 0.010 $ and the Hubble constant $H_0 = (68.63 \pm 0.79)[{\rm km\, s^{-1}Mpc^{-1}}]$.
Additionally, we measure the amplitude of clustering $\sigma_8 = 0.841 \pm 0.034$.  
The DESI DR1 galaxy clustering results are in agreement with the $\Lambda$CDM model based on general relativity with parameters consistent with those from \textit{Planck}.
The cosmological interpretation of these results in combination with DESI DR1 Ly-$\alpha$ forest data and external datasets are presented in the companion paper \cite{DESI2024.VII.KP7B}.

}
\begin{document}

\maketitle
\label{firstpage}

\section{Introduction}
\label{sec:intro}

Since the 1980s, galaxy redshift surveys have revolutionised our understanding of the large-scale matter distribution in the Universe and significantly influenced the development of the standard model of cosmology. One of the first spectroscopic surveys, the CfA Redshift Survey~\cite{1983ApJS...52...89H}, gave clear signs which pointed to the need for cold rather than warm or hot dark matter~\cite{1985ApJ...292..371D} and indicated a matter density below the Einstein-de Sitter prediction~\cite{1990Natur.348..705E}. In the 1990s and 2000s, surveys like the 2dF Galaxy Redshift Survey (2dFGRS, \citep{2001MNRAS.328.1039C}) and the Sloan Digital Sky Survey (SDSS, \cite{2006AJ....131.2332G}) significantly expanded these datasets in density and volume, enabling crucial validations of the standard model of cosmology, such as the measurement of baryon acoustic oscillations (BAO, \cite{2001MNRAS.327.1297P,2005ApJ...633..560E,2005MNRAS.362..505C}) and tests of gravity through redshift space distortions (RSD, \cite{2001Natur.410..169P,2008Natur.451..541G}). 

The Dark Energy Spectroscopic Instrument (DESI~\cite{Snowmass2013.Levi,DESI2016a.Science,DESI2016b.Instr,DESI2022.KP1.Instr}) represents the next step in this evolution. DESI is designed to collect spectra for about 40 million galaxies and quasars in the redshift range $0 < z < 4$ thanks to its 5,000 robotic fibre positioners \cite{FocalPlane.Silber.2023} and to the 3.2-degree diameter prime focus corrector \cite{Corrector.Miller.2023}. DESI targets five main classes of tracers: low-redshift galaxies from the Bright Galaxy Survey (BGS), luminous red galaxies (LRG), emission line galaxies (ELG), quasars (QSO), and the Ly$
\alpha$ forest. The BGS targets correspond to the `bright time' spectroscopic pipeline, whereas the rest of samples to the `dark time' (see \cite{Spectro.Pipeline.Guy.2023,SurveyOps.Schlafly.2023} for a detailed  description of the spectroscopic pipeline and distinct program observations).
This broad range of tracers allows DESI to probe different epochs of the cosmic expansion history. Additionally, DESI includes a high-density Milky Way Survey, which focuses on stellar evolution and kinematics within our galaxy~\cite{Cooper2022:2208.08514v4}.

The distribution of galaxies traces the underlying matter distribution, which encodes several cosmological observables. The shape of the matter power spectrum provides information on a combination of the Hubble constant $H_0$ and the matter density $\Omega_m$. 
It also encodes key information about the expansion history through the baryon acoustic oscillation (BAO) feature. The BAO feature is an imprint in the matter power spectrum that arises from sound waves propagating through the hot plasma of the early universe before recombination~\cite{1970ApJ...162..815P,1970Ap&SS...7...20S,1987MNRAS.226..655B}. These sound waves created a characteristic scale in the distribution of matter, which is observable today as a preferred separation between galaxies. This scale serves as a ``standard ruler" for cosmological measurements, which are based on observed angles and redshifts. Translating to comoving units allows us to extract information about the comoving angular diameter distance $D_M(z)$   
and the Hubble parameter $H(z)$ relative to the BAO scale, which depends on the sound horizon scale at the drag epoch. Note that the BAO can also be used as a standard ruler at the matter-radiation equality, as recently used in \cite{Zaborowski:2024wpo}.

The Alcock-Paczynski (AP, \cite{1979Natur.281..358A}) effect exploits the fact that there is a different cosmological dependence for the comoving length of an object or pattern, observed along and across the line-of-sight. Given an object or pattern that is expanding with the expansion of the Universe and that is expected to be isotropic, we will only observe the isotropy in our comoving map if we use a cosmological model with the correct $D_M(z)H(z)$ to translate redshifts into comoving radial distances. If a wrong cosmological model is assumed, the object may appear squashed or stretched along the line of sight. This effect applies to the BAO scale, but also to objects such as voids, or to the full clustering signal of galaxies. This makes it a powerful and complementary tool in a cosmological analysis.

The galaxy power spectrum is also sensitive to the total neutrino mass, the scalar amplitude of primordial fluctuations, and the scalar spectral index, $n_s$. A non-negligible neutrino mass causes a scale-dependent suppression of the power spectrum, as neutrinos, being light and fast-moving, resist clustering on small scales. In addition, neutrinos also modify the expansion history and their mass can be constrained by measuring the baryon acoustic oscillations in combination with external datasets.  The scalar amplitude, $A_s$, affects the overall normalisation of the power spectrum, directly influencing the growth of structure in the universe by setting the initial conditions for how density fluctuations evolve over time. The scalar spectral index, $n_s$, provides additional information about the early universe. It characterises the scale dependence of the primordial fluctuations. In the simplest models of inflation, a nearly scale-invariant spectrum is predicted, where $n_s \approx 1$. However, slight deviations from this value, with $n_s$ typically less than 1, indicate that the fluctuations on larger scales were slightly stronger than those on smaller scales.

Redshift-space distortions (RSD) arise from the fact that the observed redshift of a galaxy is not purely cosmological but is contaminated by the galaxy's peculiar velocity along the line of sight.
In a cosmological context, the redshift of a galaxy, $z_{\rm obs}$, is composed of both its cosmological redshift, $z_{\rm cosmo}$, and its Doppler shift due to peculiar motion, $z_{\rm pec}$, leading to an observed redshift of 
\begin{equation}
    z_{\rm obs} \simeq z_{\rm cosmo} + \frac{v_{\rm pec}}{c}\, ,
\end{equation}
where $c$ is the speed of light. If we translate from redshift to distance using only a model for $z_{\rm cosmo}$ but applied to $z_{\rm obs}$, the peculiar velocity causes galaxies to appear in positions that are shifted along the line of sight compared to their true positions in real space, creating an apparent anisotropy in the galaxy distribution~\cite{1987MNRAS.227....1K,1998ASSL..231..185H}. These distortions lead to an elongation along the line of sight on small scales where the peculiar velocities are large and close to random and a contraction on large scales where the peculiar velocities are smaller and highly correlated. 
The amplitude and shape of these distortions are sensitive to the total matter content of the universe, $\sigma_8$ as well as the growth rate of structure $f$, two quantities which are connected by the underlying theory of gravity.
Thanks to the RSD effect and the line-of-sight dependent clustering we can disentangle the effect from $f$ and the linear galaxy bias. However, at linear order both parameters are completely degenerate with an overall amplitude parameter, such as $\sigma_8$. Such degeneration is poorly broken by the power spectrum 1- and 2-loop orders, and more efficiently when adding higher-order statistics such as the bispectrum signal \cite{Gil-Marin2016:1606.00439v2,Novell-Masotet:al2025}.

While galaxies have a straightforward linear relationship with the underlying matter distribution on large scales, this connection becomes significantly more intricate on smaller scales due to the effects of non-linear processes~\cite{Desjacques2016:1611.09787v5}. One of the key complications in this regard is the influence of the local non-linear velocity field, which affects the RSD signal.
We obtain our primary cosmological science results from analytic models based on perturbation theory, matching recent analyses of smaller samples from the SDSS. There has been a strong evolution in the models used over the past decade, matching the increase in quantity and quality of the data available to be modelled; from phenomenological models built on linear theory with scale-independent bias, through a series of increasingly accurate approximations to the present day, including the TNS model~\cite{Beutler2016:1607.03150v1,Beutler2013:1312.4611v2,Mattia2020:2007.09008v3,Gil-Marin2020:2007.08994v2,Gil-Marin2016:1606.00439v2}, halo perturbation theory model~\cite{Yu2022:2211.16794v2}, Convolution Lagrangian Perturbation Theory (CLPT)~\cite{Satpathy2016:1607.03148v2,Bautista2020:2007.08993v2}, the Gaussian streaming model~\cite{Reid2012:1203.6641v1} and Galilean-Invariant Renormalised Perturbation Theory (gRPT)~\cite{Sanchez2016:1607.03147v1}. In the past decade, substantial progress has led to a renewed understanding of Perturbation Theory (PT) approaches to structure formation through techniques from effective field theory (EFT) --- within this approach, contributions to galaxy clustering obeying fundamental symmetries such as Galilean invariance and the equivalence principle can be systematically enumerated, such that theory predictions for e.g.\ the galaxy power spectrum can be computed rigorously up to a desired accuracy on quasilinear scales \cite{McDonald2009,Baumann12,2012JHEP...09..082C,Vlah2015,Senatore14,Senatore2015,Mirbabayi2015,perkoetal:2016,Chen20,Fujita2020,Chen21,Pybird_damico}. In the EFT approach uncertainties from small-scales are encapsulated in a small number of ``counter terms'' that are also constrained by the symmetries of the equations of motion with numerical values fit to the data.  As described later, for DESI we employ a full one-loop treatment with all of the terms (e.g.\ biases and counterterms) allowed by symmetries to one-loop, or next to leading order. Although the allowed contributions to galaxy clustering are uniquely specified to any given order, a few potential modelling degrees of freedom do still exist : one particularly important aspect is the modelling of the nonlinear damping of the BAO, which introduces a large parameter in perturbation theory into the theory whose effects must be resummed \cite{ESW07,Crocce08,Sugiyama14,Baldauf15}. This resummation is not unique, and within DESI we cross-check three particularly well-studied examples within Lagrangian \cite{Vlah2015,Chen21} and Eulerian perturbation theory \cite{Senatore15}, including via the wiggle no-wiggle split and Time-Sliced Perturbation Theory (TSPT) \cite{Vlah16,Blas16,Ivanov18}. While they differ in specifics, all of these approaches have been shown to be sufficient for practical applications to the BOSS and eBOSS data \cite{Ivanov20,dAmico20,Chen20b,Chen2022,Simon23b,Noriegaetal24} and for the purposes of DESI \cite{KP5s1-Maus}. Therefore, any difference between the models that could arise at two-loop order and beyond can be considered as negligible.

In general, fits using models are not performed with full freedom in the number of cosmological parameters allowed. Approximations are typically used, or limitations to particular models, such as assuming zero neutrino particle masses, or flat cosmological models. If the data are not sensitive to a particular effect, then there is usually no harm in not allowing that effect full freedom. When the freedom is severely curtailed the modelling approach is typically called a template-fitting framework. Here, a template power spectrum or correlation function model is shifted according to a set of scaling parameters that match the constraining power of the data to changes in the model. These scaling parameters allow us to compress the cosmological information into a small set of observables like e.g.\ the clustering amplitude $f\sigma_8$~\cite{Reid2012:1203.6641v1}.
When this information is combined with CMB experiments (such as Planck) the compression in terms of these scaling parameters and $f\sigma_8$ can be demonstrated to be nearly lossless for $\Lambda$CDM type of models.
Earlier template fitting approaches were limited to BAO and RSD, while more recent extensions, also exploit the shape of the matter power spectrum~\cite{brieden2020}. 

In recent years it has become popular to fit the cosmological parameters directly to the galaxy power spectrum together with the large set of nuisance parameters required by the PT models.
This Full-Modelling approach however does rely on external priors on some cosmological parameters since the current generation of galaxy surveys cannot constrain the full set of $\Lambda$CDM parameters. Additionally, in order to be fully general, the perturbative models we use have around 10 nuisance parameters in order to account for bias, non-linear contributions and stochasticity. 
When the data do not tightly constrain the model, the choice of parameterisation or prior on these parameters can significantly impact the marginalised posterior likelihoods for various cosmological parameters.  This has been an issue for both Full-Modelling and template-fitting approaches for past surveys like BOSS~\cite{Carrilho2022:2207.14784v2,Simon23a,Maus23}, and continues to be an issue for DESI for some extensions to $\Lambda$CDM (see e.g.\ ref.~\cite{KP5s2-Maus,KP5s3-Noriega} and \cref{subsec:methods-projection} for further discussion). When constraining models beyond $\Lambda$CDM, it is therefore essential to combine BAO and galaxy Full-Shape measurements, but also different cosmological probes together, in order to break degeneracies and reduce projection effects.
In this paper, we will present the galaxy Full-Shape measurements and results from both the Full-Modelling and the template-fitting methods, highlighting the impact of priors and presenting a detailed study of each potential source of systematic effect.
After validating these modelling approaches on simulations we apply them to the DESI data release 1 (DR1) galaxy samples. Our analysis yields the best constraints on the clustering amplitude to date and the most precise cosmological constraints from DESI DR1 data alone, combining our galaxy Full-Shape measurements with the galaxy and quasar BAO measurements presented in \cite{DESI2024.III.KP4} and the Ly$\alpha$ BAO presented in \cite{DESI2024.IV.KP6}. The detailed cosmological inference including the combination with other cosmological probes is covered in a companion paper \cite{DESI2024.VII.KP7B}. 
An analysis of the power spectrum to constrain potential primordial non-Gaussian signals is also presented in 
\cite{ChaussidonY1fnl}.

This paper is organised as follows, in \cref{sec:catalog} we present the DESI DR1 large-scale catalogues; in \cref{sec:mocks} the mock synthetic catalogues used in this paper, used for validating the pipeline and for producing the final covariances; in \cref{sec:methods} we describe the methods used to extract cosmological information from the DESI catalogues and describe the different type of compression we use; in \cref{sec:systematics} we describe the potential systematics of our analysis and quantify their potential impact on the cosmology results. \cref{sec:unblinding} is dedicated to describing the unblinding process, and presenting all the tests that were passed before the data was ready to be unblinded. \cref{sec:results} displays the main results of this paper and finally in \cref{sec:conclusion} we present the conclusions of this work. 

\section{An Overview of the DESI samples and the LSS catalogues}
\label{sec:catalog}

\subsection{\desidrone}

The DESI Data Release 1 (DR1; \cite{DESI2024.I.DR1}) dataset comprises observations made with the DESI instrument \cite{DESI2022.KP1.Instr} on the Mayall Telescope at Kitt Peak, Arizona, during the main survey operations from May 14, 2021, to June 14, 2022, following a period of survey validation \cite{DESIsv}. DESI simultaneously measures the spectra of up to 5,000 targets using robotic positioners to place fibres at the celestial coordinates of these targets in the focal plane \cite{Corrector.Miller.2023,FocalPlane.Silber.2023}. Fibres are grouped into ten `petals', each guiding light to one of ten climate-controlled spectrographs. The data was collected through `tile' observations, following a strategy designed to prioritise completing observations in specific areas of the sky \cite{SurveyOps.Schlafly.2023}.

Each tile corresponds to a specific sky position and a set of targets assigned to each robotic fibre positioner \cite{DESItarget}. DESI dynamically allocates its observing time into `bright' time and `dark' time programs based on observing conditions. In DR1, 2,744 tiles were observed during dark time and 2,275 tiles during bright time. The bright galaxy sample (BGS) is observed during bright time \cite{DESIbgstarget}, while luminous red galaxies (LRGs \cite{DESIlrgtarget}), emission line galaxies (ELGs \cite{DESIelgtarget}) and quasars (QSOs \cite{DESIqsotarget}) are observed during dark time. These data were initially processed by the DESI spectroscopic pipeline \cite{Spectro.Pipeline.Guy.2023} the morning after observations for immediate quality checks, and then in a uniform processing run (internally called `iron') to produce the resulting redshift catalogues used in this paper and to be released in DR1.

\subsection{DR1 Large-scale structure catalogues}
\label{sec:lss}

The large-scale structures (LSS) catalogues that are created from the redshift and parent target catalogues, together with the two-point function measurements, are described in \cite{DESI2024.II.KP3}. In total, over 5.7 million unique object spectra were collected during the first year of observations and used for galaxy and quasar BAO measurements in DR1 \citep{DESI2024.III.KP4}, a factor of $\sim3$ increase compared to SDSS DR16 \cite{SDSS-DR16-cosmology}. 
However, for the Full-Shape type of analyses presented in this paper, the ELG bin between $0.8 < z < 1.1$ (\texttt{ELG1}) was not included as it failed to pass the required tests before unblinding (see \cref{sec:unblinding} for a full description of the unblinding process). Unlike the rest of the samples, \texttt{ELG1} showed uncorrected systematic effects related to fibre collisions, which suggested that the mitigation techniques used in this paper were not sufficiently robust for an unbiased measurement of the growth of structure on this sample.  However, the impact of these uncorrected systematic on the BAO measurements was negligible, and for this reason, this bin was included in the BAO analysis presented in \cite{DESI2024.III.KP4}, in combination with the high-$z$ LRG bin. More details on the imaging systematics of this and the rest of the bins can be found in \cite{KP5s6-Zhao}.

\cref{tab:Y1data} presents the basic details of the tracer samples used in this paper. The exact definitions of $z_{\rm eff}$ and $V_{\rm eff}$, together with a detailed description of the \desidrone\,LSS catalogues, can be found in \cite{DESI2024.II.KP3}.
After removing the \elgo\ bin, we are left with over 4.7 million unique redshifts for the Full-Shape galaxy and quasar measurements in \desidrone.

\begin{table*}
\centering
\begin{tabular}{|l|c|c|c|c|c|}
\hline
Tracer & redshift range & $N_{\rm tracer}$ & $z_{\rm eff}$   & $P_0$ [($h^{-1}{\rm Mpc})^3$] & $V_{\rm eff}$ [Gpc$^3$]\\ \hline \hline
\bgs  & $0.1-0.4$ & 300,017 & 0.295  & $\sim 9.2\times10^3$ & 1.7  \\
\lrgo  & $0.4-0.6$ & 506,905 & 0.510  & $\sim 8.9\times10^3$ & 2.6 \\
\lrgt  & $0.6-0.8$ & 771,875 & 0.706  & $\sim 8.9\times10^3$ & 4.0  \\
\lrgth  & $0.8-1.1$ & 859,824 & 0.919 &  $\sim 8.4\times10^3$ & 5.0  \\
\elgt  & $1.1-1.6$ & 1,415,687 & 1.317 & $\sim 2.9\times10^3$ &  2.7\\
\qso  & $0.8-2.1$ & 856,652 & 1.491 & $\sim 5.0\times10^3$ & 1.5 \\
\hline
\end{tabular}
\caption{Statistics for each of the DESI tracer types used for the DESI DR1 Full-Shape measurements presented in this paper. Redshift bins are non-overlapping, except for the shot-noise dominated QSO sample.
The effective volume calculation, $V_{\rm eff}$ provides a rough estimate for the relative amount of cosmological information in each redshift bin. $P_0$ is the amplitude of the power spectrum at the reference scale of $k=0.14\,h{\rm Mpc}^{-1}$.  For further details on the samples and the calculation of values in this table see \cite{DESI2024.II.KP3}. }
\label{tab:Y1data}
\end{table*}

In order to build the LSS catalogues, we need to create the matched random sample (`randoms'), that covers the same region as the galaxies but with only shot-noise clustering, accounting for the survey geometry. The randoms are initially produced to match the footprint of the DESI target samples \cite{DESItarget}. These samples are then processed through the DESI \textsc{fiberassign} code\footnote{\href{https://github.com/desihub/fiberassign}{https://github.com/desihub/fiberassign}}, determining whether each input random target could be reached by a fibre based on the properties of the observed DESI DR1 tiles. All of the targets that could be reached by a fibre are kept. Note that we do not run the full fibre assignment algorithm, so no fibre assignment effects are imprinted on the randoms, other than ensuring they match the same angular region covered by the galaxies. This selection is based on individual positions, allowing the DESI \textsc{fiberassign} software to predict the focal plane position of targets with accuracy better than 1 arcsecond, far exceeding the precision of a pixel-based angular mask. For more information about the \textsc{fibreassign} procedure, see \cite{DESI2024.II.KP3}.
These potential assignments are then filtered to match the combination of `good' tiles and fibres as in the DR1 data samples. Subsequently, veto masks are applied to both galaxies and randoms to remove regions close to bright stars and nearby galaxies, regions of very bad imaging conditions, and regions which are only assignable to higher priority targets \cite{DESI2024.II.KP3}. In DR1, the randoms are normalised so that the ratio of weighted data and random counts is consistent across the distinct regions relevant to the photometry used for targeting the sample. Additionally, the redshift distributions are matched between data and randoms in each region. For all but the QSO sample, there are two distinct photometric regions: data targeted from BASS/MzLS photometry in the North Galactic Cap (NGC) with declination greater than $32.375^\circ$, and those targeted from DECaLS photometry in both Galactic caps \cite{LS.Overview.Dey.2019}. For QSOs, the DECaLS sample is further divided into DES and non-DES regions due to differences in target selection in those areas \cite{DESIqsotarget}.

There have been several supporting studies that define and correct for variations in the selection function caused by imaging systematics affecting the input target samples \cite{KP3s2-Rosado,KP3s10-Chaussidon} and variations in the DESI instrument's ability to measure redshifts accurately \cite{KP3s4-Yu,KP3s3-Krolewski}. These papers detail our approach to simulating and correcting for incompleteness in the target assignment. The effects of these issues are combined into a weight column in the data and random catalogues, intended for use in summary statistics calculations, such as the two-point statistics. The number density of the DESI DR1 sample varies significantly with both redshift and the number of overlapping tiles, due to fibre assignment incompleteness. Therefore, `FKP' weights \cite{fkp} designed to maximise the signal-to-noise ratio of clustering measurements at the BAO scale in relation to such number density variations, are included in all two-point calculations. For a full description of the weight calculations, see \cite{DESI2024.II.KP3}.
In \cref{sec:systematics}, we will discuss in detail each potential source of systematic and how to mitigate them in the Full-Shape analysis.

\subsection{Blinding}\label{subsec:blinding}

A ``blinding" procedure is used to prevent confirmation bias in our analysis pipeline and possible biases in the process of data selection. On the theory side, our pipeline requires the choice of the theoretical model, the range of scales used in the analysis, and the parameter priors. On the data side, our pipeline requires the choice of systematic weights and fibre assignment corrections and whether to include any given tracer sample in the final analysis. The analysis presented in this paper includes the first application of blinding of a Full-Shape analysis of a galaxy redshift survey. 

The blinding scheme for our core cosmological analysis consists of two essential components. The first component, which serves to blind the BAO feature, is a catalogue-level blinding implemented following the work of \cite{brieden2020}; here we modify the redshift of the observed tracers and thus imprint a blind shift in the anisotropic position of the BAO peak. The second component of our blinding procedure serves to prevent confirmation biases in the analysis of the redshift-space distortions; following also the work of \cite{brieden2020} we apply density-dependent redshift shifts at the catalogue level which induces a blind shift in the growth rate parameter, $f$. Applying the blinding at the catalogue level guarantees that all summary statistics (both power spectrum and two-point correlation function, but also higher order statistics such as the bispectrum \cite{Novell-Masotet:al2024}) are blinded consistently. More details on how this blinding technique has been tailored to DESI needs can be found in \cite{KP3s9-Andrade}. 

We adopt another --- third --- kind of blinding in cosmological analyses that test for primordial non-Gaussianity (that is when we constrain the parameter $\fNL$). Because $\fNL$ primarily affects the power spectrum by producing a scale-dependent bias that increases at large scales \cite{Dalal:2007cu,Matarrese:2008nc,KP3s10-Chaussidon}, the BAO and RSD blinding described above are insufficient. Instead, we apply to individual galaxies a set of weights that alter the imaging systematics to mimic an $\fNL$ signal.\footnote{
Both imaging and $\fNL$ blinding weight are combined into a single weight in the galaxy catalogue to protect against accidental unblinding. Since the imaging weights are key for correcting the large-scale signal, accidentally not applying this combined weight will not go undetected and will lead to a large misinterpretation of the $\fNL$ signal.}
In what follows, we briefly summarise the procedure, but the full details of the specific methodology applied can be found in \cite{KP3s9-Andrade,KP3s10-Chaussidon}. We start by randomly choosing the variables $\{w_0,w_a,\fNL \}$, where $w_0$ and $w_a$ are the parameters describing the time-variable equation of state of dark energy, from a list of pre-defined values. The first two parameters are selected in such a way that the BAO isotropic-shift parameter $\alpha_{\rm iso}$ is kept within 3\% and the growth rate $f$ within $10\%$ 
of their corresponding fiducial values chosen to be centred on \emph{Planck}'s best-fit cosmological model. This blinding technique (and catalogue) is the same as the one employed to blind the \desidrone\ BAO analysis of ref.~\cite{DESI2024.III.KP4}.
Although we do not blind for other parameters that affect the expansion history of the universe (such as matter density, curvature, or the neutrino mass), for the statistical precision of DESI DR1, blinding the expansion history parametrised by $\{w_0,w_a\}$ allows sufficient freedom that a wider range of cosmological-parameter variations would also fit within the blinding provided by varying $\{w_0,w_a\}$. This is, given this type of blind catalogue other cosmologies such as variations of $\Omega_m$ or $\Omega_k$ are reasonable good fits within the statistical uncertainty in the absence of external datasets. 
The primordial non-Gaussianity parameter is blinded within the range $-15<\fNL<15$. Although in this paper and the corresponding cosmological-analysis papers \cite{DESI2024.VI.KP7A,DESI2024.VII.KP7B} we do not constrain $\fNL$, forthcoming work will make use of the same blinded catalogue, and hence we apply the $\fNL$ blinding scheme throughout. These three types of blinding processes, BAO, RSD and $f_{\rm NL}$ are always simultaneously applied on the catalogues, and we never partially blind (or unblind) those by applying only one or two of these three blinding schemes. 

Our blinding procedure modifies the summary statistics in a way that is not guaranteed to correspond to a shift in a given cosmological model (e.g.\ a combination of cosmological parameters such as $\Om, w_0, w_a$ etc.). For example, the background expansion is not linked to the logarithmic growth of perturbations with the usual expression, $f(z)=\Omega_m(z)^{0.55}$, but both perturbations and expansion are kept independent in the blinded catalogue, partially mimicking a non-GR signature.

Therefore, we only analyse the blinded two-point statistics in terms of the compressed variables, those that describe the BAO, RSD and the shape in the ShapeFit compression technique, namely $\alpha_\parallel$, $\alpha_\perp$, $f\sigma_{s8}$ and $m$ (see \cref{subsubsec:SF} for the details of this method and the parameter definitions). This type of analysis allows for sufficient freedom to account for these blind shifts without the need to assume a specific cosmological model. During this process, the blind cosmology of the shift was kept constant with every catalogue update and never revealed. 
Additionally, the $f_{\rm NL}$ blinding through the scale-dependent bias indirectly also blinds for the shape of the power spectrum at large scales. This collateral blinding happens because both $f_{\rm NL}$ and Shapefit parameter $m$ are very sensitive to the slope of the power at large scales (see fig 4 of \cite{brieden_gilmarin_verde:2021} for their internal correlation). Thus, the shape parameter $m$ is also indirectly blinded when the data are analysed under the assumption of $f_{\rm NL}=0$, as done in this paper. 

Although the Full Modelling was not applied to blind data, we did apply it to mocks for various tests before unblinding. 
The actual unblinded two-point clustering measurements were only measured when the DESI DR1 analysis advanced to the point where all choices were final. We refer to this process as the DESI DR1 unblinding event, which took place on 12th June 2024. 
After the unblinding event, all the previous unblinding tests were repeated to check for unexpected results (hidden by the blinding process), although we did not find any significant changes. These results are presented in \cref{sec:unblinding}.

\section{Mocks}
\label{sec:mocks}

Realistic and accurate mock simulations are essential for our analysis as they enable us to evaluate the limitations of our theoretical models in handling non-linear evolution and galaxy-halo physics. They also help us assess our ability to mitigate imperfections in our survey due to inhomogeneities in the target selection, atmospheric conditions, foreground astrophysical systematics, and instrument constraints. Due to the extensive computational resources required, constructing a single set of mock simulations that covers all these aspects is impractical. Therefore, we developed various types of DESI mock simulations that are used for different studies of systematic effects. \cref{tab:tasksims} summarises the set of simulations used for each task.

Two types of simulations were used for different purposes: 1) The \texttt{AbacusSummit} suite of high-resolution N-body simulations, which are called \abacus\ in the paper, were used to perform high-precision tests of systematic effects. 2) The \ezmocks\ simulations which were computationally cheap (and therefore were less precise) were used to produce large volumes for covariance matrix estimation. 

All mocks are based on the \emph{Planck} 2018 $\Lambda$CDM cosmology, specifically the mean estimates of the \emph{Planck} TT,TE,EE+lowE+lensing likelihood chains: $\Omega_c h^2 = 0.1200$, $\Omega_b h^2 = 0.02237$, $\sigma_8 = 0.811355$, $n_s = 0.9649$, $h = 0.6736$, $w_0 = -1$, and $w_a = 0$ \citep{Planck2018}. For testing the impact of the fiducial cosmology on Full-Shape measurements, we also considered \abacus\ simulations with other cosmologies, see \cite{KP5s8-Gsponer} and \cref{subsec:sys-fiducialcosmo} for more details.

\begin{table}
\centering
\small
    \resizebox{\columnwidth}{!}{%
    \begin{tabular}{|l|r|r|}
    \hline
         Task/Systematics &  Mocks & Ref  \\
    \hline\hline
     Comparison of EFT models  &  \abacusfirst\ cubic, noiseless DR1 & \cite{KP5s1-Maus}	\\
    Theoretical systematics for EFT model \velocileptors &\abacusfirst\ cubic, noiseless & \cite{KP5s2-Maus} \\
    Theoretical systematics for EFT model \folps  &\abacusfirst\ cubic, noiseless & \cite{KP5s3-Noriega} \\
    Theoretical systematics for EFT model \pybird &\abacusfirst\ cubic, noiseless & \cite{KP5s4-Lai}  \\
     Theoretical systematics for EFT model \eftgsm  &\abacusfirst\ cubic, noiseless &  \cite{KP5s5-Ramirez} \\
     HOD-dependent systematics for DESI 2024 & \abacusfirst\ cubic, DR1 &  \cite{KP5s7-Findlay} \\
     Fiducial-cosmology-dependent systematics for DESI 2024 & \abacussecond\ DR1  & \cite{KP5s8-Gsponer}  \\
     Impact and mitigation of the imaging systematics for DESI 2024 &\abacussecond\ DR1, blinded data & \cite{KP5s6-Zhao} \\
     Impact and mitigation of fibre collisions for DESI 2024 & \abacussecond\ DR1, blinded data & \cite{KP3s5-Pinon} \\ 
     Impact and mitigation of the spectroscopic systematics for DESI 2024 &\abacussecond\ DR1, blinded data & \cite{KP3s4-Yu,KP3s3-Krolewski} \\
     Comparison between analytical and \ezmocks\ covariance matrices  & \ezmocks\ DR1 & \cite{KP4s6-Forero-Sanchez} \\
     Analytical covariance matrices of DESI 2024 correlation function multipoles &\abacussecond/\ezmocks\ DR1 & \cite{KP4s7-Rashkovetskyi} \\
     Analytical covariance matrices for power spectrum multipoles for DESI 2024 &\abacussecond/\ezmocks\ DR1  & \cite{KP4s8-Alves}   \\
Tests of the catalog-level blinding method for DESI 2024 & \abacussecond/\ezmocks\ DR1 &  \cite{KP3s9-Andrade}  \\
    \hline
    \end{tabular}
    }
\caption{Summary of the supporting papers for Full-Shape measurements for \desidrone\ and the simulations used for each of them.  
\abacussecond\ and \ezmock\ DR1 simulations are matched to the footprint of the \desidrone\ and account for the variation of completeness with target number density, and an approximate fiber assignment effect. These \desidrone\ mocks were used for the study of systematics that require to simulate the \desidrone\ survey realism.    
The tasks that test the theoretical systematics with the least level of noise use the cubic simulations; in this case, we use \abacusfirst\ cubic instead of \abacussecond, as the former was completed first and was sufficient for our analysis. }\label{tab:tasksims}
\end{table}

\subsection{\abacus }

The large suite of high-resolution gravity-only N-body \texttt{AbacusSummit} simulations \citep{AbacusSummit} were produced with the \textsc{Abacus} N-body code \citep{abacusnbody}. The outputs of the simulations feature realisations of the density field and dark matter halos in cubic boxes with 97 different cosmologies. We used the 25 ``base'' boxes of the \planck\ cosmology, containing $6912^3$ particles within a $(2h^{-1}$Gpc$)^3$ volume which corresponds to a particle mass of $2.1 \times 10^9$~M$_\odot/h$. \footnote{More details can be found at \url{https://abacussummit.readthedocs.io/en/latest/abacussummit.html}} Dark matter halos were identified using the {\sc CompaSO} halo finder \citep{compaso}. We also implemented a post-processing cleaning procedure, following the procedure of \citep{2021Bose}, to eliminate the over-deblended halos in the spherical overdensity finder and to intentionally merge physically associated halos that have initially merged and subsequently separated.
We then populated the dark matter halo catalogues using the halo occupation distribution (HOD) model: for the dark time tracers we used the extended HOD framework implemented in the \textsc{AbacusHOD} code \citep{abacushod} and for bright time tracers we used a halo tabulation method to fit HODs for different absolute magnitude threshold samples \citep{2024Smith}.

We produced a first generation of the \abacus\ mocks (called \abacusfirst) that is based on a very early version of the DESI early data release (DESI-EDR \cite{DESI2023b.KP1.EDR}) to find the best fit HOD model. Having this version early allowed us to start testing the analysis pipeline as we were collecting more data. For dark time tracers, the HOD parameters were found by fitting the galaxy two-point correlation function averaged in angular bins at small scales. More details about the HOD models can be found in \citep{AlamMulti}. Satellite galaxies were distributed using NFW profile fits to the density profile of each halo in the simulation. For the QSO mocks an additional velocity dispersion to account for the non-negligible QSO redshift errors was also included. In total, for a given redshift and galaxy sample, we populated the 25 base $\Lambda$CDM available realisations of 8$h^{-3}$Gpc$^{3}$ box size, which yields a total volume of 200$h^{-3}$Gpc$^{3}$.

We then produced a second version of \abacus\ mocks (called \abacussecond) whose HOD have been fitted to the final DESI-EDR, and some versions of which use the final DESI DR1 survey geometry and account for all the corrections for systematic effects and includes a detailed model of the DESI focal plane effects. We used the final DESI EDR redshift-space two-point correlation function measurements to fit the HOD parameters. We refer the readers to \cite{EDR_HOD_LRGQSO2023, EDR_HOD_ELG2023, 2024Smith} for the exact HOD models and calibration. The final lightcone mocks based on the \abacussecond\ cubic boxes contain the redshifts bins
at $z = 0.1,\, 0.2,\, 0.4$ for the BGS sample; $z = 0.5,\, 0.8,\, 1.1$ for the LRG  sample;  $z = 0.95,\, 1.10,\, 1.325$ for the ELG sample; and $z = 1.10,\, 1.40,\, 1.70$ for the QSO sample. As for the \abacusfirst\ mocks, each single redshift snapshot consists of 25 independent realisations with the same total volume of \abacusfirst.

\subsection{\ezmocks\ }
We have generated large simulation volumes for covariance matrices and pipeline validations using the \ezmocks\ code \citep{Chuang:2014vfa}. The code was calibrated to reproduce two- and three-point clustering of the DESI DR1 tracers on the relevant scales for this analysis, without incurring the expense of a full N-body simulation.
The \ezmocks\ technique has been already used in eBOSS \citep{Zhao2021} and DESI \citep{Zarrouk:2020hha}.  

The first stage of the method consists of generating a dark matter density field and populating galaxy catalogues. The dark matter density field is based on the Zel'dovich approximation \citep{Zeldovich:1969sb}. Then, the resulting density field is populated with galaxies using an effective bias model to account for non-linear evolution and galaxy bias. We refer the reader to \cite{Zhao2021} for a description of the effective bias model. Similarly, as for the \abacus\ mocks,  we produced two generations of DESI \ezmocks\ by fitting the two-point clustering of \abacusfirst\ and \abacussecond\  in order to give matched covariance matrices. 
For the BGS, we produced 1000 realisations of each generation of \ezmocks\ with a box side of 2$h^{-1}$Gpc. For the LRG, ELG and QSO tracers, we use a box side of 6$h^{-1}$Gpc to produce 1000 realisations in order to fit the volume occupied by the \desidrone\ data without any repetition of structure to validate our covariance matrices for the full survey volume.

\subsection{Simulations of DR1} \label{subsec:dr1mocks}

Several types of mock simulations were used to simulate the \desidrone\ LSS dataset \cite{KP3s8-Zhao,hernandez-aguayoetal2020}. 
In order to tailor those mocks to \desidrone\ data, first the box coordinates were transformed into angular sky coordinates and redshifts. Then, the data were sub-sampled as a function of redshift such that the total projected density matches that of the given target sample and the $n(z)$ (after accounting for redshift failures) matches that of the observed DR1 sample \cite{DESI2024.II.KP3}. The corresponding simulated DESI target sample was then cut so that it covers the same sky area as the real data. The same process was also applied to randoms as described in \cref{sec:lss}. The last step was running the mocks through the DESI \textsc{fibreassign} code that determines the assignments for each simulated target, using all of the properties of the observed DESI DR1 tiles. These potential assignments were cut to the same combination of `good' tiles and fibres as the DR1 data samples. Moreover, veto masks were applied following the same process as applied to DR1 data described in \cite{DESI2024.II.KP3}.

Following such a procedure allowed us to reproduce the small-scale structure of the DESI DR1 footprint, however, it does not account for the incompleteness within it. In order to test it, we produced three variations in the fibre assignment completeness of the \abacussecond\ mocks:
\begin{itemize}
    \item The `{\bf complete}' mocks have no assignment incompleteness added and thus can be used as a baseline comparison for understanding the effect of the incompleteness. 
    \item The {\bf `altmtl'} mocks represent our most realistic simulations of the DR1 data. They were created using the process described in \cite{KP3s7-Lasker} to apply the DESI \textsc{fibreassign} code to tiles in the same ordering and cadence in a feedback loop to the target list as occurred for the observed data. The process was demonstrated to perfectly reproduce DESI fibre assignment on real DESI targets, with no approximations.
    \item The `{\bf fast-fibreassign}' mocks emulate the fibre assignment process by sampling from the average targeting probability of the galaxies multiple times, learned from the data as a function of the number of overlapping tiles and local (angular) clustering. The final sample was obtained by recombining the multiple realisations in a way that deliberately creates a small-scale exclusion effect, which approximately reproduces the fibre-collisions pairwise incompleteness. The process is much faster than the \altmtl\ and is described and validated in \cite{KP3s6-Bianchi}.
\end{itemize}
The computation time required for the \altmtl\ mocks prohibits it from being run on all 1000 realisations of the \ezmocks. Thus, we applied only the fast-fibreassign process to these mocks.

All flavours of mocks go through the process of assigning redshifts and weights to randoms in the same way as for the real data samples and are normalised within the same regions (e.g. all integral constraints effects described in \cite{deMattia19IC}, are the same between data and mock LSS catalogues) following the prescription in \cite{DESI2024.II.KP3}.

\section{Methodology}
\label{sec:methods}

This section summarises the various methods used to fit the Full-Shape galaxy clustering. We refer the reader
to the referenced supporting papers for more detail and validations. 

\subsection{Two-point function codes}
\label{subsec:methods-2pt}
The Full-Shape measurements derive from the two-point clustering statistics of the data, the power spectrum in
Fourier space as a baseline and the correlation function in configuration space. The power spectrum measurements are based on an FKP estimator \citep{yamamoto2006,Bianchietal:2015,Hand2017:1712.05834v1} and the correlation function is estimated using the Landy-Szalay estimator \citep{Landy1993}. We refer the reader to \cite{DESI2024.II.KP3} for a more detailed description of the implementation of two-point function codes. The specific codes used are
\textsc{pycorr}\footnote{\url{https://github.com/cosmodesi/pycorr}}  for
correlation functions and
\textsc{pypower}\footnote{\url{https://github.com/cosmodesi/pypower}} for power
spectra. The angular dependence to the line of sight was compressed into Legendre multipoles and we used $\ell=0$ (monopole)
and $\ell=2$ (quadrupole) components for our baseline analysis. For some tests on mocks, we also included the $\ell=4$ (hexadecapole).
The power spectrum and correlation function estimators rely on random catalogues, where the larger the number of random points we have, the smaller the statistical fluctuations are in the random catalogue. We used more than 100$\times$ random points than data points for all power spectrum measurements and more than 50$\times$ for all correlation functions measurements.
A large number of elements in the random catalogue makes the computational cost of the correlation function (based on pair counts) more expensive, unlike for the power spectrum, for which the computational cost increases with the number of grid elements and not catalogue elements. For this reason, we can use a larger number of random elements in the power spectrum computation than in the correlation function. Moreover, as mentioned in \cref{sec:catalog}, the redshift distribution of randoms matches exactly that of data by construction, meaning that radial modes in the measured power spectrum or correlation function are nulled, requiring the radial integral constraint (RIC) \cite{deMattia19IC} to be included in the modelling. The modelling of the RIC effect using \ezmocks\ realisations is described in \cite{DESI2024.II.KP3}.

Given that the clustering statistics are consistent for both Galactic caps, we combine the clustering measurements in order to determine the average clustering of the full sample, including all components. In Fourier space, the combination was performed by averaging the two power spectra and weighting them by their respective effective volume (see \cite{DESI2024.II.KP3} for more details). 
In configuration space, the measurements were combined by summing their weighted pair counts computed in each region independently. 

We weigh the galaxies to account for the selection function and to optimally measure two-point statistics. The corresponding weights are summarised in \cref{sec:catalog} and we refer the reader to \cite{DESI2024.II.KP3} for a more detailed description, including the effect of an angular integral constraint (AIC) introduced
by the use of imaging systematic weights. In \cref{subsec:sys-imaging} and \cref{subsec:sys-spectro}, we quantify the impact of the imaging and spectroscopic weights, respectively, on the cosmological parameters 
and determine the systematic error budget associated with these observational systematic effects.

In order to mitigate the imprinted signals on small scales due to the DESI fibre assignment, we remove the contribution of galaxy pairs at small angular separations ($<0.05$ deg) from the two-point statistics for both the correlation function and power spectrum following the methodology in \cite{KP3s5-Pinon}. This $\theta$-cut method affects the survey window, giving a window matrix that is very non-diagonal and thus mixes small and large-scale modes. Although this did not have a noticeable effect on the DESI analysis \cite{KP3s5-Pinon}, it is generally undesirable and so we implement a procedure of \emph{rotating} the data vector, window matrix and covariance to obtain a more diagonal window. 
The procedure is described in \cite{DESI2024.II.KP3} for the DESI DR1 analyses.
In \cref{subsec:sys-fibre}, we study the performance of the $\theta$-cut method in recovering unbiased cosmological constraints from our realistic \abacus\ \desidrone\ mocks and we determine its corresponding contribution to the total systematic error budget.
For a full description of the two-point clustering measurements, we refer the reader to \cite{DESI2024.II.KP3}. Throughout this paper (and also those following the cosmology interpretation \cite{DESI2024.VII.KP7B,ChaussidonY1fnl}) we employ the power spectrum measurements named as `$\theta$-cut+rotated+RIC+AIC' in Section 10.1.2 of \cite{DESI2024.II.KP3}.

\subsection{Compression vs Full Modelling}
\label{subsec:methods-compression}

We consider two approaches to perform a Full-Shape cosmological inference: Full Modelling and compression.
The Full-Modelling approach constrains the cosmological parameters of a model by creating a complete model of the signal. Given a cosmological model with a set of free parameters (e.g., $A_s$, $H_0$, $\Omega_m$, etc.) a linear matter power spectrum is generated using a Boltzmann code and is then used to generate the non-linear redshift-space galaxy power spectrum that is fitted to the data.
The compression approach instead defines a set of parameters that directly affect features in the clustering (eg, the BAO position along and across the line-of-sight, the growth of structure parameter, and the tilt of the power spectrum at some pivot scale), which are usually defined relative to a chosen fiducial (fixed) cosmological model. This set of physical (or compressed) parameters affects several features of the shape of the fiducial power spectrum (its amplitude, its BAO peak position, etc), converting it into a power spectrum with revised cosmological parameters. This set of compressed parameters can be interpreted as combinations of the usual cosmology parameters of the model. 

The Full-Shape galaxy clustering measurements performed by the BOSS \cite{SDSS-DR12-cosmology} and eBOSS \cite{SDSS-DR16-cosmology} collaborations were obtained by first compressing the cosmological information contained in the power spectrum or two-point correlation function into three parameters per redshift bin: the BAO peak position along and across the line-of-sight and the logarithmic growth of structure parameter times the amplitude of dark matter fluctuations, $f\sigma_8$. These sets of parameters per redshift bin were in turn translated into constraints on cosmological parameters that were reported in the cosmology interpretation papers.

Several works over the last years have shown that avoiding this intermediate compression step and fitting directly the full power spectrum signal or correlation function can lead to much tighter constraints on cosmological parameters \cite{Ivanov20,dAmico20,Chen2022,Simon23a,Chudaykin23} if no external dataset was added. In order to bridge the gap between the classic approach ({Standard compression}) and the direct-fitting (or Full-Modelling) approach, an extended compression technique has been proposed called ShapeFit \cite{brieden2021}, which has obtained a statistical constraining power comparable to Full-Modelling analysis when applied to BOSS and eBOSS data \cite{Brieden2022}. In this section, we review the three techniques and show a comparison performed on \desidrone\, Abacus cut-sky mocks.
This section heavily relies on five supporting papers \cite{KP5s1-Maus, KP5s2-Maus,KP5s3-Noriega,KP5s4-Lai,KP5s5-Ramirez}.

\begin{figure}[htb]
    \centering
    \includegraphics[width=1.0\linewidth]{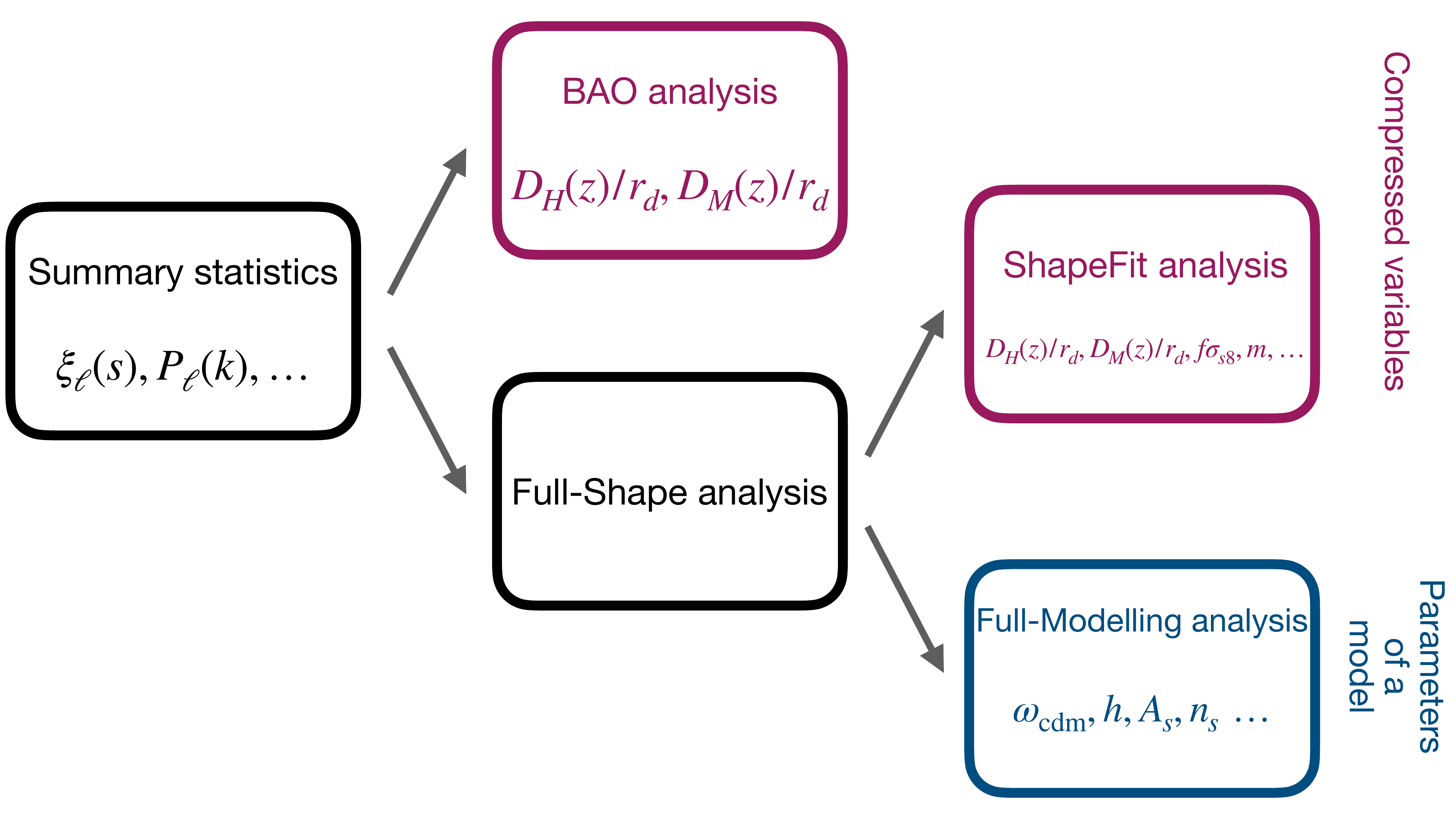}
    \caption{Schematic diagram displaying the different types of analyses we perform or refer to in this paper. The magenta boxes represent the analyses in terms of compressed variables (BAO and ShapeFit), whereas the blue boxes represent the analysis whose variables are the ones of the model assumed (Full Modelling). The Standard compression is not explicitly included as it is considered a particular case of the ShapeFit analysis. Note that both ShapeFit and Full Modelling are considered different types of Full-Shape analyses, in contrast to the BAO analyses, which do not exploit the broadband shape for cosmology inference.}
    \label{fig:FSanalysis}
\end{figure}

\subsubsection{{Full Modelling}}\label{subsubsec:FM}

The Full-Modelling approach extracts all the cosmological information contained in the power spectrum by directly fitting models for $P(k)$ to data. The information extracted consists of the BAO and RSD signals together with other features, such as the power spectrum slope at different scales. At each step of the analysis, an Einstein-Boltzmann code is used to generate the linear power spectrum, and a non-linear theory code evaluates the one-order loop corrections to eventually produce the non-linear galaxy power spectrum. In this approach, we keep the cosmology used to convert redshifts to comoving distances fixed. In order to account for the distortions that this causes if this choice does not match the model being tested --- the AP signal --- we need to apply a coordinate transformation to the model. This is done by distorting the true scale, $k^\prime$, into the observed scales $k_\parallel$ and $k_\perp$ along and across the line-of-sight, with the scaling parameters, 
\begin{eqnarray}
   q_\parallel &=&  k_\parallel/{k_\parallel}^\prime \,, \\
 q_\perp    &=&   k_\perp/{k_\perp}^\prime \,.
\end{eqnarray}
These $q-$scaling parameters are related to the Hubble distance, $\Dh(z)\equiv c/H(z)$ with $H$ being the Hubble parameter, and the comoving angular diameter distance, $\DM$ through, 
\begin{eqnarray}
     q_\parallel(z) = \Dh(z)/\Dh^{\rm fid}(z)\, , \\
        q_\perp(z) = \DM(z)/\DM^{\rm fid}(z)\, .
\end{eqnarray}
The quantities without a superscript are determined using the model cosmology, and the quantities with `fid' are those evaluated at the fixed cosmology used to generate the 3D Cartesian galaxy catalogue. Note that for the Full-Modelling approach, the scaling parameters $q_\parallel$ and $q_\perp$ are internally derived parameters to correct for a fixed catalogue cosmology, unlike for the compression approaches, where these parameters are parameters of the compression.

\subsubsection{Standard compression} \label{subsubsec:standard_compression}

We call the approach that was widely used by BOSS and eBOSS collaborations {Standard compression}; it compresses all the summary statistics data into three parameters: the position of the BAO peak along and across the line of sight, and a redshift-space distortion parameter. 

The observed location of the BAO peak can be expressed as a re-scaling of that in the fiducial cosmology used to create the template. We need to allow for the intrinsic change in the sound horizon distance at drag epoch, $\rd$, between the cosmologies and the stretching due to the AP effect. We define two scaling parameters,
\begin{eqnarray}
    \alpha_\parallel(z) &=& \frac{\Dh(z)/\rd}{[\Dh(z)/\rd]^{\rm fid}}\,,\\
        \alpha_\perp(z) &=& \frac{\DM(z)/\rd}{[\DM(z)/\rd]^{\rm fid}}\,.
\end{eqnarray}
Although these scaling parameters regulate the position of the BAO scale, they are sometimes applied to the full shape (and not only to the BAO oscillatory part), as the shifts due to $\rd^{\rm fid}/\rd$ are small compared with those due to $D/D^{\rm fid}$ for cosmologies limited by the statistical precision of actual surveys (see discussion in section~5.2 of \cite{brieden2023}).
These two scaling parameters are usually re-parametrised into the following alternative basis,
\begin{eqnarray}
    \alpha_{\rm iso}(z)&\equiv& [\alpha_\parallel(z) \alpha_\perp(z)^2]^{1/3} = \frac{\DV(z)/\rd}{[\DV(z)/\rd]^{\rm fid}}\,,\\
    \alpha_{\rm AP}(z)&\equiv&{\alpha_\parallel(z)}/{\alpha_\perp(z)}=\frac{\Dh(z)/\DM(z)}{[\Dh(z)/\DM(z)]^{\rm fid}}\,,
\end{eqnarray}
where $\DV$ is usually referred as the isotropic BAO distance, $\DV(z)\equiv [z\Dh(z) \DM(z)^2]^{1/3}$, and  $\alpha_{\rm AP}$ is the Alcock-Paczynski parameter \cite{1979Natur.281..358A}. Throughout this paper, we will often use this alternative parametrisation, unless we state otherwise, as in \cite{DESI2024.III.KP4}.

The third parameter defined in the Standard compression is actually composed of two variables: the logarithmic growth rate of structure $f$, and the amplitude of dark matter fluctuations smoothed on a $8\hMpc$ scale, $\sigma_8$. When only the two-point statistic is used in the analysis these two variables are highly correlated and it is common practice to combine them into a single parameter, $f\sigma_8$.

\subsubsection{ShapeFit}\label{subsubsec:SF}

Standard compression analysis has been demonstrated to be a powerful method for extracting and analyzing relevant information contained in the redshift galaxy catalogues, as it focuses on the strongest signals in the clustering, coming from BAO and RSD.  However, this approach is not able to capture other information within the transfer function. This limitation is mitigated by adding CMB external datasets, which constrain this information more strongly than LSS data can. However, when the LSS analysis is considered alone (or with just a few CMB-like priors), Standard compression is not lossless, and the constraints significantly differ from those obtained by the Full-Modelling analysis. 

ShapeFit compression is based on the same idea as Standard compression: take a fiducial template and modify its shape through a series of parameters. In this case, in addition to the scaling parameters, $\alpha_\parallel$ and $\alpha_\perp$ and the logarithmic growth of structure parameter, $f$, Shapefit includes two shape parameters, $m$ and $n$, which account for modifications of the broadband shape of the linear matter power spectrum in the following way,
\begin{equation}
    P^\prime_{\rm lin}(k) =P^{\rm fid}_{\rm lin}(k)\exp\left\{\frac{m}{a}\tanh\left[a\ln\left(\frac{k}{k_p}\right)\right]+n\ln\left(\frac{k}{k_p}\right) \right\}.
\end{equation}
where $k_p$ and $a$ values have been kept fixed at $k_p=\pi/r_d^{\rm fid}\simeq0.03\;h^{\rm fid}{\rm Mpc}^{-1}$ and $a=0.6$. Although other values for these parameters could have been chosen, this specific choice optimises the information compression in state-of-the-art cosmologies\footnote{In this sense, if the true cosmology is very different from the standard $\Lambda$CDM model by choosing this particular choice for $k_p$ and $a$ we would not be biasing the shape parameters $m$ and $n$, but make them less optimal to capture all the shape information.}.

Note that the $m$ (or Shape) parameter is the maximum slope at the chosen pivot scale $k_p$. This value has been selected approximately at the matter-radiation equality scale, so $m$ is able to capture the physics at that epoch. The $n$ parameter typically is able to capture information on the primordial scaling of the power, and its deviation from scale-invariance, playing the role of the spectral index parameter, $n_s$, in most common cosmology interpretations.
In this paper, we will only consider varying one of the two shape variables: we will vary $m$ keeping $n$ fixed. Later, in the interpretation step $m$ can be seen as if it were $m+n$. This re-interpretation is an excellent approximation due to the strong anti-correlation displayed by these two variables (see fig~13 of \cite{KP5s3-Noriega}).

When comparing the standard compression analysis and the ShapeFit analysis, we find that the cosmological interpretation of the scaling parameters, $\alpha_\parallel$ and $\alpha_\perp$, is the same for both approaches. However, the interpretation of $f\sigma_8$ is slightly different, as for ShapeFit the amplitude of dark matter perturbations is not regulated by $\sigma_8$, but rather by a slightly different parameter, $\sigma_{s8}$. This is in part because $\sigma_8$ becomes a function of $m$, but also because of the ambiguity of defining a scale of $8\hMpc$ when the cosmology that defines $h$ is in principle unknown.\footnote{See also an alternative approach through the definition of the smoothing scale in units of Mpc, instead of $\hMpc$ \cite{Sanchez:2020}} Thus, we refer to the definition of $\sigma_{\rm s8}$ as \cite{brieden2021,brieden2023},

\begin{equation}
    \sigma_{s8}=\sigma_{\rm s8}^{\rm fid} A^{1/2} \exp\left\{\frac{m}{2a}\tanh\left[a\ln\left(\frac{\rd^{\rm fid}}{s_8}\right)\right]\right\}\, ,
    \label{sigma_s8_m_correction}
    \end{equation}
where $\sigma_{\rm s8}^{\rm fid}$ is the usual definition of $\sigma_8$ at the fiducial cosmology of the fixed template, $s_8\equiv 8\,h^{-1}{\rm Mpc}$ and $A\equiv A_{sp}/A_{sp}^{\rm fid}$, where $A_{sp}$ is the fluctuation amplitude given by\footnote{Note that the argument of $P_{\rm lin,nw}$ does not depends on $r_d^{\rm fid}$ as it cancels out with the $r_d$ within $k_p$. Also, the pre-factor $(r_d^{\rm fid}/r_d)^3$ comes from the fact that ultimately we are interested in the amplitude of the scale-invariant power spectrum ($\Delta(k)\propto k^3P(k)$) whose amplitude remains invariant to a rescaling of $k$ with the factor $r_d/r_d^{\rm fid}$.},
\begin{equation}
    A_{sp}=\left(\frac{\rd^{\rm fid}}{\rd}\right)^3 P_{\rm lin\, nw}\left(k_p \frac{\rd^{\rm fid}}{\rd}\right)\, ,
\end{equation}
where the sub-index `nw' stands for the non-wiggle (or no-BAO) linear dark matter power spectrum. Note that both $r_{\rm d}^{\rm fid}$ and $s_8$ are expressed in units of $h^{-1}{\rm Mpc}$, where $h$ stands for the value of the chosen fiducial template cosmology. We follow the usual practise for the power spectrum analysis of keeping the amplitude of the template fixed and only fitting for $f$. The obtained $f$ parameter is reparametrised later by being multiplied by the fixed $\sigma_{s8}^{\rm fid}$ parameter, which provides an estimate for $f\sigma_{s8}$. This procedure is exact at first-order of perturbation theory, where $f$ and $\sigma_{s8}$ only enter as a product. At higher-order corrections, this is no longer exact but turns out to be a very good approximation for the precision being considered here. For this reason, when reporting the raw results on mocks and data, we usually will quote $f/f^{\rm fid}$, keeping in mind that this parameter will need to be parametrised into $f\sigma_{s8}$ (and eventually in $f\sigma_8$) for any cosmological interpretation. If we want to transform this $f\sigma_{s8}$ parameter into the usual $f\sigma_8$ we need to keep in mind the rescaling of scales described by the $q_{\rm iso}$ parameter and follow a procedure similar to the one described in equation~40 of \cite{Gil-Marin2020:2007.08994v2}.
Table 1 in \cite{KP5s3-Noriega} presents a compact comparison of the main characteristics of these three methodological approaches.

Both the Standard compression and ShapeFit schemes (and any fixed-template approach) assume a fiducial cosmology to compute the non-linear corrections to the linear power spectrum (the one-loop terms), unlike the full modelling approach. In practice, in the ShapeFit case, we apply an approximation to include a cosmology dependence on these correction terms, so they do not need to be recomputed at each cosmology step. However, this correction cannot be included in an exact way without fully recomputing the full one-loop integrals.
The approximation we apply is described in appendix D of \cite{brieden2020}, and more specifically for the formalism of \folps\ in appendix B of \cite{KP5s3-Noriega}. This approach has been demonstrated to be an excellent approximation, and to have a negligible effect on the final parameter constraints (see fig.~20 of \cite{KP5s3-Noriega}).

All the perturbation theory codes described above, with the compression or direct-fit options, have been included in the \texttt{desilike} package\footnote{\href{https://github.com/cosmodesi/desilike}{https://github.com/cosmodesi/desilike}}, which enables an easy switch between codes and setups for comparison.

\cref{fig:FSanalysis} displays a chart of the different types of analyses described above. The standard compression technique is not included in the diagram as it is considered a particular case of the ShapeFit compression technique (the one setting $m=n=0$). Also, note that we refer to both Full Modelling and ShapeFit techniques as part of a more general group of analyses we name `Full-Shape'.

\subsection{Perturbation Theory models}
\label{subsec:methods-eftmodel}

Each of the analysis methods presented above, Standard compression, ShapeFit and Full Modelling, needs to be implemented in a specific perturbative model. The role of the perturbative model is to transform from a linear power spectrum of the dark matter field to a non-linear power spectrum of the galaxy field, including the redshift-space distortion effect. In this paper, we focus on four different implementations of perturbative codes based on the effective field theory: three based on Fourier Space, and one in configuration space. Here, we briefly describe the main features of these codes. For more details and further tests, we refer the reader to our supporting papers \cite{KP5s1-Maus,KP5s2-Maus,KP5s3-Noriega,KP5s4-Lai,KP5s5-Ramirez}.

All the perturbation theory models used in this paper have four types of free parameters: the intrinsic cosmology parameters, the galaxy bias parameters, the stochastic parameters and the counterterm parameters. The cosmology parameters are those that define the cosmology and can be the parameters of a model (e.g. $\Lambda$CDM) or a set of compressed parameters, as for the ShapeFit compression scheme. The galaxy bias parameter defines the relation between the galaxy and the dark matter density field. In addition to those, the stochastic parameters aim to capture some extra non-linear physics in the dark matter galaxy connection, such as the halo exclusion effect, conformity\footnote{We employ the concept of conformity as a property of galaxies of a different type to tend to occupy different haloes. See for e.g., \cite{Yuan2023}.} and the Fingers-of-God effect. Finally, the counterterms aim to capture physical processes associated with dark-matter structure formation beyond the truncation scale used when evaluating the one-loop terms in the perturbative approach. Generically, we can write these models as \cite{KP5s2-Maus,KP5s3-Noriega},
\begin{equation}
P_{s,g}(k,\mu) = P^{\rm PT}_{s,g}(k,\mu)+(\alpha_0+\alpha_2\mu^2+\alpha_4\mu^4+\ldots)k^2P_{s,{\rm lin}}(k,\mu)+({\rm SN}_0+{\rm SN}_2k^2\mu^2+{\rm SN}_4k^4\mu^4+\ldots),
\end{equation}
where the term $P^{\rm PT}_{s,g}$ stands for the perturbation theory term for biased tracers in redshift-space (as it is in standard perturbation theory or similar), the second term containing the $\alpha_i$-counterterms aim to correct for the trucation scale implemented in the perturbation theory term, and the last term aims to capture the small-scale galaxy physics via the stochastic terms, $\rm SN_i$.

\subsubsection{{Velocileptors}}\label{subsubsec:velocileptors}

The Fourier-based code \velocileptors\, \cite{KP5s2-Maus} implements the redshift-space power spectrum in one-loop, Lagrangian Perturbation Theory (LPT) with resummation up to a set scale $k_{\rm IR}$ \cite{Chen21}. The biasing model includes four bias parameters: $b_1$, $b_2$, $b_s$, and $b_3$. The linear and non-linear galaxy bias, $b_1$ and $b_2$, respectively, enter the galaxy overdensity parameter by multiplying the powers of the linear density field. The non-local tidal bias parameter, $b_s$, multiplies the initial shear field and, due to degeneracies between terms, the third-order non-linear bias contributions are combined into a single operator with coefficient $b_3$. In addition, the model also includes three stochastic parameters and three counterterm parameters. 
\velocileptors\ also includes an Eulerian Perturbation Theory\footnote{This option is also referred to in some papers or codes, such as \texttt{desilike} as Resummed Eulerian Perturbation Theory (REPT).} (EPT) module, with the terms organised via an Eulerian bias expansion as in \cite{McDonald2009} and IR resummation performed via the wiggle no-wiggle split \cite{Vlah2015}. In the \desidrone\ analyses we will sample the Eulerian bias parameters as reparametrisations of the Lagrangian ones following the convention in \cite{Chen20}, such that the nonlinear bias priors are centred at their advected predictions corresponding to zero Lagrangian bias, in order to harmonise the prior choices between the EPT and LPT analyses. Note however that in other works, such as \cite{Chen20,Chen21}, the Eulerian flavour of \velocileptors\ was fit to data without this reparametrisation. Details of the priors and constraints among these parameters are discussed in \cref{subsec:methods-baseline}.

\subsubsection{{FOLPS}}\label{subsubsec:folps}

The Fourier-based code \folps\ \cite{KP5s3-Noriega} efficiently implements a one-loop EPT model in Python, using built-in \texttt{numpy} and \texttt{scipy} functions. Additionally, \folps\ provides a \texttt{JAX} implementation, which speeds up computations by a factor of 10 and enables Graphic Process Units (GPU) usage. \folps\ has the added value of accounting for the presence of massive neutrinos by modifying the standard Einstein-deSitter kernels,
which properly incorporate the effects of the free-streaming scale introduced by massive neutrinos \cite{Aviles:2021que,Noriega:2022nhf}. 
Furthermore, \folps\ provides transformation equations between different biasing schemes, enabling the use of either the bias basis $b_1$, $b_2$, $b_{s^2}$, and $b_{3nl}$ introduced in \cite{McDonald2009}, or the basis $b_1$, $b_2$, $b_{\mathcal{G}_2}$, and $b_{\Gamma_3}$ from \cite{Assassi:2014fva}.

The code also integrates EFT counterterms and stochastic parameters. In \cref{subsec:methods-baseline}, we discuss the priors and constraints applied to these parameters. 
We also note that \folps\, as well as the \velocileptors\ in their Eulerian perturbation theory mode are very similar to \texttt{ClassPT} \cite{classpt}, which we do not include in this paper.

\subsubsection{{PyBird}}\label{subsubsec:pybird}

The Fourier-space-based code \pybird\, \cite{Pybird_damico} implements a one-loop EPT model in Python  
in a similar way as the other codes described above. 
The implementation of \pybird\ used in this work \cite{KP5s4-Lai}  has a galaxy bias model which relies on the four parameters, $b_1$, $b_2$, $b_3$ and $b_4$, as described in  \cite{perkoetal:2016}, which has two different parametrization referred in the literature as `east-cost' and `west-cost' parameterization \cite{Simon23b}.
These bias terms are different to those of \folps\, and \velocileptors\ in terms of parametrisation, but equivalent upon proper mathematical description. As for the other two EFT-based codes, \pybird\ has three counterterms and three stochastic terms that describe physics beyond the cutoff scale and galaxy formation physics, respectively. Prior to Infrared (IR) resummation \pybird, \folps, and \velocileptors\ are essentially equivalent in their theoretical predictions. The IR resummation schemes differ between codes but we emphasise that these differences appear at two-loop order terms and that for the \desidrone\ precision these differences do not affect the actual cosmological-parameter posteriors.

As for the rest of the codes, we discuss the priors and constraints among all parameters in \cref{subsec:methods-baseline}.
 
\subsubsection{{EFT-GSM}}\label{subsubsec:eftgsm}

The \eftgsm\, \cite{KP5s5-Ramirez}\, is a Gaussian Streaming Model configuration-space-based code relying on LPT. The code has three bias parameters, $b_1$, $b_2$ and $b_{s^2}$, and two additional counterterm parameters, accounting for physics beyond the cutoff scale. This model does not account for the full set of non-linearities or small-scale dependencies like the other models previously described. 

\subsection{Combining BAO and Full-Shape galaxy two-point clustering analyses}
\label{subsec:methods-BAO-FS}

The Full-Shape techniques described above are able to capture information from the BAO, RSD, AP and in general any broadband shape imprinted in the two-point statistic. Conversely, the BAO analysis \cite{DESI2024.III.KP4} is focused on only extracting the BAO feature. Thanks to the reconstruction technique \cite{KP4s4-Paillas, KP4s3-Chen}, we are able to generate an additional data set, the post-reconstruction (post-recon) catalogue, where the BAO signal can be measured with greater significance in precision and accuracy~\cite{Ding2017:1708.01297v2}. The reconstruction technique is designed to undo the effect of bulk flows caused by peculiar velocities, which sharpens the significance of the BAO peak in the reconstructed two-point correlation function. This ‘Gaussianisation’ process of the original galaxy catalogue is the reason why the pre- and post-recon catalogues contain significantly different information. 
The two-point statistics of the post-recon catalogue hold information coming not only from the two-point statistics but also from its three- and four-point statistics \cite{wangetal24}. Therefore, in order to use both pre-recon full shape information and post-recon BAO information we need to account for the correlation between these two catalogues.
  
One option is to combine the pre- and post-recon information at the summary statistic level. In this case, we need to estimate the covariance matrix between the summary statistics of the pre-recon catalogue and that of the post-recon catalogue. Ref \cite{Chen22} used this approach and combined the pre-recon power spectrum with the post-recon correlation function from BOSS DR12 data to improve the cosmological constraints. This approach has the advantage that it naturally captures all the cosmological effects on the shape of the power spectrum and correlation function, without having to go through a compression step that could result in a non-Gaussian profile for the compressed parameters (such as the scaling BAO parameters).
Another option is to combine the usual compressed BAO variables extracted from the post-recon catalogue, $\alpha_{\rm iso}$ and $\alpha_{\rm AP}$ with the power spectrum (or correlation function) bins measured from the pre-recon catalogue, as done by \cite{Philcox22} also with BOSS DR12 data.  This obviates the need to compute the post-reconstruction statistics during the likelihood evaluation. A third option would be to combine the compressed post-recon BAO variables with the compressed pre-recon variables, in this case using ShapeFit. This was done in \cite{Brieden2022} in both BOSS and eBOSS data, and it gives the most compact summary of the data. In all these cases the covariance matrix could be easily estimated from mocks, although analytical approaches could in principle be applied. A detailed comparison of these different approaches was presented in \cite{Gil-Marin2022} for the eBOSS LRG sample, showing that these three approaches showed minor differences of order 5-10\% in the size of the error bars.

In this paper, we opt to combine the post-recon BAO information at the compressed parameter level, $\alpha_{\parallel}$ and $\alpha_{\perp}$, with either the power spectrum bins in the pre-recon catalogue, for the Full-Modelling case; and at the level of the compressed parameters for the  ShapeFit case.
This way we compress the full BAO post-recon information in just two parameters that are added to our pre-recon data-vector for Full Modelling, or to the four-compressed parameters for ShapeFit.
After combining them, we rotate the scaling parameter base back to $\alpha_{\rm iso}$ and $\alpha_{\rm AP}$.

\subsection{The dependence on nuisance parameter priors}
\label{subsec:methods-projection}

One of the advantages of perturbation theory models such as those presented above is that they produce robust inferences in the case of a very informative dataset (i.e., the power spectrum of a very large volume). A drawback is that they have many nuisance parameters (i.e., bias, stochastic and counterterm) that can be partially degenerate with the cosmological parameters. If a signal in the data could be explained by a complex bias model rather than, or in addition to, changes in the underlying cosmology, the models will explore this possibility in the fits.  This inevitably generates projection effects when we marginalise over the nuisance parameters. Consequently, the choices of parameterisation and priors for these nuisance parameters affect the marginal posteriors of the cosmological parameters. These projection effects are well-known in high dimensional parameter inference and have been studied in the context of galaxy clustering with the EFT formalism \cite{Handley19a,Handley19b,Lemos21,GomezValent22,Simon23b,Holm23,Maus23,Noriegaetal24}, but also in the context of photometric data for weak lensing analyses \citep{Hadzhiyska23}. We can distinguish two types of prior-dependent effects.
\begin{itemize}
    \item The \textit{prior weight effect} (PWE) appears when the location and width of the prior on some parameter pulls away from the value preferred by the data; which can shift the posteriors.
    \item The \textit{prior volume effect} (PVE) or projection effect is related to the marginalisation over nuisance parameters given the priors we impose on them; for non-Gaussian likelihoods with partially degenerate parameters this can shift the peak of the marginal posterior away from the most-likely value
\end{itemize}
 For a toy model example of the prior volume effect and how it causes the marginal posterior to be away from the truth, we refer the reader to figure 15 of our supporting paper \cite{KP5s2-Maus}. It is worth noticing that PWE only appears when the data prefer parameter values in the tails of the prior, whereas it vanishes in the case of broad, uniform priors (or very wide Gaussian priors). In order to quantify the projection effects, in addition to the marginalised constraints we also report the maximum a posteriori (MAP) value which informs both about the likelihood and the prior.\footnote{Note that the MAP can also be understood as a Maximum Likelihood (ML) estimate which also includes the additional information available through prior knowledge.} Therefore, on one hand, comparing the MAP values of different choices of priors (flat uninformative priors vs Gaussian physically-motivated priors) informs us about PWE and, on the other hand, comparing the mean of the posterior with the MAP for a given choice of priors informs us about the PVE. In particular, a very tight prior on a nuisance parameter will reduce the PVE, but will potentially increase the PWE if this prior disfavours the underlying true (but unknown) value of the parameter; also a very loose prior will reduce the PWE but will maximise a potential PVE if the data are not sufficiently informative. We choose to find a sweet spot between these two effects as we describe below. Recently, some authors \cite{EFTpriors1,EFTpriors2} have proposed the alternative approach of employing a large and diverse set of galaxy mocks to derive highly informed priors and correlation among nuisance parameters, in such a way that the PVE are reduced without introducing systematic shifts caused by PWE. However, as a drawback, the results will depend on the galaxy-halo model used to derive the prior.

In this analysis, we choose to take the conservative approach and define Gaussian priors on non-cosmological (nuisance) parameters based on tests on \abacus\ \desidrone\ complete mocks and synthetic (noiseless) data. 
Below we summarise our findings: 

\begin{figure}[htb]
    \centering
    \includegraphics[width=0.95\linewidth]{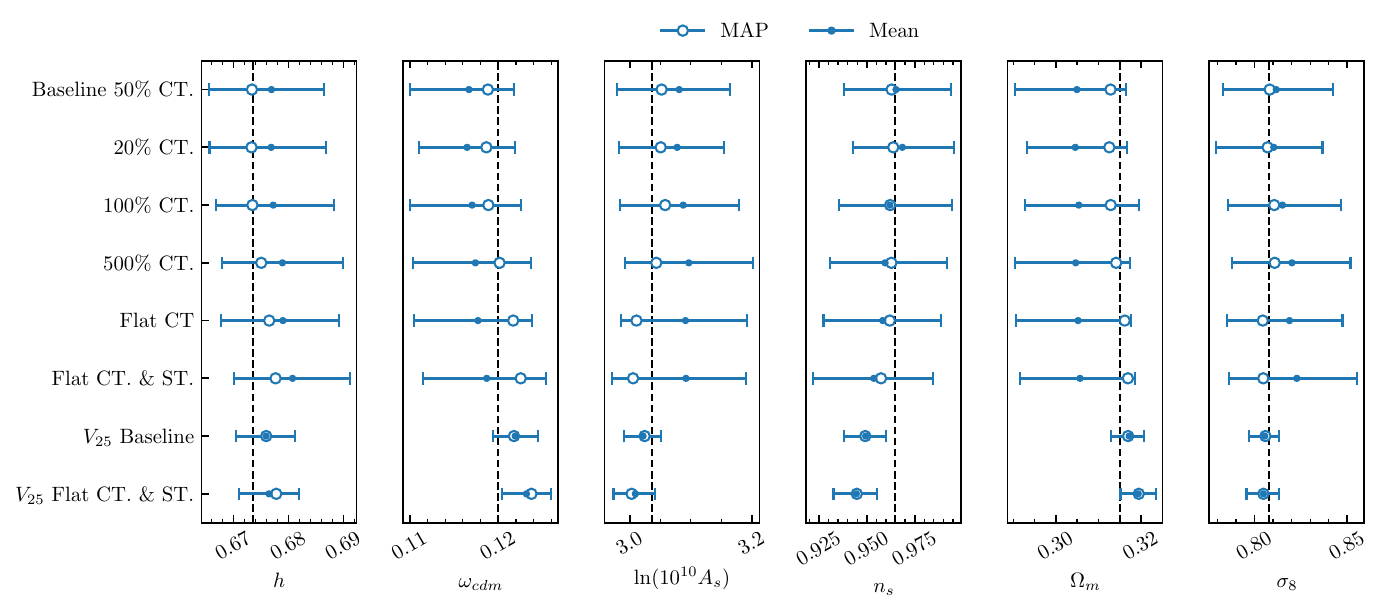}
    \caption{Constraints on $\Lambda$CDM parameters from the joint fit of the six tracers of \desidrone\, from the Abacus complete mocks (mean of 25 lightcone \desidrone\, realisations). We employ the Full-Modelling approach (\velocileptors\, with the LPT setup) using different choices of priors on counterterms (CT) and stochastic terms (ST), as displayed. For more details on the baseline setup, see \cref{subsec:methods-baseline}. The filled(empty) symbols represent the mean(MAP) value. The vertical dashed lines display the expected value given the known cosmology of these mocks. For a given case, the difference between filled and empty markers is caused by the prior volume effect (PVE) and the {\it relative} difference between MAP values across different choices of priors is caused by the prior weight effect (PWE). The first five rows correspond to analyses with the covariance corresponding to the complete \desidrone\ volume (across the six redshift bins listed in \cref{tab:Y1data}), whereas the last two rows (labelled with $V_{25}$) display the analyses corresponding to the same covariance rescaled to a volume of $25\times$\desidrone.  This allows to illustrate that when the data are highly informative (the volume is large) the projection effects (both PWE and PVE) are largely reduced.}
    \label{fig:priors}
\end{figure}

\begin{enumerate}
    \item {\bf Choice of galaxy bias parameterisation}: We use the basis $(1+b_1)\sigma_8(z)$, $b_2\sigma_8^2$, $b_s\sigma_8^2$ instead of $b_1$, $b_2$ and $b_s$.  This set of parameters is much more closely attuned to the clustering that the data can constrain and this helps to reduce the projection effects that would favour low values of $\sigma_8$ in marginalized posteriors. We refer to Refs.~\cite{KP5s2-Maus} for a further discussion on the choice of this parametrization. 
    
    \item {\bf Choice of priors on stochastic terms and counterterms}: 
    We set a Gaussian prior around the stochastic (${\rm SN}_0$ and ${\rm SN}_2$) parameters motivated by physical considerations (see eq. 3.6 of ref~\cite{KP5s2-Maus} for a definition).  We specify the stochastic parameters as multiples of the Poisson value or multiples of this times a characteristic velocity dispersion (see table 2 of \cite{KP5s2-Maus}). 
    For the counterterm parameters ($\alpha_0$ and $\alpha_2$; see also eq. 3.6 of ref~\cite{KP5s2-Maus} for a definition), we set a Gaussian prior based on the relative correction that the counterterm makes at the minimum scale considered $k_{\rm max}=0.20\,h{\rm Mpc}^{-1}$, with standard deviation set at 50\% of the correction to the linear signal of the power spectrum required to match a $1\sigma$ in the data (see \cref{app:priors}). This corresponds to a Gaussian prior centred around zero with width $12.5\,h^{-2}\text{Mpc}^2$. In \cref{subsubsec:priors} we explain the motivation for the choice of priors. 
    
    \item {\bf Removing the hexadecapole $\ell=4$}: We decided not to include the hexadecapole because it causes stronger prior weight effects. This behaviour is not unexpected; the hexadecapole requires additional modelling parameters that are very poorly constrained, and their marginalisation then leads to large projection effects in the cosmological parameters. We also find that for some cosmology models (such as $\Lambda$CDM) adding the hexadecapole does not improve the statistical errors of the parameters of interest. This is because the internal priors among parameters that the cosmology model imposes are tighter than the constraining power that the hexadecapole brings. For some other models (such as $w_0w_a$CDM) adding the hexadecapole does improve the constraints from galaxy clustering alone. However, because of projection effects, in practice, these models will be only considered when combining DESI data with external datasets, in which case the hexadecapole contribution is again minor.
\end{enumerate}

\subsubsection{Prior choices validation tests}\label{subsubsec:priors}
Our prior choices on the stochastic and counterterms derive from a series of validation tests against simulations summarised in \cref{fig:priors}, where we explore the constraints on $\Lambda$CDM parameters from the joint fit of all \desidrone\, redshift bins (6 in total, see \cref{tab:Y1data}) on the mean of the 25 \abacus\ \desidrone\, complete mocks. The empty symbols represent the MAP values and the filled symbols are the mean, as indicated. The dashed vertical lines indicate the expected value given the known cosmology of these mocks. The different rows describe different prior choices, quantifying the size of the prior in terms of the deviation with respect to the signal at the maximum scale used for fitting (see \cref{app:priors} for more details).
A close match between the MAP values of different prior choices (and the dashed vertical line) indicates a low PWE and the difference between the mean and the MAP quantifies the PVE.

For the counterterms, we choose priors that ensure that their contribution to the rest of the signal must be ``small'' on these scales. Since we do not have a robust estimate on their size, we choose a conservative prior width of $50\%$ with respect to the linear part of the signal while noting that using a narrower width (e.g.\ $20\%$) does not meaningfully change our constraints while much wider priors will tend to induce substantial projection effects in addition to allowing the model parameters to take on unphysical values. For the stochastic parameters we expect that their values must be relatively close to those motivated by halo-galaxy connection physics, and we explore different prior widths around Poissonian prediction for ${\rm SN}_0$, and a satellite velocity dispersion for ${\rm SN}_2$, as listed in \cref{tab:priors}.

The first four rows of \cref{fig:priors} compare different assumptions on widths of the Gaussian prior on counterterms (the baseline 50\%CT, 20\%CT, 100\%CT and 500\%CT) in the light of the \desidrone\, statistical precision (including the correction factors of \cref{tab:cov-corr}). We see little difference between the case 50\%CT and 20\%CT while the case 500\%CT exhibits stronger projection effects. We also compare our Gaussian physically-motivated priors with uniform (flat) uninformative priors on counterterms only (Flat CT, 5th row) and on counterterms and stochastic terms (Flat CT \& ST, 6th row). As expected, using uniform priors leads to a much stronger PVE, which corresponds to a larger difference between empty and filled symbols.
    We also show in the 7th and 8th rows of \cref{fig:priors} the constraints that come from a much bigger volume (the total volume of a 25 times \abacus\ mocks, $V_{25}$). We see for this case (with a larger volume, hence much more informative data) that the constraints using either our baseline choice (of Gaussian priors) or the uniform priors on counterterms and stochastic terms, yield very small differences (both for PWE and PVE). As expected, when the data are highly informative we do not observe projection effects (PVE or PWE) since each nuisance parameter becomes well-determined. This highlights the fact that projection effects depend on the constraining power of the data considered.
 
 \subsubsection{Beyond \texorpdfstring{$\Lambda$}{lambda}CDM models}  
\begin{figure}[htb]
    \centering
    \includegraphics[width=0.98\linewidth]{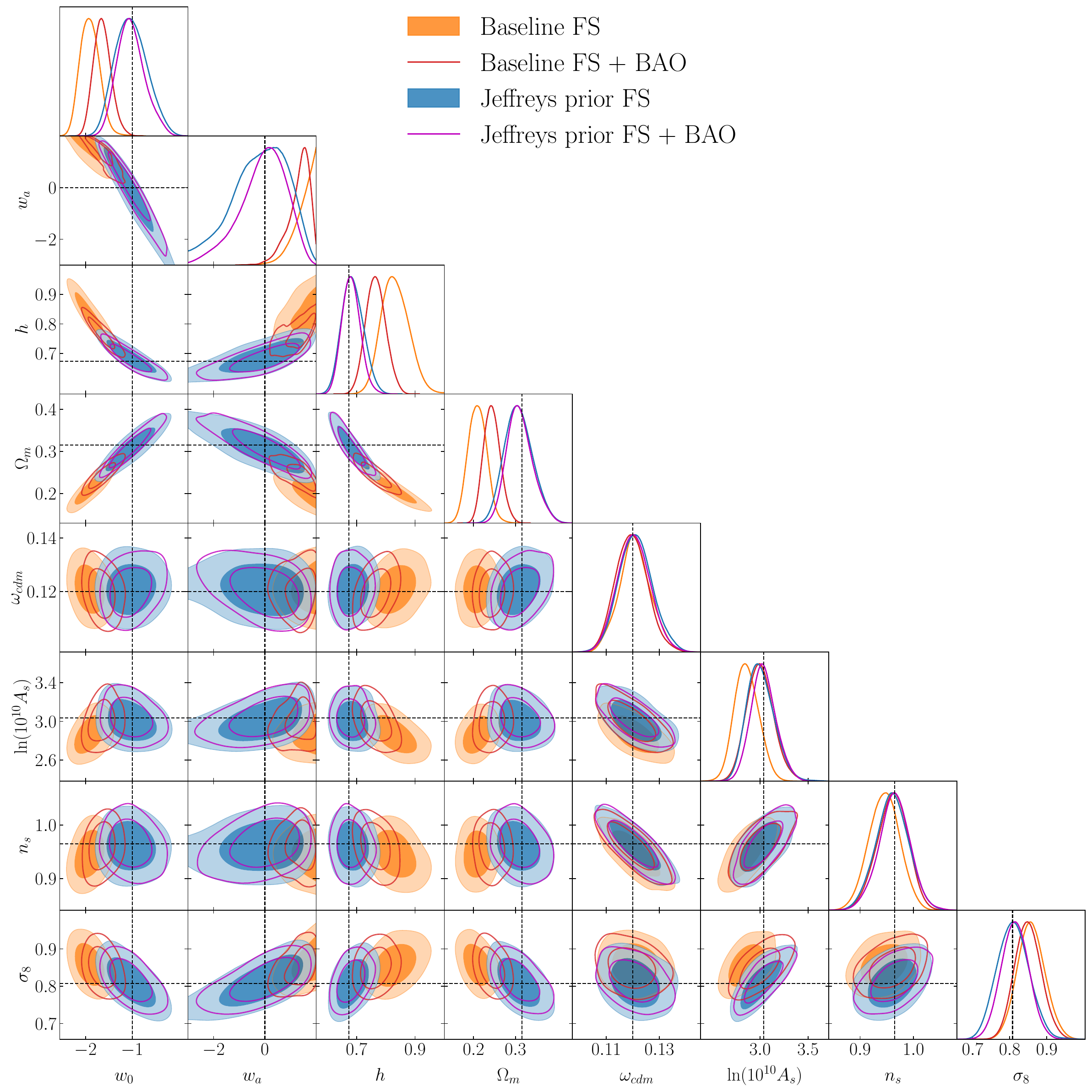}
    \caption{Toy model example for projection effects for the parameters of the $w_0 w_a$CDM cosmology. The contours display the joint fit of all six \desidrone\ types of tracers on synthetic and noiseless \velocileptors-generated power spectrum dataset employing the \desidrone\ data covariance. The orange contours display the results of the Full-Modelling analysis employing baseline choice of priors (as defined in the first item point of \cref{subsec:methods-projection} and later described in \cref{subsec:methods-baseline}). The empty red contours display the same as the orange contours, but additionally adding \desidrone\, BAO post-reconstruction information. The blue contours display the results where the Jeffreys priors on CT and ST parameters have been implemented (see text for more details), and finally, the empty pink contours display the same as the blue ones but adding the \desidrone\, BAO post-reconstruction information. The black vertical and horizontal dotted lines display the input cosmology of the mock-generated power spectrum. Since the data are noiseless we expect to obtain the same cosmology as the input one. Hence, observed differences between contours and dotted lines reflect the impact of PVE.}
    \label{fig:w0waCDM-priors}
\end{figure}

    We illustrate the effect of prior effects by fitting the parameters of the $w_0 w_a$CDM cosmology employing the covariance matrix of \desidrone\, data. We start by generating a synthetic and noiseless power spectrum data vector using \velocileptors. We choose the cosmological parameters to coincide with the fiducial cosmology (see \emph{Planck} $\Lambda$CDM in \cref{tab:fid_cosmo}), and the rest of the nuisance parameters to reproduce the observed unblinded clustering on each of the six redshift \desidrone\, bins of \cref{tab:Y1data}. \cref{fig:w0waCDM-priors} displays the fit of these mocks using the same \velocileptors\, model with the covariance matrix of \desidrone\, data for different data and prior choices. Since the data are noiseless any observed offset will be exclusively caused by PVE. 
    The orange contours correspond to a fit with the baseline choice of Gaussian priors (as described before in \cref{fig:priors}). The red empty contours display the same as the orange contours but add a synthetic and noiseless BAO \desidrone\, post-recon data vector.
    Finally, in blue (and empty pink), we display the fit to the mock-data using Jeffreys priors \cite{Jeffreys1946} (when BAO post-recon data are added), which have been employed more recently in the context of cosmological constraints \cite{DonaldMcCannetal23,Heavens18,Hadzhiyska23,Zhaoetal24} and which are more informative than the Gaussian priors\footnote{See figure 17 in \cite{KP5s2-Maus} as an example of how informative Jeffreys priors are on the $\Omega_m-\log(10^{10}A_s)$ plane}. We see that when constraining the parameters of the $w_0w_a$CDM model using \desidrone\, data alone, we are largely dominated by projection effects, even when the DESI BAO reconstruction data are added to the analysis. This effect could be mitigated by including highly informative priors, such as tighter Gaussian priors as shown previously in \cref{fig:priors}, or by Jeffreys priors. From this illustrative exercise, we come to some conclusions. 1) With the current baseline choice of Gaussian priors, \desidrone\, data without external datasets and without highly informative priors is not able to effectively constrain the parameters of some extended models, such as $w_0w_a$CDM, due to strong projection effects. 2) This cannot be mitigated by adding BAO reconstruction data, but other datasets such as CMB or uncalibrated SNe data need to be added. This latter approach is followed in our companion cosmology paper \cite{DESI2024.VII.KP7B}. 3) Alternatively, we can mitigate this effect by including more informative priors. However, tight Gaussian priors can cause PWE, and Jeffreys priors pose strong priors on cosmology parameters as shown in \cite{KP5s2-Maus}, which cast some doubts about whether the results are dominated by the priors and not by the data. Furthermore, Jeffreys priors are not able to completely remove projection effects in the presence of non-linear parameters, such as $b_3$. When allowing this parameter to be free, we find that even following the Jeffreys prior approach we observe projection effects. 

\subsection{Covariance matrix estimates}
\label{subsec:methods-cov}

Several covariance matrix estimates for galaxy two-point statistics have been proposed in the literature. For the current analysis, we focus on analytic and mock-based covariance matrix estimates but hybrid methods have also been developed recently that will be considered for future releases (e.g. \cite{carpool,fitcov}). In what follows, we briefly summarise the two types of covariance matrix estimates that we consider for \desidrone.

The analytical estimate of the covariance employs the Gaussian approximation in combination with non-linear clustering methods. It is based on works such as \cite{RascalC,Wadekar2019} for the \textsc{RascalC} and \textsc{TheCov} approaches, and is a relatively cheap way to generate covariance to any arbitrary clustering measurement. Therefore, one can either produce analytic covariances that aim at reproducing the real data or realistic mocks. For \desidrone\, analyses, we produce both data-based analytic covariance and \texttt{EZmocks}-based analytic covariance where, as discussed in \cref{subsec:dr1mocks}, the fibre assignment effect has been imprinted through the fast fibre-assignment algorithm \cite{DESI2024.II.KP3}. More details relative to how these methods are tuned to produce covariance for the DESI samples can be found in \cite{KP4s7-Rashkovetskyi,KP4s8-Alves}.

As a baseline choice, we choose the mock-based covariance, employing 1000 samples of the \texttt{EZmocks} for the different tracers described. The overall amplitude of these mock-based covariances is rescaled by the factors displayed in \cref{tab:cov-corr}. As we briefly explain in \cref{subsec:sys-cov} these factors correct for the systematic error budget associated with the potential inaccuracies of the \texttt{EZmocks} covariance.
More details about such a covariance matrix comparison for \desidrone\ can be found in \cite{DESI2024.II.KP3}.

\begin{table}
    \centering
   \begin{tabular}{|l|c|}
    \hline
       sample & 1/$\chi_{\rm red}^2$ correction factor \\
        \hline\hline
       \bgs   &  1.39  \\
        \lrgo  & 1.15  \\
        \lrgt  & 1.15  \\
        \lrgth  & 1.22  \\
        \elgt & 1.29  \\
        \qso   & 1.11 \\
        \hline
    \end{tabular}
    \caption{Multiplicative correction factor to apply to the \texttt{EZmocks} covariance to account for the differences between mock-based and analytic-based covariance matrix estimates \cite{DESI2024.II.KP3,KP4s6-Forero-Sanchez}.}
    \label{tab:cov-corr}
\end{table}


\subsection{Baseline choice}
\label{subsec:methods-baseline}

In the previous sections, we have described several choices we need to make to perform the Full-Shape analysis of the galaxy two-point statistics. Our official \desidrone\, results for galaxy Full-Shape analysis are based on the power spectrum, although for completeness we also investigate the correlation function. These choices include the compression scheme, the perturbation theory model, the galaxy bias model, the selected range for fitting the data, and other effects. In this section, we summarise the set of choices that define our Full-Shape pipeline. We highlight that the choices for defining this base pipeline were done before unblinding the data, by performing tests on DESI mocks for systematic effects but also on noiseless synthetic data for projection effects. Therefore these choices are not subject to confirmation bias based on previous experiments.

\begin{enumerate}    
    \item {\bf Compression scheme}. We have described two types of Full-Shape analyses (or compression schemes): ShapeFit and Full Modelling. Analysing the data through these two methods represents a step forward with respect to previous state-of-the-art galaxy two-point clustering analyses that were based on the standard compression described in \cref{subsubsec:standard_compression}. Therefore, for \desidrone\ galaxy Full-Shape analysis, we perform a detailed comparison and test all systematic effects in both ShapeFit and Full-Modelling bases (within the standard $\Lambda$CDM model assumption). Although both approaches are perfectly valid and compatible when performing cosmological inference (see for instance figures 4 and 12 of \cite{KP5s4-Lai} for a direct comparison), for the full cosmology interpretation and combination with other external datasets (presented in \cite{DESI2024.VII.KP7B}) we employ only the Full-Modelling approach.  
Projection effects can impact how the compressed parameters of the ShapeFit method are interpreted under a cosmological model, especially when the MAP value does not coincide with the mean of the posterior.
    
    The Full-Modelling approach can be combined at the $P(k)$ level with external datasets, thus avoiding the problem of projection effects during the combination and interpretation steps.

    \item {\bf Perturbation theory model}. We have studied three Fourier-space-based theory models as described in \cref{subsec:methods-eftmodel}. We select \velocileptors\, with its EPT option as our baseline choice for this study. However, as described above, all the models perform very consistently, and within \velocileptors\ both EPT and LPT flavours return consistent results. Hence, we do not foresee that this particular choice would have any noticeable impact on the final cosmology results.
        
    \item {\bf Bias scheme}. In order to be as agnostic as possible between the dark matter-galaxy connection, we allow $b_1$, $b_2$ and $b_{s}$ bias parameters (within the \velocileptors\, LPT scheme) to freely vary, but scale them with factors of $\sigma_8(z)$ as this parameterisation represents better the signal in the data and hence reduces volume effects. In the basis \{$(1+b_1)\sigma_8(z)$, $b_2\sigma_8^2(z)$, $b_s\sigma_8^2(z)$\} we place a uniform prior on $(1+b_1)\sigma_8(z)$ and Gaussian priors on $b_2\sigma_8^2(z)$ and $b_s\sigma_8^2(z)$ that are wide enough for PWE to be negligible.
    The freedom in the bias allows us to be agnostic about whether galaxy bias co-evolves with halo bias. Assuming that $b_s$ co-evolves with $b_1$ (as expected in models of halo bias) would add cosmological information on $A_s$ and $\omega_{\rm cdm}$ (in Full-Modelling analysis) or on $f\sigma_{\rm s8}$ and $m$ (in ShapeFit analysis). For these reasons, we decided to keep $b_s$ decoupled and independent from $b_1$ and $b_2$. However, we choose to fix the $b_3$ bias parameter to be null. This also has cosmological implications (varying it broadens the posterior of $m$ in the ShapeFit analysis).  The reason for this choice is twofold: first, the cubic bias is expected to be small for intermediate-mass haloes, and second and more importantly, $b_3$ is quite degenerate with the counterterms, as discussed in appendix D of \cite{KP5s2-Maus}, and barely modifies the posteriors within the Full-Modelling analysis of $\Lambda$CDM.
    
    \item {\bf Range of scales}. We define our fitting range of scale to be $0.02 < k\,[h{\rm Mpc}^{-1} ] <0.20$. Our supporting papers \cite{KP5s1-Maus,KP5s2-Maus,KP5s3-Noriega,KP5s4-Lai} show that the exact value for the minimum scale cut $k_{\rm max}$ do not have any significant impact, neither in the peak of the posterior nor on the size of the error-bars, suggesting that, given the adopted baseline scheme described above, there is not much cosmological information on the mildly non-linear scales. The large-scale cut value is imposed by how much we trust our imaging systematic correction (see \cref{subsec:sys-imaging}). A similar study was done in configuration space \cite{KP5s5-Ramirez}.
    
    \item {\bf Legendre multipoles}. We account for the anisotropies induced by redshift-space distortions and the Alcock-Paczynski effect by including the monopole and the quadrupole Legendre moments of the power spectrum. As explained in \cref{subsec:methods-projection}, not including the hexadecapole helps to reduce the prior-weight effects while including it only marginally improves the precision of some cosmological constraints, especially when we combine it with BAO. As we refine the analysis pipeline and manage to control the projection effects better, this choice may change for future analyses and the constraining power of the hexadecapole for dynamical dark energy models could also be investigated.
    
    \item {\bf Priors on non-cosmology parameters}. As described in \cref{subsec:methods-projection}, we set physically motivated Gaussian priors around the stochastic and counterterm parameters. We set a uniform prior on $(b_1+1)\sigma_8$ and wide Gaussian priors on $b_2 \sigma_8^2$ and $b_s \sigma_8^2$. 
    The priors on the non-cosmology parameters are summarised in \cref{tab:priors}.
    
    \item {\bf Priors on cosmology parameters}. When doing cosmology inference in either \break ShapeFit or Full Modelling, we set all the priors on cosmology parameters to be flat (except for $n_s$ and $\omega_b$), with a range sufficiently wide that the chain does not reach the imposed prior boundaries. When doing a Full-Modelling analysis using DESI data alone we follow the procedure described in \cite{KP7s1-MG} where we utilize the joint constraint
on $\omega_b$ and the number of relativistic species $N_{\rm eff}$, and fix the latter parameter to its fiducial value of 3.044; the resulting projected constraint on the physical baryon
density is $\omega_b =  0.02198 \pm 0.00053$ \cite{Schoneberg24}. This prior is slightly different from the one stated in \cref{tab:priors}.

    \item {\bf Covariance}. As described in \cref{subsec:methods-cov}, we have two methods for generating covariances in Fourier space: the analytical estimate based on Gaussian approximation and the mock-based covariance which relies on \texttt{EZmocks}. Our fiducial choice is to use the mock-based covariance, as they better include systematic effects such as the fibre assignment. In addition, the lack of post-reconstruction analytical covariance poses a limitation when combining the Full-Shape analysis with the post-reconstruction BAO analysis, which using mock-based covariance can be easily done. We account for potential inaccuracies of the mock-based covariance by including a correction factor described in \cref{subsec:sys-cov}.
    
    \item {\bf Emulator}. When doing the cosmology inference using either Full Modelling or ShapeFit, each likelihood evaluation requires calculating loop-correction terms which consist of 2D integrals, and is thus computationally costly, especially if repeated tens of thousands of times in an MCMC chain per model. To accelerate this process, we can employ a fourth-order Taylor expansion emulator, which avoids re-evaluation of the loop integrals by interpolating on a pre-evaluated grid of models. However, the emulator may be inaccurate if evaluating a model far from the initial calibration point. Therefore we only employ the emulator when analyzing mock data, for which the calibration can be performed close to the known true cosmology. When analyzing real data, we either do not employ the emulator, or validate chains run with an emulator by recalculating the likelihood with the model without an emulator at each point of the chain.
   Although this is slower, it means the results obtained from DESI data are robust against possible emulator inaccuracies. MCMC chains presented in this paper are either run with the No U-Turn Sampler (NUTS~\cite{2011arXiv1111.4246H}, as implemented in blackjax~\cite{cabezas2024blackjax}\footnote{\url{https://github.com/blackjax-devs/blackjax/}}), when emulating the galaxy power spectrum model or with the MCMC sampler~\cite{Lewis:2002ah, Lewis:2013hha} implemented in \texttt{Cobaya}\footnote{\url{https://github.com/CobayaSampler/cobaya/}} when running without an emulator
    
    \item {\bf Fiducial cosmology}. We must choose a fiducial cosmology to create the 3D galaxy catalogues. This is the cosmology used to transform the redshifts into comoving distances expressed in $h^{-1}{\rm Mpc}$. This cosmology is often referred to as catalogue or grid cosmology. In addition, when performing the ShapeFit analysis we also need a fiducial cosmology to generate the linear cold dark matter + baryon power spectrum of reference that is later modified through the ShapeFit parameters ($\alpha_{\rm iso}$, $\sigma_{s8}$, $m$ and $n$).\footnote{Note that since the universe is assumed to be isotropic, $P_{\rm lin}(k)$, the anisotropic parameters $\{\alpha_{\rm AP},\, f\}$, do not depend on the arbitrary choice of the fiducial cosmology of the linear template.} For simplicity, this cosmology, namely template cosmology, is chosen to be the same as the grid cosmology. When analyzing the data we set this cosmology to be consistent with the \emph{Planck} cosmology. The details of this cosmology can be found in the first row of \cref{tab:fid_cosmo}. 
\end{enumerate}

\begin{table}[htb]
    \centering
    \begin{tabular}{|c|c|}
        \hline
        Cosmological parameters (SF) & Priors \\\hline
                $\alpha_{\rm iso}$ & $\mathcal{U}$[0.8, 1.2] \\
                $\alpha_{\rm AP}$ & $\mathcal{U}$[0.8, 1.2] \\
                $f/f_{\rm fid}$ & $\mathcal{U}$[0.0, 2.0] \\
                $m$ & $\mathcal{U}$[-0.8, 0.8] \\
\hline\hline
        Cosmological parameters (FM) & Priors \\\hline
        $\omega_\mathrm{cdm}$ & $\mathcal{U}$[0.01,0.99] \\
        $\omega_\mathrm{b}$ & $\mathcal{N}[0.02218,0.00055^2]$ \\
        $h$ & $\mathcal{U}$[0.2,1] \\
        ln($10^{10}A_s$) & $\mathcal{U}$[1.61,3.91] \\
        $n_s$ & $\mathcal{N}[0.9649,0.042^2]$ \\
        \hline \hline
        Non-cosmological parameters & Priors \\
        \hline
        $(1+b_1) \sigma_8$ & $\mathcal{U}$[0,3] \\
        $b_2 \sigma_8^2$ & $\mathcal{N}[0,5^2]$ \\
        $b_s \sigma_8^2$ & $\mathcal{N}[0,5^2]$ \\
        $\alpha_0$ & $\mathcal{N}[0,12.5^2]$ \\
        $\alpha_2$ & $\mathcal{N}[0,12.5^2]$ \\
        SN$_0$ & $\mathcal{N}[0,2^2] \times 1/\bar{n}_g$ \\
        SN$_2$ & $\mathcal{N}[0,5^2] \times f_{\rm sat} \sigma_{1\,\rm eff}^2/\bar{n}_g$ \\
        \hline
    \end{tabular}
    \caption{Priors on the cosmological and non-cosmological parameters used in \desidrone\ galaxy Full-Shape analysis for the baseline choice. Here, $\mathcal{U}$ stands for a uniform prior, and $\mathcal{N}(\mu,\sigma^2)$ for a Gaussian prior with mean $\mu$ and standard deviation $\sigma$. The cosmological parameters of the ShapeFit (SF) are listed in the top sub-panel; the cosmological parameters of the Full Modelling (FM) are listed in the middle sub-panel; and the non-cosmological parameters are listed in the bottom sub-panel. The bias parameters $b_1,b_2,b_s$ are defined in the Lagrangian basis. For the stochastic parameters the listed prior width on SN$_0$ is multiplied by the Poissonian shot-noise ($1/\Bar{n}_g$) while the prior width for SN$_2$ is multiplied by the expected fraction of satellite galaxies times the mean velocity dispersion
of the satellites in question times the Poissonian shot-noise, $f_{\rm sat} \sigma_{1\,\rm eff}^2/\bar{n}_g$, which represents the characteristic velocity dispersion (see \cite{KP5s2-Maus} for details).} 
    \label{tab:priors}
\end{table}

\section{Systematic error budget}
\label{sec:systematics}

This section provides an overview of the systematic errors in the Full-Shape galaxy two-point clustering analysis, their mitigation, and their impact on the cosmological parameters of interest, for both ShapeFit and Full-Modelling approaches, as determined through the tests performed in our supporting papers \cite{KP5s1-Maus,KP5s2-Maus,KP5s3-Noriega,KP5s4-Lai,KP5s5-Ramirez,KP5s6-Zhao,KP5s7-Findlay,KP5s8-Gsponer,KP3s3-Krolewski,KP3s4-Yu,KP4s6-Forero-Sanchez,KP4s7-Rashkovetskyi,KP4s8-Alves,KP3s9-Andrade} (see \cref{tab:supportingpapers} for the specific task covered by each of these works). As summarised in \cref{subsec:methods-baseline}, the baseline choice for the analysis of the Full-Shape of \desidrone\ is the Full Modelling.
The main reason for this choice is the impact of projection effects on the compressed set of parameters. When this compressed set is interpreted in terms of a cosmology model using the usual Gaussian approximation, we observe a systematic offset on the derived cosmology parameters caused by projection effects, mainly in $\alpha_{\rm AP}$ and $m$, in those redshift bins with less constraining power.  
This systematic offset is the main reason for our choice of not using ShapeFit parameters for cosmological inference. Another reason is the potential residual dependence on the assumed cosmology of the template, although this can be mitigated by an iterative run on reference cosmologies not disfavored by the data. 
However, for completeness, in this paper we also present the main determination of the systematic contribution to the total error given in terms of ShapeFit parameters, which is also used to determine whether the corresponding systematic effect can be neglected or not with respect to the statistical precision of \desidrone. The ShapeFit approach helps to isolate systematic errors on physically-motivated parameters which determine the $P(k)$, and we do not need to assume a cosmological model to decide whether a given systematic is relevant for the data.

Following \cite{DESI2024.III.KP4}, we account for a systematic effect in the total error budget when its effect on the cosmological parameter is above $\sfrac{1}{5}$ of the \desidrone\ precision. Since we work within the Full-Modelling approach, we decide to apply this systematic contribution (once identified as relevant) at the level of the power spectrum, $P(k)$ bins. That is, we add a contribution to the statistical covariance matrix, instead of increasing the statistical error of the derived parameter within a specific model. We will describe later which systematics are added at the power spectrum level, and which are not considered or are included in a different way.
This allows us to automatically propagate that systematic budget to any cosmology fits using our power spectrum data vector. 

Both approaches of quoting a systematic error budget, at the parameter or at the power spectrum covariance matrix level, are compared in the context of $\Lambda$CDM in \cref{subsec:sys-total}, yielding very consistent results, and thus validating our approach.

\begin{table}
    \centering
    \small
    \resizebox{\columnwidth}{!}{%
        \begin{tabular}{|l|r|r|}
            \hline
              Task & Section & Ref\\
            \hline\hline
             Comparison of EFT models & \cref{subsec:methods-compression,subsec:methods-eftmodel,subsec:sys-th} & \cite{KP5s1-Maus}\\
            Theoretical code description for EFT model \velocileptors & \cref{subsec:methods-compression,subsec:methods-eftmodel,subsec:sys-th} & \cite{KP5s2-Maus} \\
           Theoretical code description for EFT model \folps\, &	\cref{subsec:methods-compression,subsec:methods-eftmodel,subsec:sys-th} & \cite{KP5s3-Noriega} \\
           Theoretical code description for EFT model \pybird\,	 & \cref{subsec:methods-compression,subsec:methods-eftmodel,subsec:sys-th} & \cite{KP5s4-Lai}	\\
            Theoretical code description for EFT model \eftgsm\ & \cref{subsec:methods-compression,subsec:methods-eftmodel,subsec:sys-th} &  \cite{KP5s5-Ramirez}	\\
             HOD-dependent systematics for DESI 2024   & \cref{subsec:sys-hod} & \cite{KP5s7-Findlay} \\
             Fiducial-cosmology-dependent systematics for DESI 2024  & \cref{subsec:sys-fiducialcosmo} & \cite{KP5s8-Gsponer}  \\    
            Impact and mitigation of the imaging systematics for DESI 2024  & \cref{subsec:sys-imaging} & \cite{KP5s6-Zhao} \\
            Impact and mitigation of fibre collisions for DESI 2024  & \cref{subsec:sys-fibre}  & \cite{KP3s5-Pinon} \\  
            Impact and mitigation of the spectroscopic systematics for DESI 2024  & \cref{subsec:sys-spectro} & \cite{KP3s4-Yu,KP3s3-Krolewski} \\
           Comparison between analytical and \texttt{EZmocks} covariance matrices  & 	\cref{subsec:methods-cov,subsec:sys-cov} &  \cite{KP4s6-Forero-Sanchez} \\
             Analytical covariance matrices for correlation function for DESI 2024& \cref{subsec:methods-cov}  &  \cite{KP4s7-Rashkovetskyi} \\
             Analytical covariance matrices for power spectrum multipoles for DESI	& \cref{subsec:methods-cov,subsec:sys-cov}  & \cite{KP4s8-Alves}  \\
             Tests of the catalog-level blinding method for DESI 2024 & \cref{subsec:blinding} &  \cite{KP3s9-Andrade}  \\
            \hline
             \end{tabular}
    }
    \caption{\label{tab:supportingpapers}
    The list of the studies supporting this paper and the corresponding sections 
    where their results are discussed.}
\end{table}

\subsection{Theoretical Systematics}\label{subsec:sys-th}

\begin{figure}
    \centering
    \includegraphics[width=0.8\linewidth]{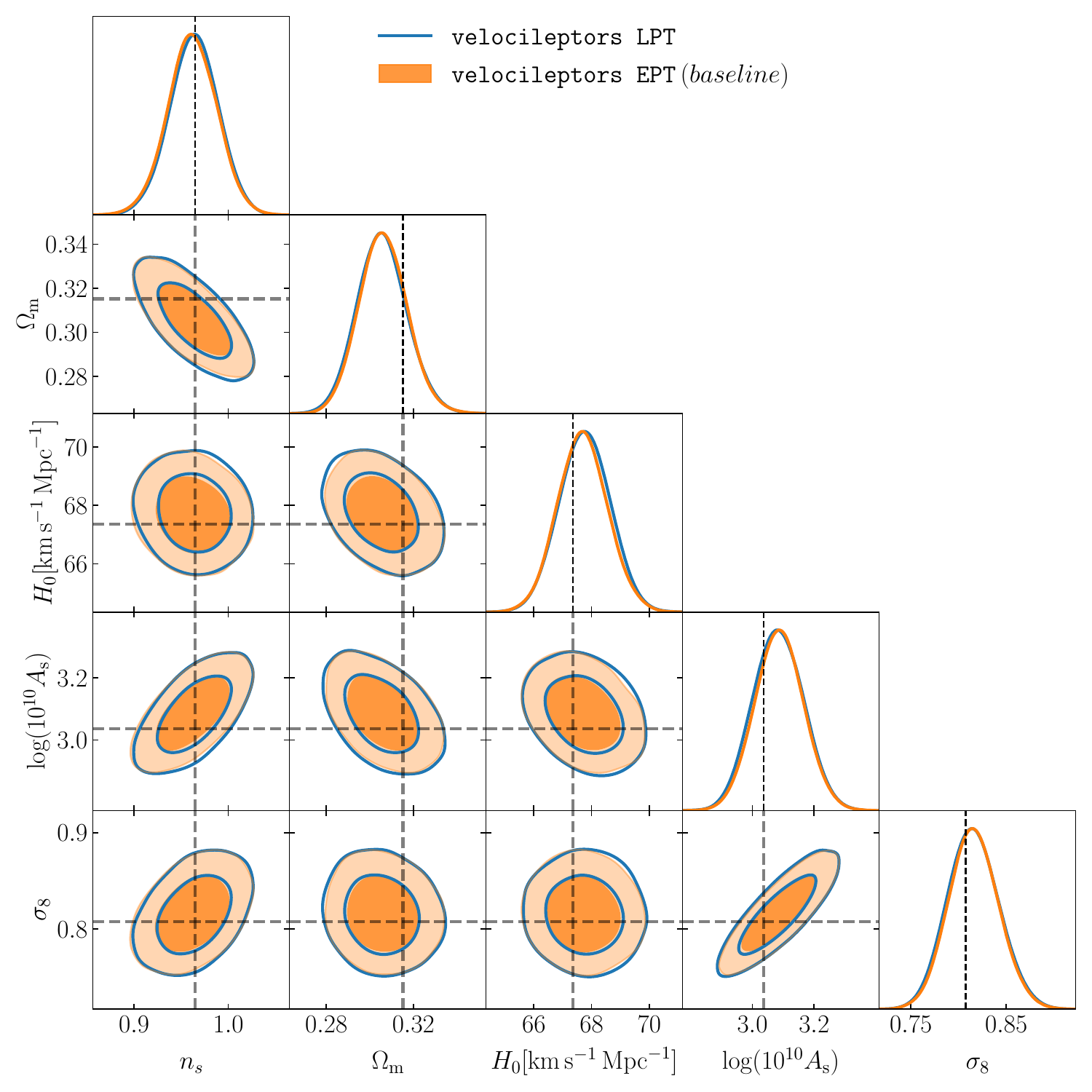}
    \caption{Posteriors obtained when analyzing the mean of 25 Abacus DESI mocks with the \desidrone\ sky geometry covering the six redshift bins of \cref{tab:Y1data}. The contours show the results obtained using \velocileptors\ assuming the Full-Modelling analysis, with the Lagrangian PT mode (blue open contours) or Eulerian PT mode (orange solid contours)
   on a $\Lambda$CDM model with a BBN-like prior. The covariance corresponds to a single \desidrone\ realisation volume. }
    \label{fig:th-syst}
\end{figure}

\begin{figure}
    \centering
    \includegraphics[width=1\linewidth]{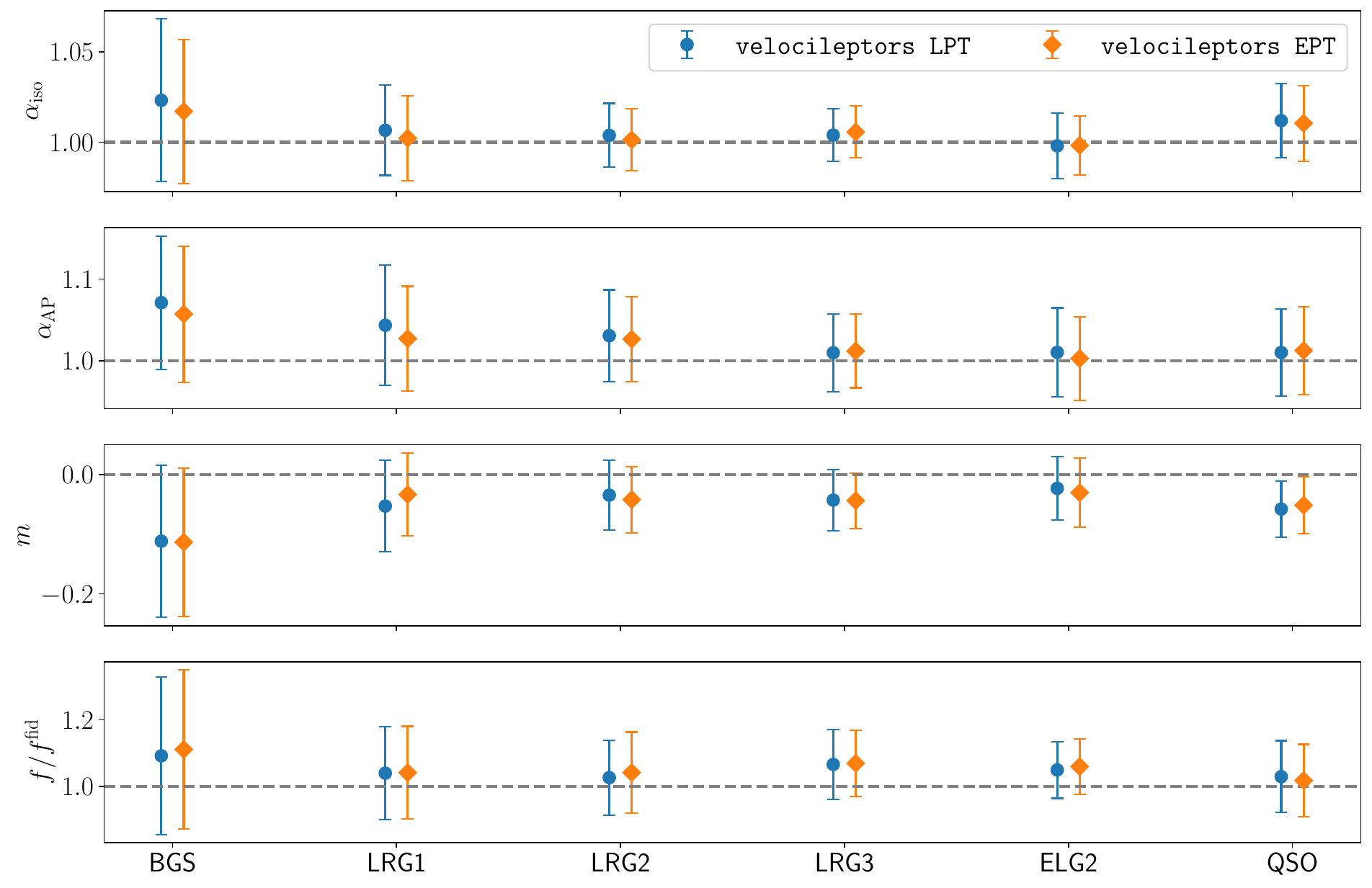}
    \caption{
    Constraints on key parameters from each of the six redshift bins obtained from the mean of 25 Abacus DESI mocks with the \desidrone\ sky geometry.
    The orange and blue contours display the results obtained using the \velocileptors\, ShapeFit option (with both Eulerian and Lagrangian flavours as indicated). The covariance corresponds to a single \desidrone\ realisation volume. The error bars display the 68\%  credible intervals. The horizontal dashed grey lines display the expected value for these mocks.}
    \label{fig:th-syst-SF}
\end{figure}

We refer to the systematics associated with the model (and setup) chosen to analyse the data as theoretical systematics. In \cite{KP5s2-Maus,KP5s3-Noriega,KP5s4-Lai,KP5s5-Ramirez}, we presented a thorough study of these systematics for each of the Perturbation Theory models presented here. \cref{fig:th-syst} displays the posteriors resulting from the analysis of the Abacus mocks with the \velocileptors\ EPT and LPT pipelines, with the baseline choice described before. The mocks represent the average of 25 independent realisations with the DESI DR1 geometry imprinted, across the six redshift bins containing BGS, LRG, ELG and QSO. We employ the covariance matrix associated with the volume of a single realisation of \desidrone\ data. Therefore, the contours are comparable to those expected to be obtained by the actual data although the significance of the signal is much higher. As our baseline choice, we employ the Full-Modelling analysis option of \velocileptors, along with the rest of the baseline options described in \cref{subsec:methods-baseline}. Blue and orange represent the two options of Lagrangian and Eulerian frameworks within \velocileptors\ as indicated. We see that none of the variables present a significant bias to their true value (denoted with the dashed and dotted black lines). We additionally show in \cref{fig:th-syst-SF} constraints on mocks using the ShapeFit method with the \velocileptors\ LPT and EPT modules. As in the case of Full-Modeling, we find very close agreement between the two modeling pipelines.

Further and more detailed studies on the Abacus cubic boxes can be found in \cite{KP5s2-Maus}, including a ShapeFit analysis, with similar results. Thus we conclude that the systematic errors associated with the model are negligible given the DESI DR1 statistical precision.
Additionally, we also make a relative comparison of the three EFT models described before \cite{KP5s1-Maus}, and find that the difference in the parameters of interest was below 0.1\%. We then conclude that the effects beyond the one-loop corrections are not significant for our inference when using any of these models.

\subsection{HOD-dependent systematics}\label{subsec:sys-hod}

The HOD-dependent systematics refer to those associated with the inability of the galaxy bias model of the non-linear $P(k)$ to capture realistic variations of the halo-galaxy connection physics in the data. Note that in the theoretical systematics section, we have performed a detailed study of different models at a fixed HOD, the one associated with the \texttt{Abacus-2} mocks. Instead, here we aim to consider a variety of HOD models for BGS, LRG, ELG and QSO samples that are consistent (within a certain uncertainty threshold) with the two-point clustering measurement of DESI.

The considered HODs are fully described in \cite{KP5s7-Findlay} and references therein. All of them consist of changes which span up to $3\sigma$ variations around the fiducial or reference HOD baseline model (except for some of the ELG models) that matches the small-scale clustering and number density of the DESI One-Percent Survey \cite{DESI2023b.KP1.EDR}. Below we summarise the variations in the HOD we have explored.

\begin{itemize}
    \item For the LRG sample, we explored variations of the mean number density of central and satellite galaxies through the parameters of the HOD model of \cite{EDR_HOD_LRGQSO2023,KP4s10-Mena-Fernandez,Zheng2007} around the fiducial baseline model. These variations are parameterised through changes in the minimum mass of the halo able to host a central galaxy, the typical mass of a halo hosting one single satellite galaxy, a velocity galaxy bias with respect to the host halo, assembly bias effects through two environmental-dependent parameters, as well as a parameter to include baryonic effects which is able to modulate the radial distribution of galaxies. Variations of seven different types of models were explored in total, in addition to the baseline model. 
  
    \item For the ELG sample, we explored variations that are tuned both to reproduce the 2D correlation function alone and to jointly fit the projected correlation function and 2D correlation function monopole and quadrupole.
    These variations \cite{EDR_HOD_ELG2023,Yuan2023,Garcia-Quintero2024} result in 21 different models which include mass quenching effects for high-mass haloes, variations in the concentration of the halo (motivated by assembly bias effects), modified satellite density profile and galactic conformity. 
    \item The variations of the BGS sample, detailed in \cite{2024Smith}, follow a similar form of the HOD model of LRGs but are instead parameterised by 17 meta-parameters that introduce a luminosity dependence to the standard HOD parameters to account for BGS being a magnitude-limited sample. No assembly or velocity bias effects are included for BGS. In total, we explored 11 variations, tuned to the projected correlation function, that sample the posterior around the best-fit model.
    \item The QSO sample HOD variations, described in \cite{EDR_HOD_LRGQSO2023}, are very similar to baseline LRG `A models', including three variations around the Baseline. These models are tuned to the 2D correlation function.
\end{itemize}

\begin{figure}
    \centering
    \includegraphics[width=\linewidth]{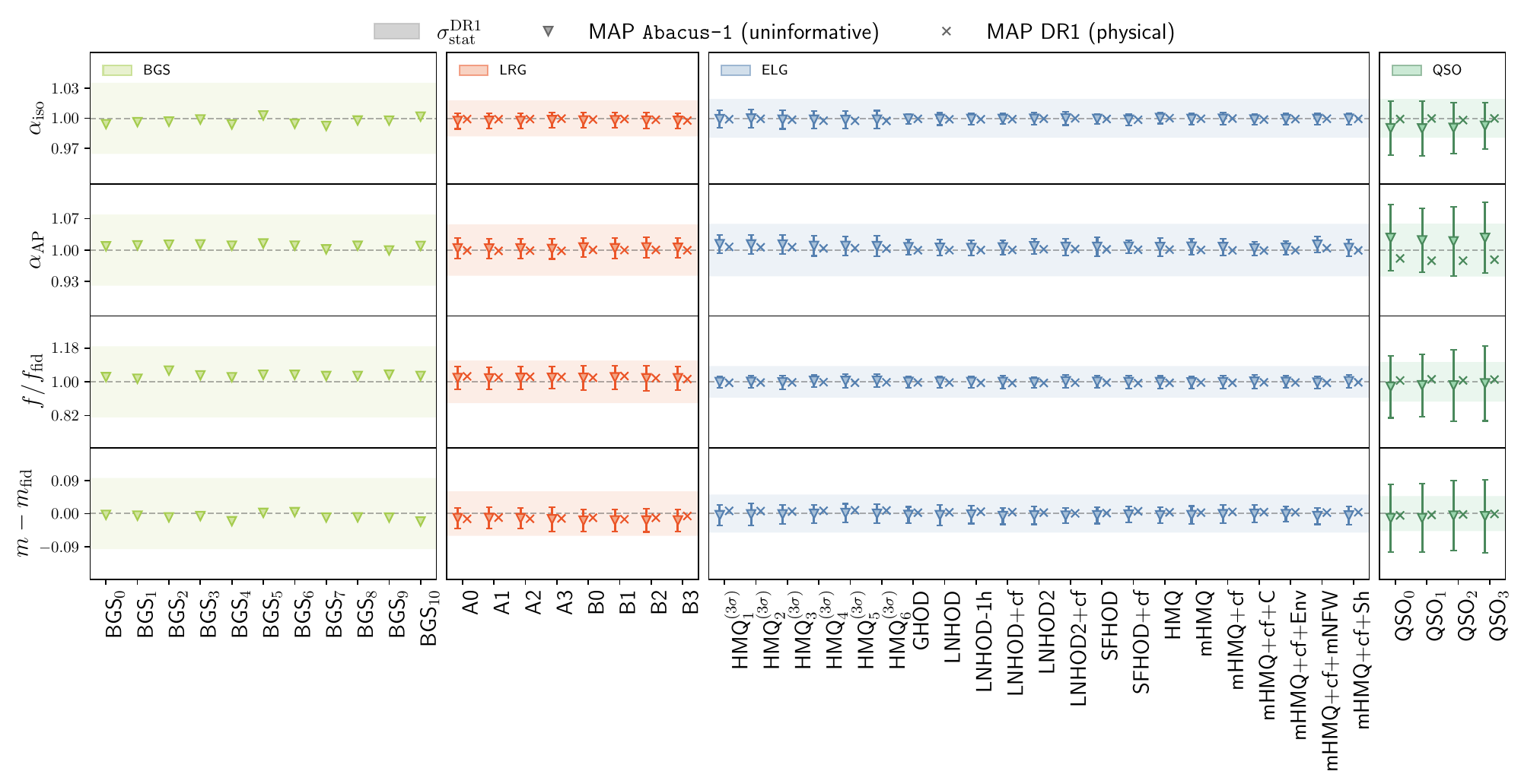}
    \caption{
    Parameter fits for several HOD models in the ShapeFit analysis of \texttt{Abacus-2} cubic mocks (triangles) and DR1-like data (crosses).
    Fits to \texttt{Abacus-2} have flat ``uninformative" priors on nuisance terms such that differences between these points and those in the baseline ``physical" parameterisation are due to the miscentring of priors---referred to as the prior weight effect (see \cref{subsec:methods-projection}). The coloured bands show the DR1 statistical uncertainty for each tracer at the redshift of the data (including \texttt{EZmock} rescaling discussed in \cref{subsec:methods-cov}). Error bars represent the standard deviation of the recovered parameter values over 25 individual mock realisations and are not included in the case of DR1-like data as we fit to an averaged measurement. Only a single mock realisation is available for the BGS sample so no error bars are provided in this case.
    }
    \label{fig:HOD_shifts}
\end{figure}

By fitting to \texttt{Abacus-2} mocks populated using a range of different HOD models listed above, we find that the standard deviation of the shifts in the recovered cosmological parameters can be larger than $>0.2\sigma_{\rm DR1}$. We conservatively interpret these shifts as an upper bound on the size of the systematic error from uncertainty in the galaxy-halo connection, due to the possible inability of the EFT model to perfectly absorb changes in the halo-scale physics across the range of HOD models tested. This additional uncertainty propagates to all of the Full-Modelling parameters, but in ShapeFit manifests primarily in shifts in the $f/f^{\rm fid}$ parameter.
While all tracers are affected, we find that the effects to be strongest for the ELG and QSO samples (see Table 2 in \cite{KP5s7-Findlay}). The prior weight effect (PWE) was also investigated for each of the HOD models using DR1-like power spectrum measurements generated from a window convolution of the \texttt{Abacus-2} measurements and corresponding DR1-like covariance matrices. We find this effect to be subdominant for DR1 (except for QSO), shown as a difference between the ``uninformative" prior case with flat nuisance priors and the baseline ``physical" prior case in \cref{fig:HOD_shifts} but is still included in our systematic estimation to be conservative.

In order to include the systematic contribution, we encapsulate the effect of such
shifts in terms of a systematic covariance term, $\tens{C}_{\rm HOD}$, measured at the $P(k)$-level. Given that the values of nuisance parameters preferred by the data are HOD-dependent, we include an additional diagonal contribution that quantifies the ``goodness-of-fit" to the set HOD models given the baseline priors. The diagonal contribution is measured on DR1-like data to ensure our results are robust to the realistic PWE expected for DR1. We validate our method against adding the HOD+PWE systematic at the level of the parameters and find excellent agreement. The inclusion of the HOD-dependent systematic contribution at the level of the data vector has minimal effect on the DR1 posteriors for all tracers.

In this work we have not performed a detailed study on how more exotic effects such as the intrinsic alignments between galaxies and the tidal field, the relative velocities between baryons and dark matter, or line-of-sight selection biases could impact the systematics in terms of cosmological parameters
\cite{Desjacques2018,Schmidt2016,Lamman2024}. Most of these effects could be better quantified including higher-order statistics, such as the bispectrum, which goes beyond the scope of this work. In the upcoming DESI full-shape analyses special attention should be paid to determine these and other systematics related to the halo-galaxy relation.

\subsection{Systematics due to the assumption of the fiducial cosmology} \label{subsec:sys-fiducialcosmo}

We next study the impact of the assumption of a fixed fiducial cosmology both for Full-Modelling and ShapeFit. In addition to the fiducial choice (consistent with \emph{Planck} $\Lambda$CDM), we study four extra choices as a fiducial cosmology: low-$\Omega_m$, thawing Dark Energy, high $N_{\rm eff}$ and low $\sigma_8$ (see \cref{tab:fid_cosmo}). When performing the analysis using the ShapeFit methodology, the fiducial cosmology choice enters at the level of constructing the catalogue in terms of comoving distances -- by transforming redshift into comoving distances in $\hinvmpc$ units -- and at the level of the fiducial $P_{\rm lin}$ fixed template, i.e. the baseline linear power spectrum that the SF variables modify to adjust to the data. On the other hand, for the Full-Modelling type of analysis, this assumption only enters at the level of constructing the catalogue.

\begin{table}[!ht]
    \centering
    \begin{tabular}{|c|c|c|c|c|c|c|c|c|}
    \hline
        Name  &  $\omega_{\rm b}$ & $\omega_{\rm cdm}$ & $h$ & $10^{9}A_s$ & $n_s$ & $N_{\rm ur}$ & $w_0$ & $w_a$\\
       \hline\hline
       \emph{Planck} $\Lambda$CDM&  $0.02237$ & $0.1200$ & $0.6736$ & $2.0830$ & $0.9649$ & $2.0328$ & $-1$ & $0$\\
       Low-$\Omega_m$ & $0.02237$ & $0.1134$ & $0.7030$ & $2.0376$ & $0.9638$ & $2.0328$ & $-1$ & $0$\\
       Thawing DE&  $0.02237$ & $0.1200$ & $0.6278$ & $2.3140$ & $0.9649$ & $2.0328$ & $-0.7$ & $-0.5$\\
       High-$N_{\rm eff}$  & $0.02260$ & $0.1291$ & $0.7160$ & $2.2438$ & $0.9876$ & $2.6868$ & $-1$ & $0$\\
       Low-$\sigma_8$  & $0.02237$ & $0.1200$ & $0.6736$ & $1.7949$ & $0.9649$ & $2.0328$ & $-1$ & $0$\\
\hline       
    \end{tabular}
    \caption{Parameters of the fiducial cosmology models used to quantify the associated systematic effects of this arbitrary choice. The physical density of baryons and cold dark matter, $\omega_{\rm b}$ and $\omega_{\rm cdm}$, respectively, the dimensionless Hubble parameter $h$, the primordial amplitude, $A_s$, the spectral index, $n_s$, the number of ultra-relativistic species, $N_{\rm ur}$ and the dynamical dark energy parameters according to the CPL parametrization, $w_0$ and $w_a$ \cite{Chevallier:2001, Linder2003}. These models correspond to the \textsc{c000}-\textsc{c004} cosmologies of the \texttt{AbacusSummit} suite \cite{AbacusSummit}. These are also the same fiducial cosmologies used in the companion BAO paper \cite{DESI2024.III.KP4}.}
    \label{tab:fid_cosmo}
\end{table}

We study the systematics related to the assumption of the fiducial cosmology by fitting \desidrone\ \texttt{Abacus-2} complete mocks with analytic covariance. We first determine whether there is a detection of a systematic effect by computing the maximum shift observed between different cosmologies for all tracers and redshift bins. 
We express the shifts in terms of the precision of the complete 25 realisations of \texttt{Abacus}. This way we have enough sensitivity (of the volume of 25 realisations) to detect potential systematic shifts that would otherwise be hidden by the statistical precision of just 1 realisation.
For shifts that are above the 3$\sigma$ threshold, we report a detection of a systematic. Then, we look at both the relative shifts with respect to the baseline \emph{Planck} $\Lambda$CDM cosmology and with respect to the truth. We quantify those shifts in terms of DR1 error so we convert the shifts expressed in terms of $\sigma_{\rm V25}$ into shifts expressed in terms of $\sigma_{\rm DR1}$. The analysis and the results are further described in our supporting paper \cite{KP5s8-Gsponer}. For the Full-Modelling approach, we did not detect any systematic effect for any of the tracers if the detection limit is set to 3 $\sigma_{\rm V25}$ (all of them are well below 1 $\sigma_{\rm V25}$). Moreover, the maximal shifts for each tracer is below 0.2 $\sigma_{\rm DR1}$.
For ShapeFit parameters, we found that the shape parameter $m$ is the most affected by changes in the fiducial cosmology, with observed shifts that are often above 0.2 $\sigma_{\rm DR1}$. This is consistent with what was reported in \cite{brieden2021} using another set of simulations with a volume that would roughly correspond to the lowest DESI LRG redshift bin. However, for the parameters of interest ($\alpha_\parallel$,$\alpha_\perp$ and $f\sigma_8$), the observed maximal shifts are below 0.2 $\sigma_{\rm DR1}$. Therefore, we report a common systematic error contribution that is below 0.2 $\sigma_{\rm DR1}$ for both Full-Modelling and ShapeFit in \cref{tab:systematicbudget}.

\subsection{Fibre-collisions}\label{subsec:sys-fibre}

As mentioned in \cref{subsec:methods-2pt}, in order to mitigate the impact of fibre incompleteness on the Full-Shape galaxy two-point clustering analysis, we cut angular scales smaller than 0.05 deg. This method called `$\theta$-cut' was proposed in \cite{KP3s5-Pinon} where it was validated against \abacus\ mocks for the ELG sample.

\begin{figure}
    \centering
    \includegraphics[width=1.\textwidth]{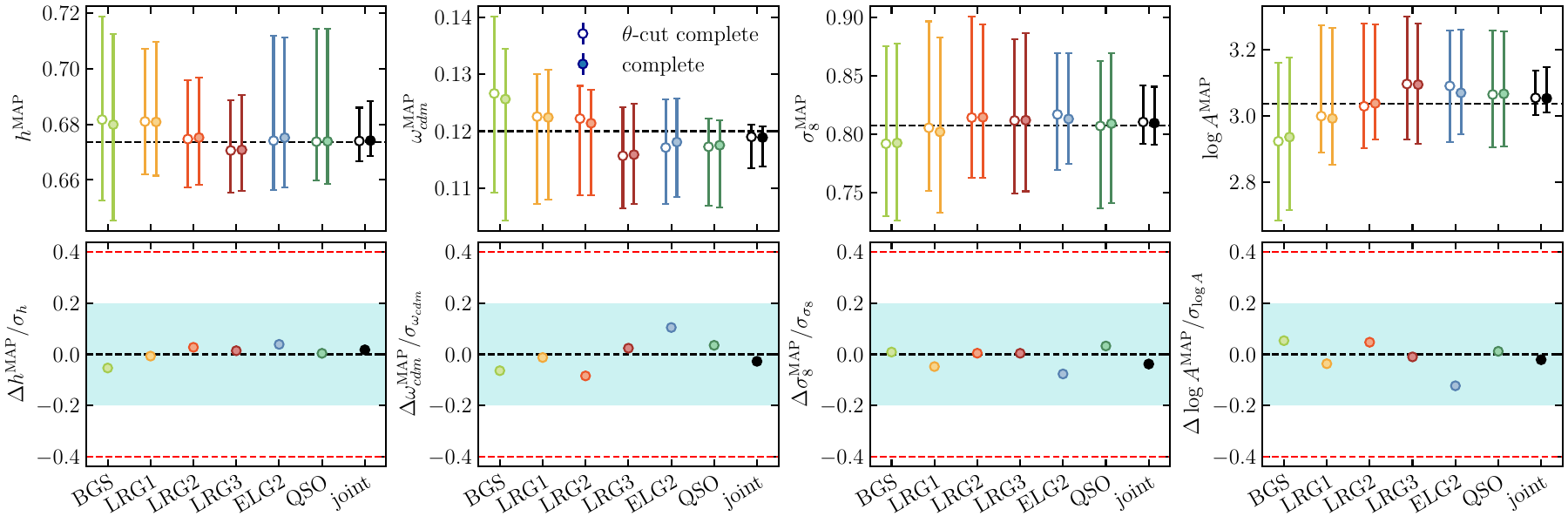}
    \includegraphics[width=1.\textwidth]{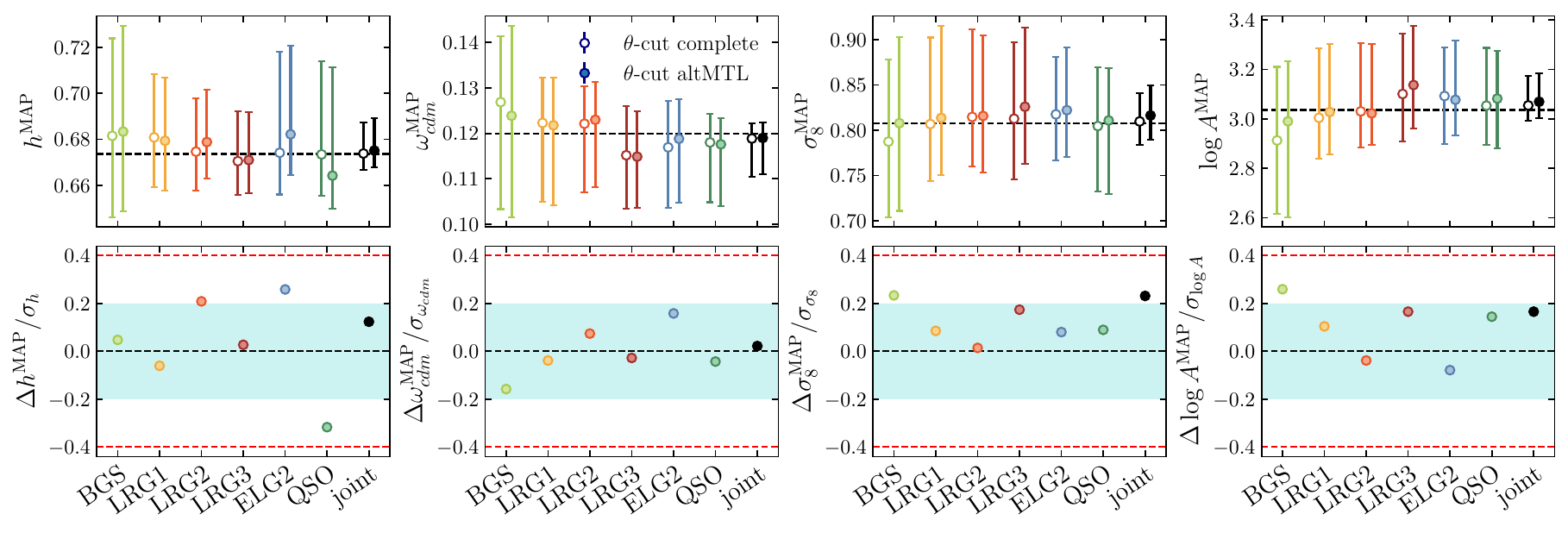}
    \caption{\textit{Top figure:} Full-Modelling results on each individual tracer and corresponding redshift bin(s) and the joint fit of the mean of the 25 \texttt{Abacus-2} complete mocks with or without the $\theta$-cut method. \textit{Bottom figure:} Comparison between complete and `altmtl' Abacus mocks with $\theta$-cut on $\Lambda$CDM parameters using Full Modelling for all individual tracers and for the joint fit. The blue bands display the $\sfrac{1}{5}$ of the \desidrone\ precision, whereas the red dashed lines the $\sfrac{2}{5}$. The symbols display the MAP value and the errorbars the $68.3\%$ confidence level around the mean. In all cases, the bottom panel shows the difference in the MAP of $\Lambda$CDM parameters.}
    \label{fig:fa-FM}
\end{figure}

\begin{figure}
    \centering
    \includegraphics[width=1.\textwidth]{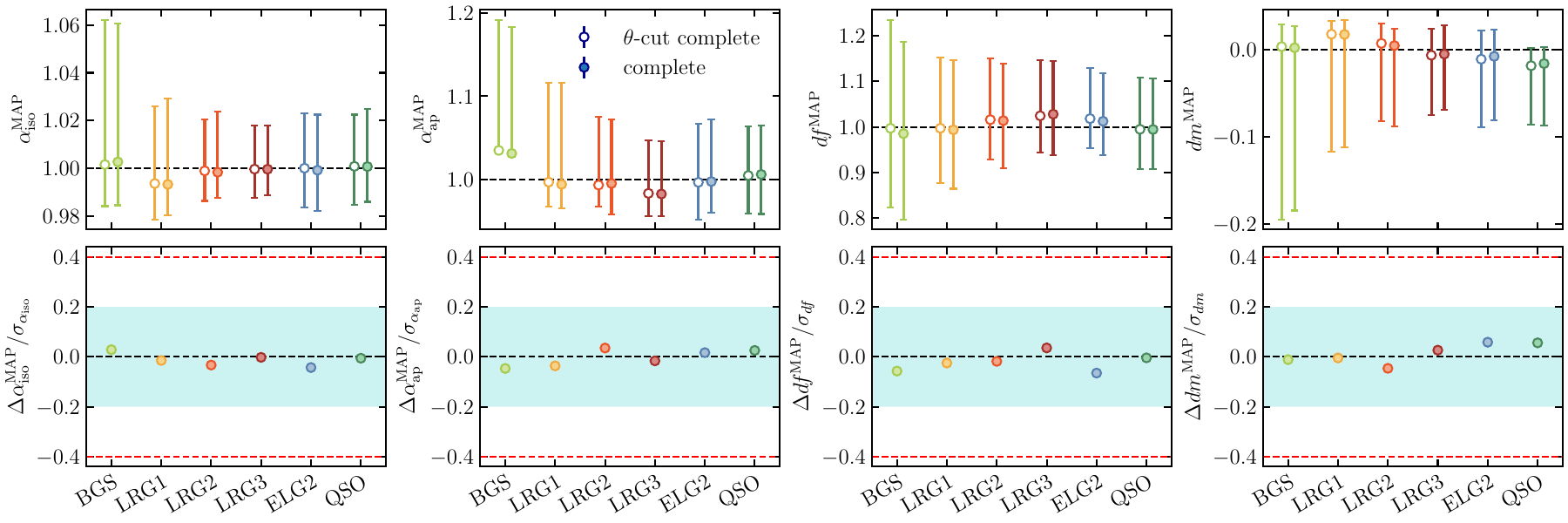}
    \includegraphics[width=1.\textwidth]{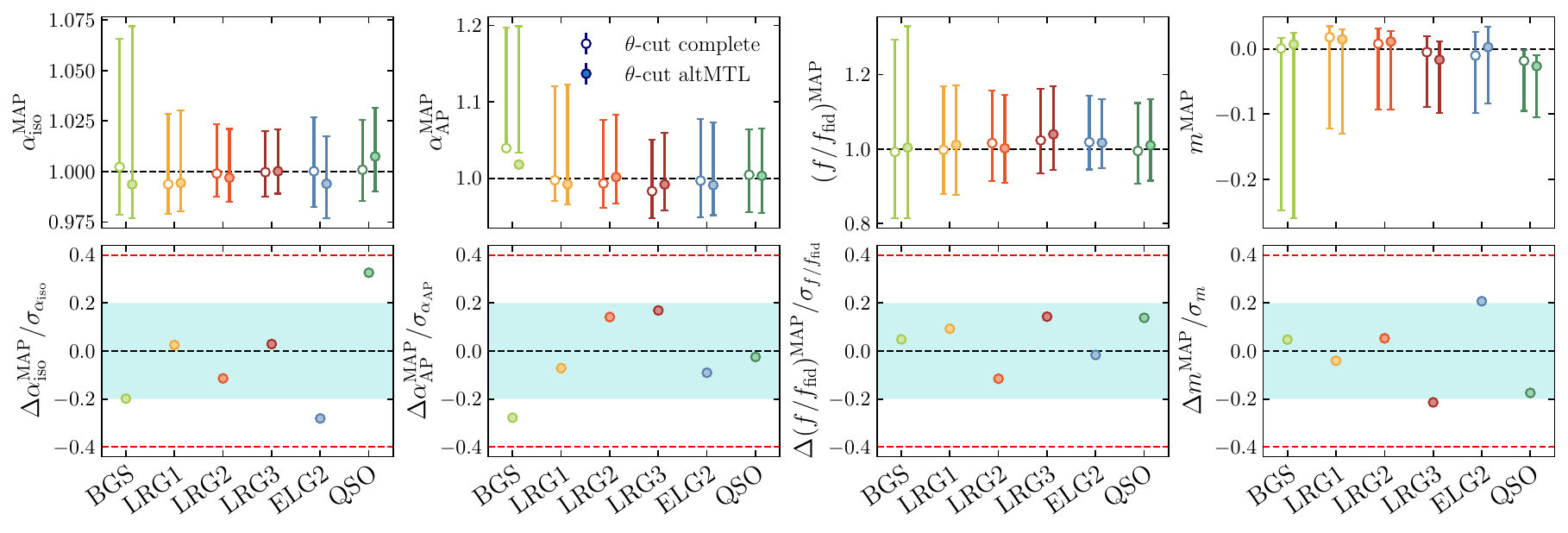}
    \caption{Same as \cref{fig:fa-FM} but for ShapeFit compressed parameters, $\{\alpha_{\rm iso},\,\alpha_{\rm AP},\,f/f^{\rm fid},\,m\}$. In this case, we do not include a joint fit as the compressed parameters in ShapeFit are in general $z$-dependent. Strong projection effects (PVE) causes in some bins that the centre of the error-bars do not coincide (or even overlap) with the MAP value. 
    }
    \label{fig:fa-SF}
\end{figure}

To further validate the $\theta$-cut method for Full-Shape galaxy two-point clustering analysis, we first look at the \desidrone\ \texttt{Abacus-2} complete mocks that do not have fibre incompleteness and fit the mean of the 25 mocks with and without a $\theta$-cut applied. Then, we look at the relative difference in cosmological parameters inferred from the \desidrone\ \abacus\ `complete' (without fibre incompleteness) and `altmtl' mocks (with fibre incompleteness as implemented into the data). We also fit the mean of the 25 mocks for each tracer and redshift bin. The results of the fits on the mean of 25 mocks for each tracer and redshift bin are shown in \cref{fig:fa-SF} for the ShapeFit parameters and in \cref{fig:fa-FM} for the Full-Modelling results on $h$, $w_\mathrm{cdm}$, $\sigma_8$ and $\log A_s$. In both cases, the figures on the top correspond to the difference between without and with $\theta$-cut on the complete \abacus\ mocks and the bottom figures show the difference between complete and `altmtl' mocks with $\theta$-cut. For the top and bottom figures, we show the MAP results and the bottom panel shows the difference between the two methods divided by the error with the blue shaded region representing $\pm$ 1/5 of DR1 error and the red dashed lines $\pm$ 2/5 of DR1 error. Given that we only have 25 realisations, the intrinsic statistical error is 5 times smaller than the \desidrone\ error, meaning that it is very likely to see a deviation within the shaded region, or in other words we cannot detect a systematic below 0.2$\sigma$ with only 25 mocks. If we focus on the top figures first, the effect of applying (or not) the $\theta$-cut method on complete mocks is negligible, which validates the accuracy of the $\theta$-cut method. Then, if we look at the bottom figures, the comparison between complete and `altmtl' mocks with $\theta$-cut informs us about any potential residual fibre-assignment effect which would not be fully mitigated by the $\theta$-cut method. As expected, we see a larger scatter in that case with some variations within the two red lines but with no systematic trend as a function of a given parameter or tracer. Therefore such deviations do not necessarily correspond to systematic offsets. For example, we see a consistent over-estimation of $\sigma_8$ across all the redshift bins. Moreover, the results from the joint fit, which is the case used for the final cosmological constraints, lie mostly within 1/5 of DR1 error. Therefore, we conclude that the $\theta$-cut method works well enough to mitigate the effect of fibre assignment in DR1 galaxy Full-Shape analysis, and we do not report any residual systematic contribution.
For future analysis of the impact of DESI fibre assignment, we will produce mocks with a more realistic fibre-assignment effect that will allow us to test the $\theta$-cut method further and also include the Probability Inverse Pair (PIP) weighting \citep{PIP,Smith19} which should allow us to also fit smaller scales. Although this was not practical for DR1 samples given the large fraction of the sample with just one-pass regions, it would be possible for DR2, where the one-pass fraction is much smaller. 

\subsection{Imaging Systematics}\label{subsec:sys-imaging}

As presented in \cite{KP3s15-Ross} and explained in detail in \cite{DESI2024.II.KP3}, the standard way to remove the spurious fluctuations in the target density with imaging observing conditions is to apply weights that correct for these dependencies.
There are three different regression methods that lead to different imaging weights: 1) linear method, as used in the eBOSS LSS catalogues \citep{eBOSSlss}, 2) \textsc{Regressis} random forest (RF) \cite{KP3s10-Chaussidon}, 3) non-linear \textsc{SYSNet} neural net (NN) \citep{sysnet} as used in \cite{Zarrouk:2020hha,2024MNRAS.532.1902R}. Linear weights are sufficient to capture most of the trend in target density with imaging observing conditions for the BGS and LRG samples, which are caused by bright objects. For the ELG and QSO samples, we found that different choices of imaging weights (linear, RF, NN) yield dissimilar clustering measurements, which can impact the cosmological constraints. To handle this we have developed a template correction, with one extra free parameter ($s_p$), allowing us to marginalise over some of the remaining imaging systematic effects in the data. Further, when we ran null tests on mocks uncontaminated by systematics, we found that methods beyond linear regression tend to remove some modes in the signal, which could bias our estimated two-point functions. Thus we also apply a mode removal correction, which is estimated from mocks. For consistency, we decided to test these two corrections on all tracers and study their impact on the clustering and the cosmological results.

The effective model used to fit the data is defined by:
\begin{eqnarray}
\nonumber\mathrm{model}_\mathrm{eff} &=& \mathrm{theory} + s_p \mathrm{Poly}[(X^{\mathrm{data}}_{\mathrm{weight}} - X^{\mathrm{data}}_{\mathrm{noweight}}) - (X^{\mathrm{mock}}_{\mathrm{weight}} - X^{\mathrm{mock}}_{\mathrm{noweight}})] +\\
 &+& \mathrm{Poly}[X^{\mathrm{mock}}_{\mathrm{weight}} - X^{\mathrm{mock}}_{\mathrm{noweight}}]
\label{eq:imaging-corr}
\end{eqnarray}
where $s_p$ is a free parameter, $\mathrm{Poly}$ is a polynomial function of third order\footnote{We adopted a form of $c_{-5}k^{-5} + c_{-3}k^{-3} + c_{-2}k^{-2}$ for the power spectrum to get a smooth template.}, $X$ is either the power spectrum or the correlation function.
The third term corresponds to the mode removal correction for angular integral constraints associated with the weighting scheme and the second term corresponds to the polynomial template correction that corrects for any remaining systematic effect caused by the imaging weighting method itself. The level of added noise and large scale angular modes removal depends on the regression method used and in order to estimate it, we look at the difference between with and without weights in the mocks. For a more detailed description of these effects, we refer the reader to section 10.1.2 of \cite{DESI2024.II.KP3}.

\begin{figure}
    \centering
    \includegraphics[width=1.\linewidth]{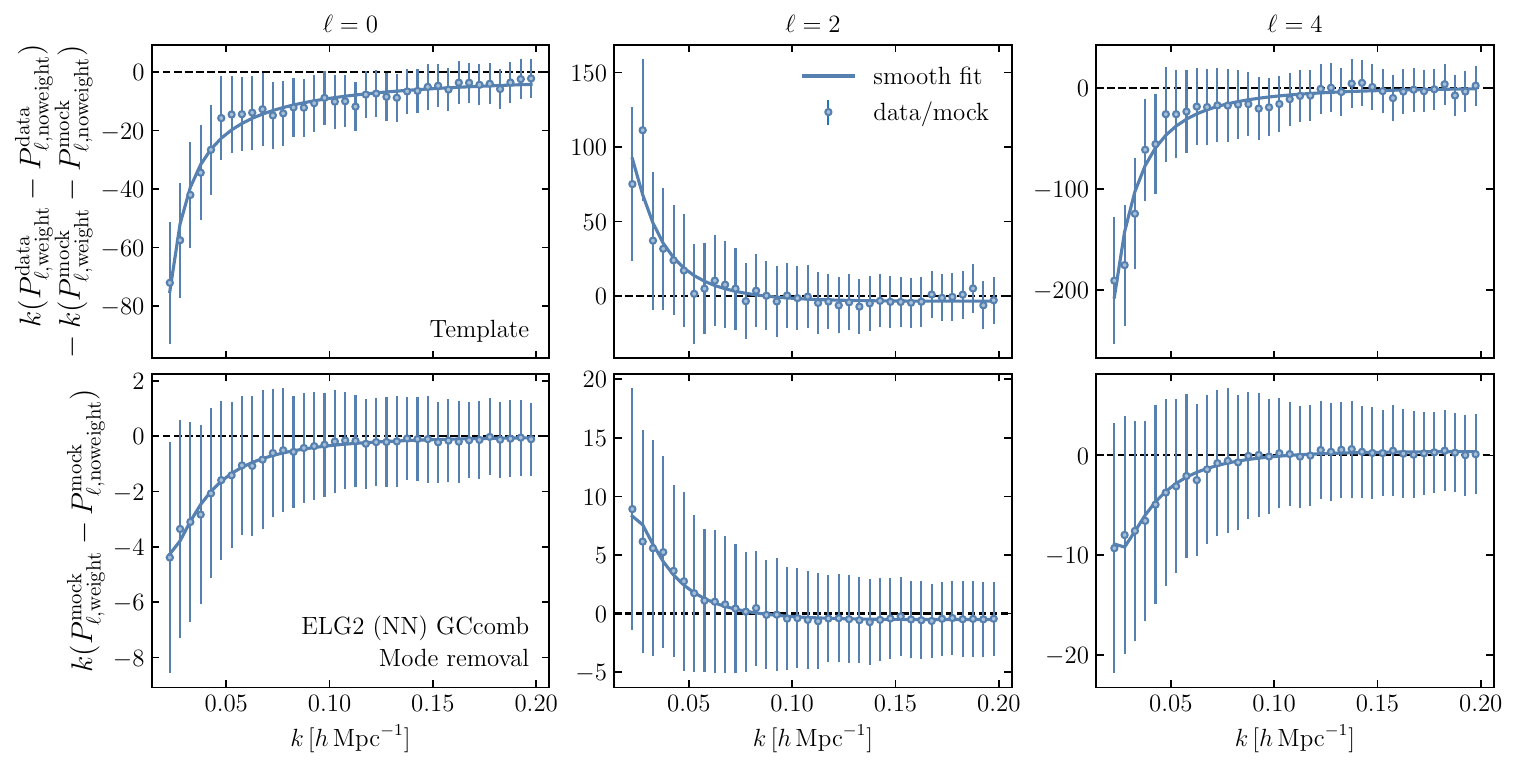}
   \caption{Monopole, quadrupole and hexadecapole residuals of the ELG sample in the redshift bin $1.1 < z < 1.6$. The upper panels show polynomial corrections derived from data and Abacus mocks, lower panels show mode removals derived from mocks. Smooth fits are used as templates in the fits of the ELG and QSO samples only as these corrections are found to be small for BGS and LRG. The covariance matrix is computed from the 1000 DR1 \texttt{EZmocks} and is rescaled by 1/5 in the bottom panel for visibility.}
    \label{fig:im-syst-ELG}
\end{figure}

The polynomial template correction is determined for each multipole by plotting $(P^{\mathrm{data}}_{\mathrm{weight}} - P^{\mathrm{data}}_{\mathrm{noweight}}) - (P^{\mathrm{mock}}_{\mathrm{weight}} - P^{\mathrm{mock}}_{\mathrm{noweight}})$ as a function of $k$.
The second correction for mode removal is directly computed from the mocks. Both effects are illustrated in \cref{fig:im-syst-ELG} for the ELG sample.
We implement this effective model in \texttt{desilike} and we validate it on uncontaminated \texttt{Abacus-2} mocks for the ELG to make sure adding those corrections does not bias the cosmological results. In particular, we tested different priors on the free parameter $s_p$ and showed its impact on the cosmological parameters.
For more information about the imaging systematic effect and its mitigation for \desidrone\ Full-Shape galaxy two-point clustering analysis, we refer the reader to our supporting paper \cite{KP5s6-Zhao}. 

The conclusions of this study are the following: 1) For BGS and LRG, we do not detect any systematic bias due to imaging systematics that the linear weights would not capture, 2) For ELG and QSO, using the effective model given by equation~\eqref{eq:imaging-corr} in addition to the \textsc{SYSNet} weights for ELG and \textsc{Random Forest} weights for QSO adds a systematic error contribution that represents up to 1/4 of \desidrone\ statistical error. 
In our fiducial analysis, we adopt the model of ~\eqref{eq:imaging-corr} at the level of the two-point statistics. We do so by adding the constant angular mode removal piece $\mathrm{Poly}[X^{\mathrm{mock}}_{\mathrm{weight}} - X^{\mathrm{mock}}_{\mathrm{noweight}}]$ to the measured (rotated) power spectrum (as presented in~\cite{DESI2024.II.KP3}), and adding to the statistical covariance matrix C the systematic covariance matrix $(s_p X_\text{poly})(s_p X_\text{poly})^T$, where $s_p = 0.2$ is the chosen scale of the Gaussian prior on $s_p$ and $X_\text{poly}$ is the polynomial template correction. Adding the systematic covariance is equivalent to analytically marginalizing over the $s_p$ parameter in the effective model.

\subsection{Spectroscopic Systematics}\label{subsec:sys-spectro}
Spurious non-cosmological density fluctuations can arise from variations in the galaxy density due to the spectroscopic observing conditions. These spurious fluctuations can impact the redshift success rate by lowering the redshift efficiency in some regions of the sky. Therefore, by studying the uniformity of the redshift success rate, we can build a model to capture its dependence on observing conditions and develop redshift failure weights ($z_\mathrm{fail}$). For more details, see \cite{KP3s3-Krolewski} which presents the characterisation of the spectroscopic systematics and its impact on the large-scale structure measurements. They find that the small remaining trends with spectroscopic observing conditions lead to less than $0.2\sigma$ shifts in the cosmological parameters, which can be considered negligible. \cite{KP3s4-Yu} focused on the ELG sample which is the most challenging to measure accurate redshifts for. They found that the redshift failure weights also have a minor influence on the galaxy clustering and they also investigated two sources of systematic effects for ELGs with a secure redshift: 1) catastrophic redshifts which correspond to a large shift in the radial position of a small fraction of objects and 2) redshift uncertainty which affects all galaxies. The former represents 0.26\% of the total DR1 ELG sample and its effect has been simulated on \texttt{Abacus-2} DR1 ELG mocks, together with other cases of catastrophic redshifts. \cite{KP3s4-Yu} showed that the impact of catastrophic redshifts on Full-Shape galaxy clustering, both with ShapeFit and Full Modelling, leads to a systematic contribution less than 0.2$\sigma$ of the statistical error, again making this systematic effect negligible. The latter effect can be measured using repeated observations and modelled with a Lorentzian profile. The redshift uncertainty for ELG DR1 is more than twice smaller than the one from eBOSS ELGs, highlighting the improvement of the spectrographs and redshift pipeline of DESI in measuring accurately the [O II] doublet of ELGs.

\subsection{Systematics related to the covariance matrix estimate}\label{subsec:sys-cov}

To estimate the systematics associated with our choice of the covariance matrix, we perform both Full Modelling and ShapeFit fits on \desidrone\, \texttt{Abacus-2} mocks using either analytic or \texttt{EZmocks} covariance.
The results presented in this paper are based on the supporting papers exploring the impact of the \textsc{RascalC} code for generating analytical covariance in configuration space  \cite{KP4s7-Rashkovetskyi}, \textsc{TheCov} code in Fourier space \cite{KP4s8-Alves}, and their application and comparison with mock-based covariance in the context of DESI \cite{KP4s6-Forero-Sanchez}. 

Below we summarise the key findings that set the methodology to account for systematic effects in our baseline \texttt{EZmocks} covariance:
\begin{enumerate}
    \item When performing the fits on \abacus\ mocks without fibre-assignment effect, \texttt{EZmocks} and analytic covariances agree very well, both in Fourier and configuration spaces.
    
    \item When performing the fits on \abacus\ mocks with `altmtl' fibre assignment\footnote{This is, using the `altmtl' version as it closely corresponds to the realistic data processing, whereas the fast fibre-assignment procedure is approximate but more efficient, and thus can be applied to hundreds of mocks.}, there is a $\sim 20\%$ mismatch between \texttt{EZmocks} (with fast-fibre assignment) covariance and analytic covariance tuned on \texttt{EZmocks} clustering in Fourier space. However, in configuration space, both \texttt{EZmocks} and analytic covariance match well within the expected statistical scatter, when the analytic covariance is calibrated on the \texttt{EZmocks}.
     
    \item When we compare the elements of the \textsc{RascalC} analytical covariance matrix (tuned to reproduce the \desidrone\ data) with the \texttt{EZmocks} (with fast-fibreassign) prediction we find that \textsc{RascalC} matrix elements present a higher variance than \texttt{EZmocks} matrix elements. In other words, we find that the minimum $\chi^2$ in fits to the data is higher when employing the  \texttt{EZmocks} covariance than when employing the \textsc{RascalC} one.    
\end{enumerate}

Thus, we conclude that \texttt{EZmocks} with fast-fibreassign are not able to capture all aspects of the \desidrone, which results in an underestimation of the error bars. This is in part caused by the limited fast-fibre assignment method applied to the mocks, but also in part by the nature of these approximate mocks. Therefore, we can use the \texttt{RascalC} analytic covariance matrix in configuration space tuned on the data to re-calibrate the EZmock covariance. The correction corresponds to a rescaling of all the covariance elements by a tracer-dependent factor previously introduced in \cref{tab:cov-corr}. These are computed by comparing the covariance matrix scaled by the inverse of the reduced chi-squared of \texttt{RascalC} analytical covariance, and the mock-based covariance, both evaluated in configuration space over the range $20 < s < 200 \hinvmpc$. For a more detailed discussion about the correction and its robustness with respect to the scale range and number of multipoles used to compute it, we refer the reader to section~10.2 and equation~10.12 of \cite{DESI2024.II.KP3}.

\subsection{Total systematic error budget}\label{subsec:sys-total}

As discussed in \cref{subsec:sys-hod}, the baseline Full-Modelling approach for constraining the cosmological parameters makes the contributions from systematic effects at the parameter level more difficult to interpret beyond $\Lambda$CDM models. In order to avoid repeating the study of systematics for each cosmological model considered, we include the systematic error budget at the data vector level. Further details about how the systematic and statistic contributions of the covariance can be added are discussed in \cref{app:invcov}.

The final covariance matrix for Full-Shape analysis of the galaxy power spectrum is inferred as follows. The covariance matrix of the galaxy power spectrum monopole and quadrupole elements, ($N_{\rm bins}$), is estimated for each tracer and redshift bin with $N_\mathrm{obs} = 1000$ \texttt{EZmocks}. Then, the corrective factors described in \cref{subsec:sys-cov} are applied and are multiplied by the (inverse) Hartlap \cite{Hartlap2007} factor. At this point, we could have also followed the Sellentin-Heavens approach \cite{Sellentin-Heavens}  to obtain an unbiased likelihood, however, for this large number of mock realizations the difference between these two approaches is not significant (see \cite{Gil-Marin2022} for a comparison). The Hartlap inverse matrix which gives an unbiased estimate of the inverse covariance matrix is,

\begin{equation}
C_\mathrm{Hartlap 2007}^{-1} = \frac{N_\mathrm{obs} - 1}{N_\mathrm{obs} - N_\mathrm{bins} - 2}
\label{eq:hartlap}
\end{equation}
and the \cite{Percival2014} correction factor (to propagate uncertainty in the covariance matrix to final cosmological constraints):

\begin{equation}
C_\mathrm{Percival 2014} = \frac{1 + B (N_\mathrm{bins} - N_\mathrm{params})}{1 + A + B (N_\mathrm{params} - 1)}
\end{equation}
with
\begin{equation}
A = \frac{2}{(N_\mathrm{obs} - N_\mathrm{bins} - 1) (N_\mathrm{obs} - N_\mathrm{bins} - 4)}, B = \frac{N_\mathrm{obs} - N_\mathrm{bins} - 2}{(N_\mathrm{obs} - N_\mathrm{bins} - 1) (N_\mathrm{obs} - N_\mathrm{bins} - 4)}
\label{eq:percival}
\end{equation}
assuming $N_\mathrm{params} = 7$ parameters (corresponding to bias parameters $(1 + b_1) \sigma_8$, $b_2 \sigma_8^2$, $b_s \sigma_8^2$, counter-terms $\alpha_0$, $\alpha_2$ and stochastic parameters $\mathrm{SN}_0$, $\mathrm{SN}_2$).\footnote{This results in a conservative estimate: considering 12 parameters instead (e.g. including cosmology parameters) would reduce this factor by $1\%$.}
Both Hartlap and Percival corrections are small and represent less than the rescaling factor we applied to the EZmock covariance, therefore we can assume a Gaussian likelihood for the DESI DR1 analysis. Other recent works also showed that cosmological measurements from large scale structures are robust with respect to the statistical assumptions about the likelihood distribution \citep{Krywonos:2024fqm} and that a Gaussian likelihood with appropriate corrections to the covariance is similar to a t-distribution \citep{Percival:2021cuq}. The advantage of assuming a Gaussian likelihood is that we can use analytic marginalisation to speed up the cosmological inference. The validity of the Gaussian likelihood assumption has to be evaluated for each statistic of interest and for future analyses with additional data, though.

Eventually, to the galaxy power spectrum covariance, we add the systematic covariance matrices, namely:
\begin{itemize}
\item the marginalisation of the window matrix rotation parameter $\vec{s}$, as detailed in~\cite{DESI2024.II.KP3, KP3s5-Pinon}
\item the HOD+PWE systematic covariance matrix, as explained in \cref{subsec:sys-hod} and detailed in~\cite{KP5s7-Findlay}
\item the imaging systematic covariance matrix, as explained in \cref{subsec:sys-imaging} and detailed in~\cite{KP5s6-Zhao} 
\end{itemize}

The total systematic error budget at the parameter level is summarised in \cref{tab:systematicbudget} for both Full Modelling and ShapeFit.
Systematic contributions at the level or above 0.2$\sigma$ of DR1 error are added in quadrature to obtain the total systematic error contribution at the parameter level. This includes the effect due to imaging systematics, fibre-assignment, and HOD+PWE, and it results in a total systematic error budget of 0.46$\sigma$ of DR1 error.

\begin{table}[!ht]
    \centering
       \resizebox{\columnwidth}{!}{
    \begin{tabular}{|l|c|l|}
        \hline
        \textbf{Systematic} &  \textbf{Methodology} & \textbf{Contribution} \\
        & & \textbf{(units of $\sigma$ DR1)} \\
        \hline\hline
        Theoretical & Comparison between 4 EFT models & not detected for DR1  \\ 
        & $\kmax = 0.20\ihMpc$ & ($<0.1$) \\
        \hline
        Observational &  &  \\
        a. Imaging & Imaging weights per tracer, mode removal & $\sim$ $0.2$ (ELG, QSO) \\
        & and polynomial correction (ELG, QSO) & $<0.1$ (BGS, LRG) \\        
        b. Spectroscopic & Tested with mocks and & $<0.2$ (ELG)\\
        & repeated observations & $<0.1$ (BGS, LRG, QSO) \\
        c. Fiber & $\theta$-cut method tested on mocks & $\sim0.2$ \\
        assignment & with and without fibre-assignment & \\ 
        \hline
        HOD+PWE & Varying HOD in Abacus-1 cubic & HOD: $\sim 0.3$ (Table 2 of \cite{KP5s7-Findlay}) \\
        & and DR1-like mocks & PWE: $\sim 0.2$ \\
        \hline
        Fiducial & Varying catalogue cosmology & $<0.2$ (FM, SF)  \\
        cosmology & Varying catalogue and template cosmology &   \\
        \hline
        Covariances & Based on comparisons between analytic and & $<0.2$ \\
        & mock covariances, rescaling factor &  \\
        \hline
        \textbf{Total} & All contributions above 0.2$\sigma$ of DR1 & $\sim0.46$ (FM, SF) \\
        & error are added in quadrature &  \\
        \hline
    \end{tabular}
    }
    \caption{Summary of individual systematic errors. When there is no significant difference between each tracer or between Full Modelling (FM) and ShapeFit (SF), we just quote a single systematic contribution. Only the systematic contributions which are equal to or above 0.2$\sigma$ of DR1 error are included in our total systematic error budget. In some cases, we do not have any statistically significant detection of a systematic bias (below 0.1$\sigma$ of DR1 error).}
    \label{tab:systematicbudget}
\end{table}

\begin{figure}
    \centering
    \includegraphics[width=0.98\linewidth]{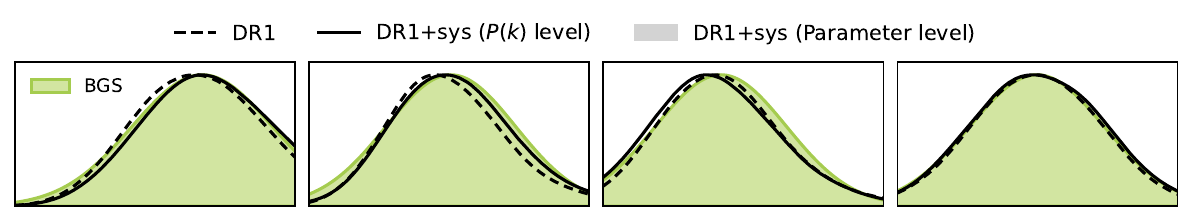}
    \includegraphics[width=0.98\linewidth]{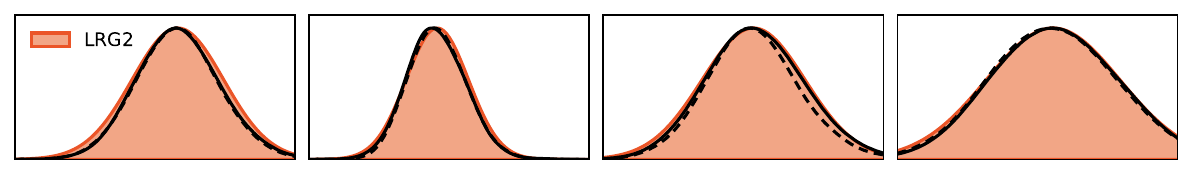}
    \includegraphics[width=0.98\linewidth]{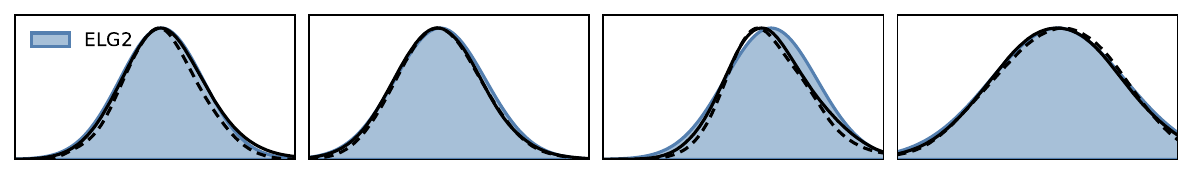}
    \includegraphics[width=0.98\linewidth]{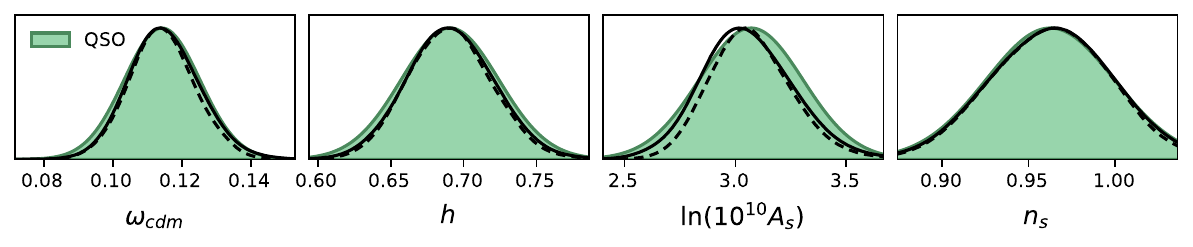}
    \caption{    Validation of the inclusion of systematics at the $P(k)$-level on DR1 mocks for the \bgs, \lrgt, \elgt\ and \qso\ samples. 
    The dashed line displays the 1D profile obtained without any systematic component. The solid line displays the 1D profile when the systematic component (imaging, fibre assignment and HOD+PWE systematics) is added at the $P(k)$-level. The coloured region displays a Gaussian profile centred at the mean of the dashed profile with a standard deviation obtained by adding the parameter-level systematic contribution to that of the dashed profile. Imaging and fibre assignment systematic components have been added according to the values quoted in \cref{tab:systematicbudget},  the HOD systematic component has been added according to the values in table 2 of \cite{KP5s7-Findlay} and the maximum difference between MAP values with flat and baseline nuisance priors has been added to account for the PWE.}
    \label{fig:pk_cov_vs_Omega_cov}
\end{figure}

\cref{fig:pk_cov_vs_Omega_cov} shows the comparison between including the systematic contributions at the parameter and data vector levels for the joint fit of the six \desidrone\ galaxy redshift bins. We remind that the systematic contributions at the parameter level directly include the effect of fibre-assignment while at the data vector level, we only include the marginalisation of the window matrix rotation. Despite those different treatments, as one can see in \cref{fig:pk_cov_vs_Omega_cov}, both approaches yield very similar results, which justifies our decision to include systematic errors at the data-vector level. 

Eventually, in order to combine the Full-Shape analysis of the pre-reconstruction galaxy power spectrum with the post-reconstruction BAO measurements, the joint covariance between the power spectrum and post-reconstruction BAO parameters ($\alpha_\mathrm{iso}$ for BGS and QSO, $\alpha_\parallel, \alpha_\perp$ for LRG and ELG) is first estimated for each tracer and redshift bin with $N_\mathrm{obs} = 1000$ \texttt{EZmocks}. Then, it is also corrected by the rescaling factors described in \cref{subsec:sys-cov}, where the size of the joint matrix is $N_\mathrm{bins} = 2 \times 36 + 2$ (resp. $N_\mathrm{bins} = 2 \times 36 + 1$)\footnote{Two Legendre multipoles, with 36 $k$-bins each, and 1 or 2 post-reconstructed BAO parameters.} for LRG and ELG (resp. BGS and QSO), and by the factors defined by equations \ref{eq:hartlap} and \ref{eq:percival}, respectively. Then, we use for BAO parameters the covariance obtained from the DR1 BAO data. Specifically, we split the joint covariance in terms of the correlation matrix and diagonal variance and replaced the part corresponding to BAO parameters with the corresponding BAO parameter correlation and variance estimated from the posterior of BAO fits to the DR1 data post-reconstruction correlation function. To the auto-BAO block of the covariance, we add systematic contributions as described in~\cite{DESI2024.III.KP4}.

\section{The unblinding tests}
\label{sec:unblinding}

From the initial stages of the survey, even before the data was taken or inspected, DESI collaboration decided to imprint a post-processing on the redshifts of the galaxies to distort the true underlying cosmological signal, which we refer as the blinding process, as previously described in \cref{subsec:blinding}. As far as we are aware, this represents the first ever blinded Full-Shape analysis of a galaxy redshift survey. 

The creation of the large-scale structure catalogues for the six tracer samples, took place over six iterations over 13 months as described in \cite{DESI2024.II.KP3}. During this evolution, the catalogue remained blinded to avoid confirmation bias, limiting our ability to unconsciously make any changes in the process to reproduce results we feel comfortable with (i.e., results supported by previous observations). The evolution leading to the final six-tracer catalogues of the \desidrone\, included performing a series of tests on our data pipeline, resulting in the inclusion of the latest techniques for correcting for the DESI-related systematics, the imaging and fibre-assignments (see \cref{subsec:sys-imaging} and \cref{subsec:sys-fibre}). In addition, during this time, the modelling pipeline evolved to include the final prior choice, the range of scales that we fit, and a free parameter which encodes the uncertainty in the imaging weights that is marginalised over in the analysis (see equation~\ref{eq:imaging-corr}). 

We instigated a series of tests and conditions that need to be passed before we would consider unblinding the catalogue. These include sanity and consistency checks on the data catalogues, freezing the model pipeline including the perturbation theory model, the priors on the parameters, the fitting ranges, the way that the data covariance is estimated, and the way the systematic error budget is defined. Most of the data tests were performed using ShapeFit instead of the baseline Full-Modelling approach, due to the following: 1) We demonstrated that for a large range of models ShapeFit and Full Modelling can retrieve the same information; 2) ShapeFit does not rely on the assumption of a specific choice of cosmology; 3) ShapeFit is more flexible in identifying potential systematic effects in the shape of the power spectrum, as each compressed variable is associated to a particular range of scales; 4) Since at this point we do not need to make the cosmological interpretation, we can analyse the ShapeFit results in terms of compressed parameters and not suffer from systematic shifts due to the prior volume effects. The full conditions to unblind are listed below and summarised in \cref{tab:Y1checklist}.

\begin{enumerate}
\item  The official \desione\ galaxy Full-Shape analysis base pipeline should be decided before unblinding. This includes the choice of compressed or full-modelling approaches, the choice of EFT model, the definition of nuisance parameters, their priors and the baseline set-up (fitting range, binning, multipoles considered). The resulting galaxy Full-Shape analysis baseline has been summarised in \cref{subsec:methods-baseline}.

\item The different options for combining the BAO post-reconstruction and the Full-Shape pre-recon measurements should be identified and described before unblinding. Although the final decision on which model to implement can be decided after unblinding, this choice was not based on the unblinded catalogue performance. As previous studies have demonstrated, the different available methodologies perform very consistently. This has been described in \cref{subsec:methods-BAO-FS}. 

\item The upper limit of systematic errors has to be determined before unblinding in order to avoid inflating the final error bars such that the final results align better with expectations influenced by confirmation bias. \cref{tab:systematicbudget} summarises the list of systematic effects and their maximal effect on the cosmological parameters that we constructed before unblinding. However, the decision and the exact methodology on how to include the systematics at the power spectrum $k$-bins level was developed after unblinding because of its novelty. Each individual source of potential systematic effects is described in detail in \cref{sec:systematics}. Further details on how the power spectrum systematic covariance is defined are in \cite{KP5s7-Findlay}.

\item  The minimal set of consistency tests described in the checklist (listed in \cref{tab:Y1checklist}) has to be completed on the blinded data and repeated on unblinded data as a first validation after unblinding. Only ShapeFit fits are allowed to be performed on blinded data, as performing Full-Modelling fits would inform us about the cosmological interpretation of the results and thus could be broken or be inconsistent with the blinding procedure. These consistency tests include i) changing the baseline covariance with an analytic covariance estimate, ii) checking the consistency between NGC+SGC and NGC only\footnote{The size of SGC only is too small to be compared on its own.} and the Northern and Southern photometric areas, iii) comparing the ShapeFit results on $\aiso$ and $\alap$ with BAO post-recon blinded results based on $\xi(s)$, iv) comparing the Fourier space ShapeFit results with their configuration space counterparts.
\end{enumerate}

\begin{table*}
    \centering
   \resizebox{\columnwidth}{!}{
        \begin{tabular}{|l|c|}
        \hline
        {\bf Test} & {\bf Results} \\\hline\hline
        $\chi^2$ acceptable for a ShapeFit fit? & Yes  \\\hline
        $\alpha_{\parallel,\,\perp}$ from ShapeFit consistent with BAO post-recon? & Yes \\\hline
        $\alpha_{\parallel,\,\perp}$ consistent between NGC+SGC and NGC? & Yes \\\hline
        Consistency between North and South? & Yes (at the level of the clustering, see section~\ref{sec:catalog}) \\\hline
        ShapeFit results consistent between $\xi(r)$ and $P(k)$? & Yes \\\hline
        \end{tabular}
       }
    \caption{\label{tab:Y1checklist} Unblinding checklists using the baseline choices of section~\ref{subsec:methods-baseline} except for the choice of ShapeFit instead of Full-Modelling (see text). For each entry in the 
    checklist, if the difference observed in the blinded data is within the full range covered by the 25 \abacussecond\ DR1 mocks, we consider that the test was passed.}
    
\end{table*}

In addition, before unblinding, we decided which tests and changes we would allow ourselves to do after unblinding without breaking the blinding rules. These are listed below:
    
\begin{enumerate}  
    \item We allow ourselves to update the covariance matrix if a more precise covariance matrix estimate is available after unblinding. This choice is independent of unblinding and not based on unblinding results. This includes a covariance derived from more realistic mocks or a better theory model. 
    \item We allow ourselves to update the fiducial template cosmology as the effect of varying the template can be mitigated by a re-analysis (after unblinding) in a new cosmology such that the $\alpha_\parallel$ and $\alpha_\perp$ parameters are very close to 1, and $m$ close to 0. This point only applies to the ShapeFit analysis, and not to the Full-Modelling. However, if the underlying catalogue cosmology is found to be very different from the obtained cosmology, the catalogue cosmology would be also updated.
    
    \item We prioritised the determination of the HOD systematics for the LRG and ELG samples, the LRG redshift bins being the most constraining ones and the galaxy-halo connection and its impact on cosmological parameters for ELG being less investigated than the other tracers. Therefore, we allow ourselves to determine the HOD systematics for BGS and QSO after unblinding. This systematic error estimation does not rely on the data, only on mocks, so we are confident that this preserves the spirit of the blinding procedure.
\end{enumerate}

The unblinding event took place on 12th June 2024. A few hours before the internal presentation the unblinding catalogues (i.e. those catalogues without the blinding redshift shift applied) were produced, and the ShapeFit and Full Modelling pipeline ran on them to produce a final version of the results. Some time after the unblinding event we added the systematic error budget to the data and also explored its cosmological impact in the light of external datasets, as its is described in \cite{DESI2024.VII.KP7B}. Since then, the data and model pipelines have remained frozen.

\subsection{Consistency tests on unblinded data}\label{sec:consistency_unblindeddata_SF}

\begin{figure}
    \centering
    \includegraphics[width=1.\textwidth]{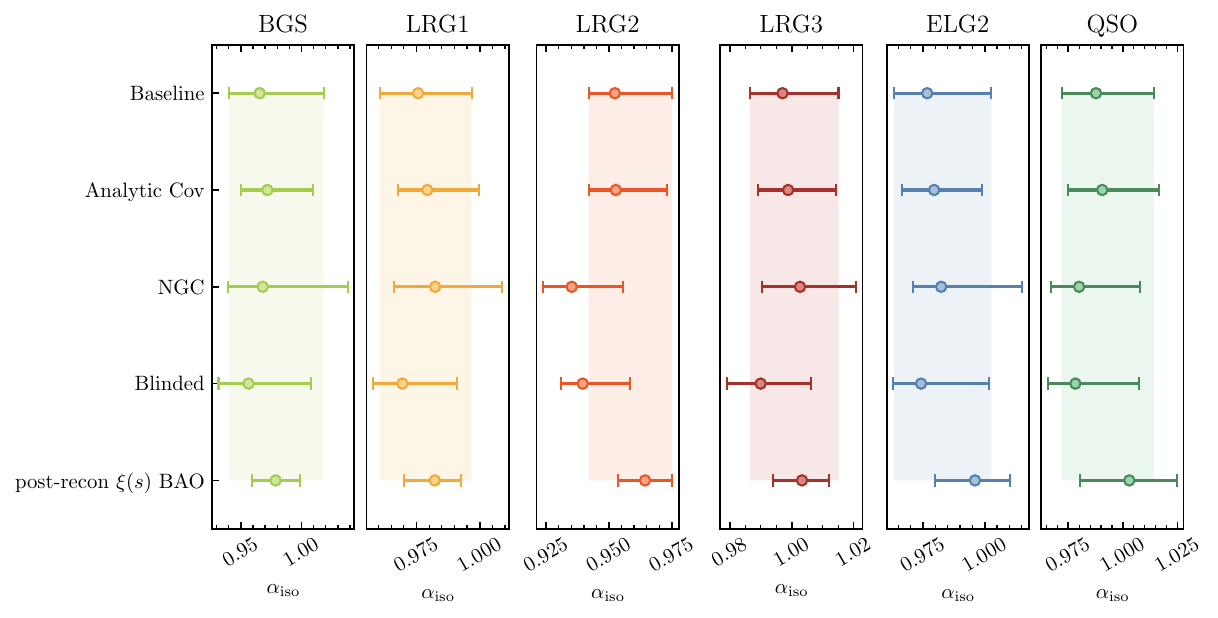} \\
    \includegraphics[width=1.\textwidth]{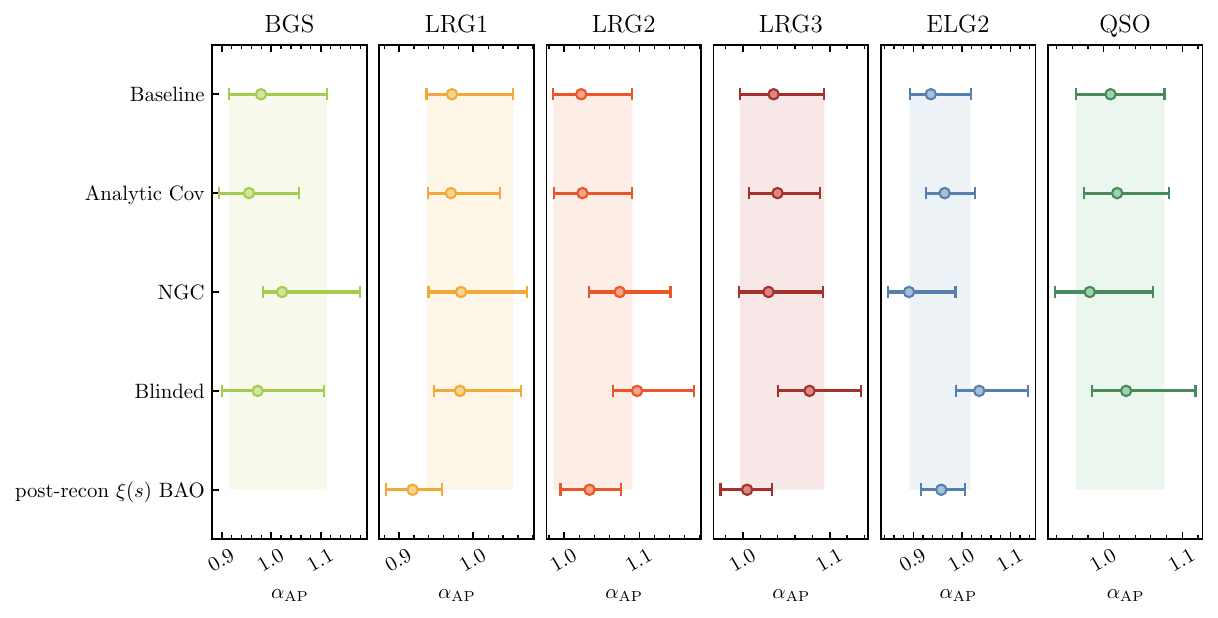}
    \caption{Consistency tests performed on the unblinded catalogues for each tracer and redshift bins (each of the six tall vertical panels) for the isotropic dilation parameter (top panels) and for the Alcock-Paczynski parameter (bottom panels). Circles represent MAPs and the error bars show the 68\% credible interval. We tested the consistency between two covariance matrix estimates (mock-based and analytic), between NGC+SGC and NGC only, and between post-recon BAO for the alphas.  We also show the consistency between the unblinded and blinded results for the baseline. The shaded area corresponds to the 68\% credible intervals of the baseline result (first row in each panel).}
    \label{fig:unblind-tests-alpha}
\end{figure}

\begin{figure} 
    \centering
    \includegraphics[width=1.\textwidth]{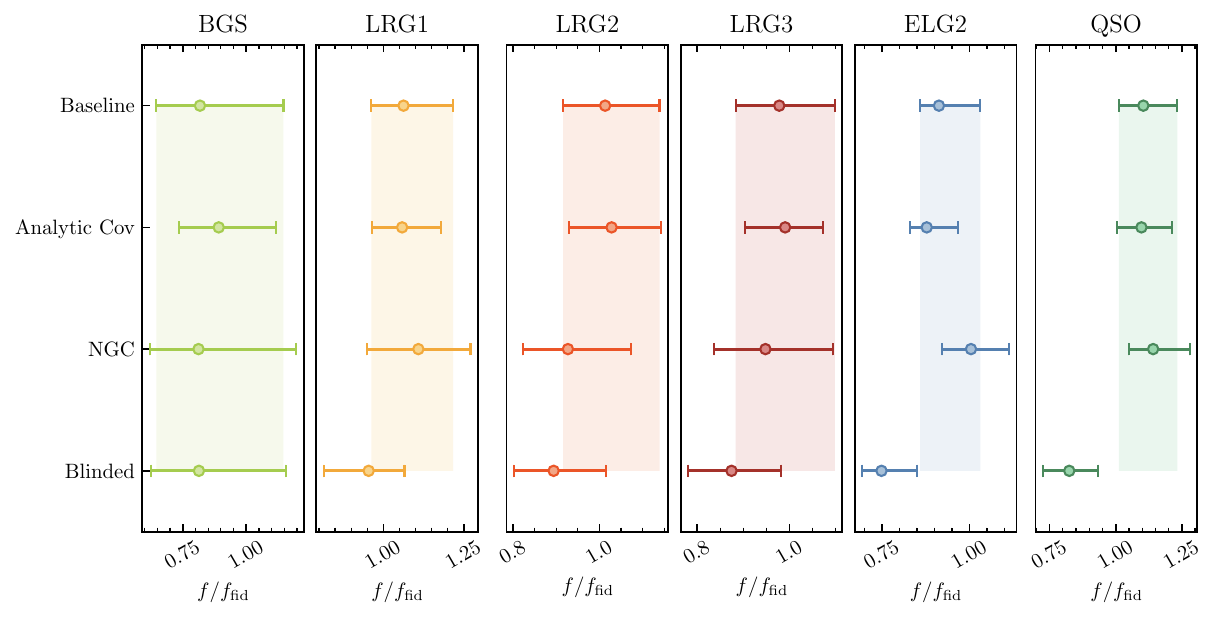} \\
    \includegraphics[width=1.\textwidth]{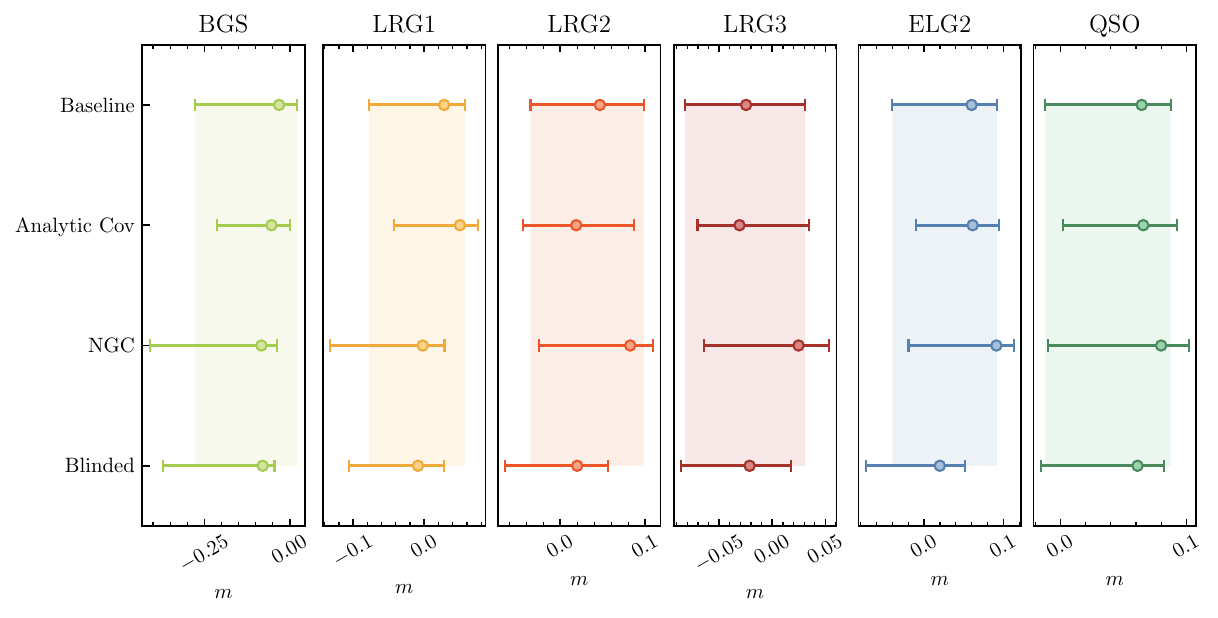} \\
    \caption{Same as \cref{fig:unblind-tests-alpha} but including the growth of structure parameter normalised by the template fiducial cosmology value, $f/f^{\rm fid}$ (top panel) and by the shape parameter $m$ (bottom panels).}
    \label{fig:unblind-tests}
\end{figure}

\cref{fig:unblind-tests-alpha} and \cref{fig:unblind-tests} summarise the consistency tests performed on unblinded data for each of the six \desidrone\, tracers and redshift bin considered as displayed in \cref{tab:Y1data}. The first row describes the results obtained from the baseline setup as described in \cref{subsec:methods-baseline}. As commented before, the only exception in this baseline setup is that we employ the ShapeFit scheme instead of Full-Modelling. In short, these are fits to the power spectrum multipoles using the \ezmock\ covariance matrix to analyse the NGC+SGC catalogue and with a set of priors described in \cref{tab:priors}. The other rows describe variations of this baseline setup:  we test the consistency with another covariance matrix based on analytical perturbation theory and with configuration space results; we test the consistency between NGC+SGC and NGC; and for the parameters $\aiso$ and $\alap$ we also check the consistency with the BAO post-reconstruction results which rely on the two-point correlation function. For the BGS and QSO samples, only the isotropic BAO measurement $\aiso$ was performed, so we do not report any $\alpha_{\rm AP}$. As expected, the BAO isotropic measurement, $\alpha_{\rm iso}$ is more precise in focussed BAO measurements that use reconstructed catalogues where the BAO peak is identified with a better precision due to the removal of bulk flows \cite{DESI2024.III.KP4}. This also applies partially to $\alpha_{\rm AP}$, but in this case ShapeFit extracts extra information from scales other than those hosting the BAO, so the change in error in this parameter is more complex, although for LRG and ELG samples, $\alpha_{\rm AP}$ remains better measured in the post-recon catalogues. For the rest of the cases, the $\alpha_{\rm iso,\, AP}$ are consistent with each other for all tracers within less than 1$\sigma$. For the $f$ parameter ($f$ is measured at a fixed amplitude of the template of the power spectrum), the biggest shift observed is between the unblinded and blinded results for ELG and QSO, with a difference between 2-2.5$\sigma$. This shift is fully expected from the blinding procedure (see \cite{KP3s9-Andrade}). For the $m$ parameter, we only see moderate $\sim1\sigma$ shifts with respect to the blinded catalogue for some of the tracers. We expect that the $f_{\rm NL}$ blinding applied can bias $m$ in these catalogues and therefore explain those moderate shifts. Hence, we conclude that all the tests show consistent results with respect to the baseline and to each other which validates the unblinding.

\section{Results}
\label{sec:results}

In this section, we present the results of the Full-Shape analysis using the unblinded \desidrone\ catalogues, in terms of the ShapeFit compressed variables, and the Full-Modelling variables. Results are only given for the standard flat-$\Lambda$CDM model with a BBN prior on $\Omega_b h^2$, and a wide prior on $n_s$, as listed in \cref{tab:priors}. We consider adding the post-recon \desidrone\,BAO information from \cite{DESI2024.III.KP4}. For further models and external dataset combinations, we refer the reader to \cite{DESI2024.VII.KP7B}.

\subsection{Constraints on compressed parameters from the \desidrone\ galaxies }\label{sec:SF-results}
We start by presenting the results in terms of the compressed parameters obtained from the ShapeFit compression approach for each of the redshift bins of \cref{tab:Y1data}. These results are listed in \cref{tab:SF-results}, both for the Full-Shape fits to the pre-reconstructed \desidrone\ catalogues, and when the \desidrone\ BAO post-reconstructed information is added to each of these redshift bins. For the combination of ShapeFit and BAO post-recon results, we follow the standard ShapeFit type fit on the pre-recon $P_\ell(k,z)$ datavector, adding two extra data-elements: $\alpha_{\rm iso}(z)$ and $\alpha_{\rm AP}(z)$ from the post-recon BAO fit reported in \cite{DESI2024.III.KP4}.\footnote{For the BGS sample, we only add $\alpha_{\rm iso}(z)$, as the $\alpha_{\rm AP}(z)$ is not reported in that case. See Ref.~\cite{DESI2024.III.KP4} for more details.} The ShapFit+BAO fits are performed with the proper covariance estimated from the \texttt{EZmocks}, cross-correlating each $P(k_i,z)$ element, with the best-fitting BAO post-recon $\alpha_{\rm iso}$ and $\alpha_{\rm AP}$ parameters obtained for each mock realisation. We have performed sanity checks and found that this joint fit returns very consistent results with an {\it a posteriori} combination of the ShapeFit compressed parameters with the BAO post-reconstructed parameters.

\begin{table}
    \centering
            \small
    \resizebox{\columnwidth}{!}{%
    \begin{tabular}{|c|c|c|c|c|c|}
    \hline
    Dataset & $\aiso$ & $\alap$ & $f\sigma_{s8}/(f\sigma_{s8})_{\mathrm{fid}}$ & $m$ & $\chi^2/{\rm ndof}$ \\ \hline\hline
    {\bf ShapeFit only} & & & & & \\
    \bgs   & $0.966^{+0.053}_{-0.025}$ & $0.979^{+0.13}_{-0.064}$ & $0.80 \pm 0.20$ & $-0.031^{+0.053}_{-0.25}$ & $49 / (72 - 11)$ \\
    \lrgo  & $0.976^{+0.021}_{-0.015}$ & $0.971^{+0.082}_{-0.034}$ & $1.09^{+0.12}_{-0.14}$ & $0.028^{+0.029}_{-0.11}$  & $57 / (72 - 11)$ \\
    \lrgt  & $0.952^{+0.023}_{-0.010}$ & $1.023^{+0.068}_{-0.037}$ & $1.05 \pm 0.12$ & $0.047^{+0.051}_{-0.082}$ & $48 / (72 - 11)$\\
    \lrgth & $0.997^{+0.018}_{-0.010}$ & $1.035^{+0.059}_{-0.039}$ & $0.96^{+0.11}_{-0.10}$ & $-0.025 \pm 0.059$ & $45 / (72 - 11)$ \\
    \elgt  & $0.977^{+0.026}_{-0.014}$ & $0.936^{+0.082}_{-0.044}$ & $0.95^{+0.11}_{-0.08}$ & $0.060^{+0.032}_{-0.10}$  & $46 / (72 -11)$\\
    \qso   & $0.988^{+0.027}_{-0.016}$ & $1.009^{+0.069}_{-0.044}$ & $1.16 \pm 0.12$ & $0.065^{+0.023}_{-0.077}$ & $58 / (72 -11)$ \\
    \hline 
    {\bf ShapeFit+BAO} & & & & & \\
     \bgs & $0.982^{+0.021}_{-0.017}$ & $0.967^{+0.10}_{-0.076}$ & $0.84\pm 0.19$ & $-0.049^{+0.080}_{-0.15}$ & $50/(73-11)$ \\
    \lrgo & $0.984^{+0.012}_{-0.0097}$ & $0.934^{+0.036}_{-0.031}$ & $1.16\pm 0.13$ & $0.020^{+0.046}_{-0.082}$  & $59/(74-11)$\\
    \lrgt & $0.967^{+0.012}_{-0.0090}$ & $1.037\pm 0.034$ & $1.04^{+0.11}_{-0.092}$ &  $0.022^{+0.072}_{-0.055}$ & $50/(74-11)$ \\
    \lrgth & $1.0019\pm 0.0088$ & $1.011\pm 0.029$ & $0.997^{+0.10}_{-0.084}$ & $-0.025^{+0.058}_{-0.043}$ & $46/(74-11)$ \\
    \elgt & $0.992\pm 0.014$ & $0.967^{+0.048}_{-0.037}$ & $0.945^{+0.097}_{-0.077}$ & $0.042^{+0.043}_{-0.077}$ & $47/(74-11)$ \\
    \qso & $0.990^{+0.023}_{-0.016}$ & $1.007^{+0.060}_{-0.052}$ &  $1.16\pm 0.12$ & $0.062^{+0.030}_{-0.070}$ & $58/(74-11)$ \\
    \hline
    \end{tabular}
    }
    \caption{Results from the ShapeFit baseline fit each of the six \desidrone\, redshift bins using the MAP values. We include both the results when fitting to only the power spectrum pre-reconstructed multiples (top sub-panel) and when the BAO post-reconstruction information from \cite{DESI2024.III.KP4} is added (bottom sub-panels). We consider each redshift bin independent, but the variables are correlated among them. In \cref{app:SFcov} we provide the full covariance under a Gaussian approximation for the case where the BAO is also added, in terms of the physical distances, the $f\sigma_{\rm s8}$ parameter and the shape. In the BGS, the result on the $\alpha_{\rm AP}$ is dominated by the prior bounds between 0.8 and 1.2.
    The reported $f\sigma_{\rm s8}/(f\sigma_{s8})_{\rm fid}$ parameter does include the $m$-correction factor of \cref{sigma_s8_m_correction}. The fiducial values for the BAO distances and $f\sigma_{s8}$, as well as for the isotropic BAO distance, $D_V/r_d$, and the $D_H/D_M$ distance ratio can be found in \cref{tab:fiducial_values}.}
    \label{tab:SF-results}
\end{table}

\begin{figure}
    \centering
    \includegraphics[width=1.\textwidth]{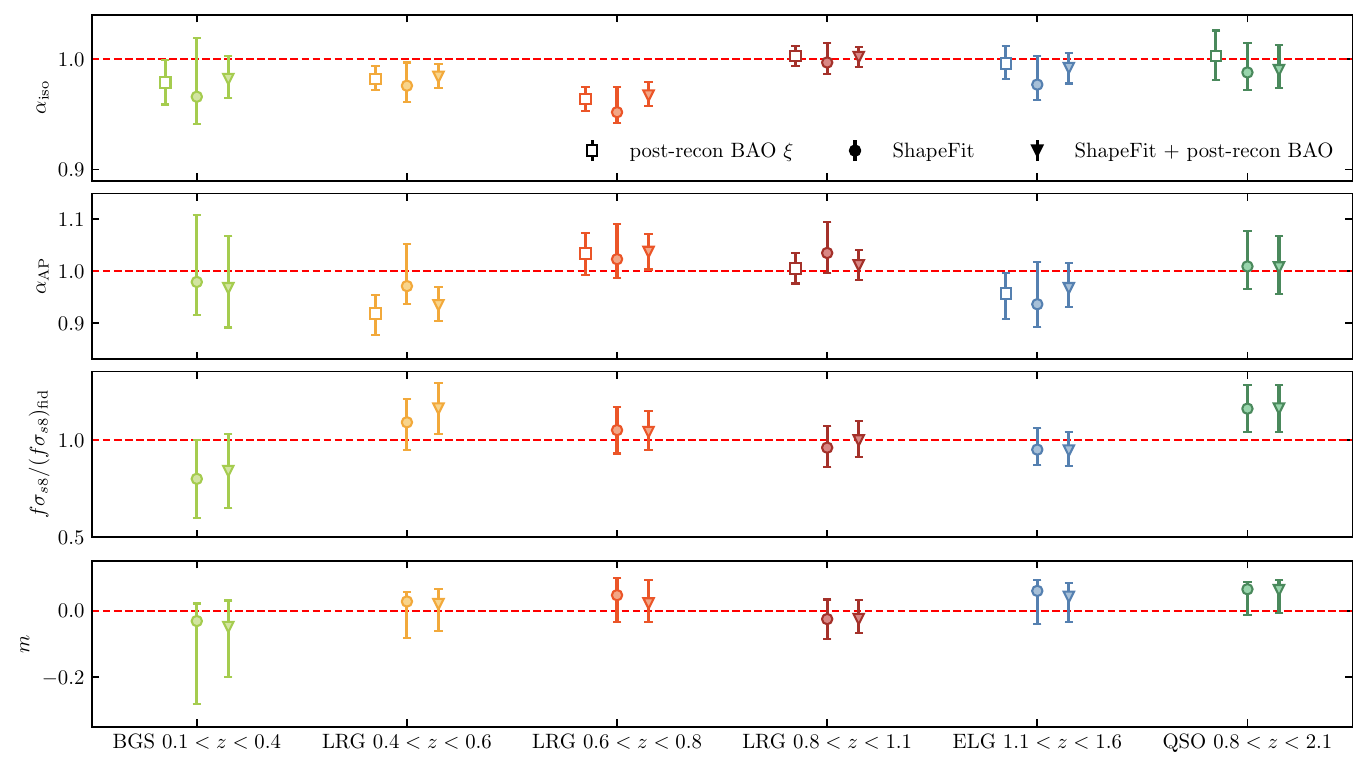}
    \caption{
    Constraints on the BAO parameters, $\alpha_{\rm iso}$ and $\alpha_{\rm AP}$, the growth parameter combination $f\sigma_{\rm s8}$ relative to its fiducial value, and the ShapeFit parameter $m$. The open circles show the ShapeFit fits to the pre-reconstructed power spectrum, the empty symbols show the BAO post-reconstruction fits on $\xi(s)$ and the open triangles show the combination of both (only applicable to the alphas), while the horizontal red lines display the fiducial values.}
    \label{fig:SF-unblind}
\end{figure}

\cref{fig:SF-unblind} displays the results listed in the top sub-panel of \cref{tab:SF-results}, where each colour corresponds to a different redshift-bin. For each of these, the error-bar interval represents the 68\%, and the symbol displays the MAP value. Filled symbols represent the results of the ShapeFit fit to the pre-reconstructed power spectrum, whereas empty symbols show the BAO post-recontruction fits on $\xi(s)$. The red dashed horizontal lines display the fiducial cosmology employed for performing these fits.  As a general trend, we observe a good consistency between the ShapeFit and BAO results, as already noted in \cref{sec:unblinding}. Also, the consistency with the fiducial cosmology choice is good, which validates the choice for this arbitrary cosmology for performing the ShapeFit fits. We do observe that the \texttt{LRG2} displays a low value of the isotropic scale parameter, $\aiso$, of about $2-3\sigma$. This low value was already reported on the BAO-only analysis in \cite{DESI2024.III.KP4}, and is consistent with a statistical fluctuation.

\begin{figure}
    \centering
\includegraphics[width=1.\linewidth]{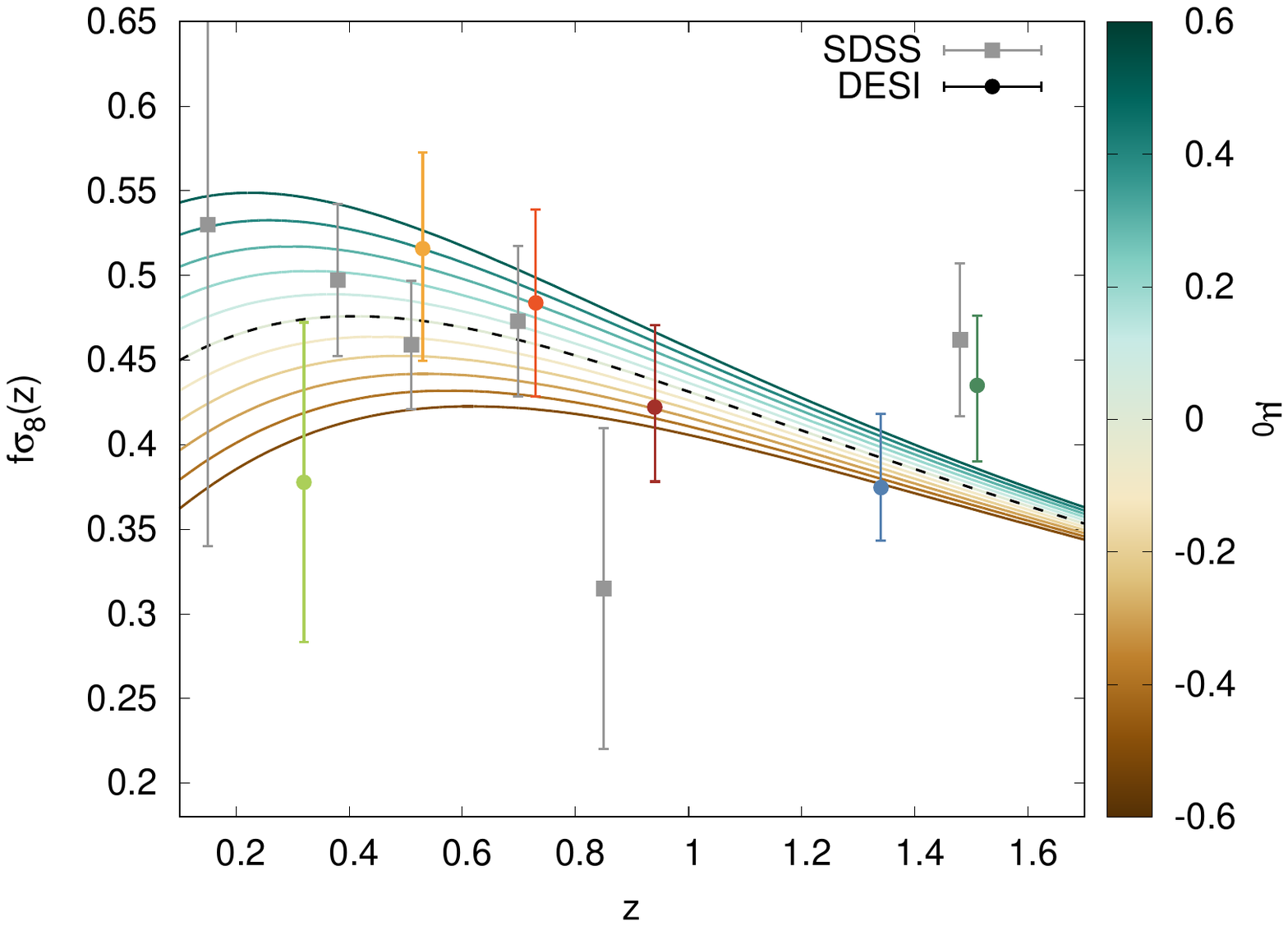}
    \caption{The growth of structure measurement parameter as a function of redshift, $f\sigma_{8}(z)$ obtained from the six \desidrone\ redshift bins using the ShapeFit compression scheme (coloured symbols), and from the whole SDSS program using the standard compression technique (grey symbols).  In all cases, the symbols represent the MAP and the errorbars the 68.3\% credible interval. The $f\sigma_8(z)$ values are derived from the reported $f\sigma_{\rm s8}(z)$ with the additional assumption that the assumed fiducial sound horizon scale is the true one. Additionally, for comparison, the coloured lines represent a range of $-0.5<\mu_0<+0.5$, a parametrisation of modified gravity models that changes the strength of the gravitational interaction. The case $\mu_0=0$ corresponds to general relativity, represented with black dashed lines. 
    }
    \label{fig:fs8}
\end{figure}

In \cref{fig:fs8} we report the $f\sigma_{8}(z)$ parameter extracted from the $f\sigma_{s8}(z)$ using ShapeFit as listed in \cref{tab:SF-results}. The coloured symbols display the results for the six redshift bins of \desidrone. For comparison, we also report in grey the measurements of SDSS using the standard compression technique. 
In order to do so we need to transform the smoothing scale, from $8\,h^{-1}{\rm Mpc} $ in the cosmology of the template to $8\,h^{-1}{\rm Mpc}$ in the best-fitting cosmology. In order to do so we assume that the isotropic dilation parameter, $\alpha_{\rm iso}$, exclusively accounts for a shift in the isotropic distance, $D_V$, and do not include any shift of the horizon scale parameter, $r_d$. This is equivalent to set the sound horizon scale parameter to its fiducial value, $r_d=r_d^{\rm fid}$, which makes $q_{\rm iso}=\alpha_{\rm iso}$\footnote{Where, $q_{\rm iso}\equiv (q_{\parallel}q_{\perp}^2)^{1/3}$}. Under this assumption, we are setting a physical scale for our system, which allows us to define absolute scale, such as the smoothing scale of $8\,h^{-1}{\rm Mpc}$, and where $f\sigma_{s8}$ becomes $f\sigma_8$. 
For a further discussion on this, we refer the reader to \cite{KP5s8-Gsponer}, where the impact of the template is explicitly studied in the case where fiducial and true sound horizon scales are not equal. In any case, we stress that the $f\sigma_8$ results should only serve for visualization purposes, and they should never be used to infer cosmology, for which one should use the actual $f\sigma_{s8}$ results. We note that the DESI measurements only account for the perturbation (or growth) information of redshift-space distortions, compressed into the $f\sigma_{s8}$ parameter in the model-agnostic ShapeFit analysis. Alternatively, one could have assumed a gravity model, performed a Full-Modelling analysis and reported $f\sigma_8$ as a derived parameter. In this case, the error bars would be much tighter as the measurements would also be informed by the background expansion history through the matter density parameter, $f(z)=\Omega_m(z)^\gamma$, with $\gamma=0.55$ for general relativity. We do not follow this approach because we aim to report the growth of structure measurements without assuming {\it a priori} any gravity model and keep these measurements independent of the background expansion history. 

We report these measurements along with a parametrisation of modified gravity which changes the strength of the gravitational interaction (coloured solid lines), and thus the growth rate of structure, through the function $\mu(a)$ such that \cite{Simpson:2012ra,DES:2018ufa}:
\begin{equation}
\mu(a)=1+\mu_{0}\frac{\Omega_{\text{DE}}(a)}{\Omega_\Lambda}
\label{eq:explicit-form_muSigma}
\end{equation}
where the modified gravity parameter $\mu_0$ is equal to zero in general relativity (black dashed line) and the modification of gravity through the $\mu(a)$ function is connected to the between the observed cosmic acceleration through the dark energy content $\Omega_{\text{DE}}$ (coloured solid lines for different values of $\mu_0$).
In this approach, we ignore the scale dependence of the $\mu$-function. A more in-depth discussion of the results and their implications for modified gravity is presented in section 5 of our companion paper \cite{DESI2024.VII.KP7B} and in the supporting paper dedicated to modified gravity \cite{KP7s1-MG}.
We see that the growth of structure measurements by DESI are in excellent agreement with the Planck measurements under the assumption of general relativity and a $\Lambda$CDM model.

\subsection{Constraints on \texorpdfstring{$\Lambda$}{L}CDM parameters from Full-Shape \desidrone\ galaxies}\label{sec:FSBAO-results}

\begin{figure}
    \centering
    \includegraphics[width=1.\textwidth]{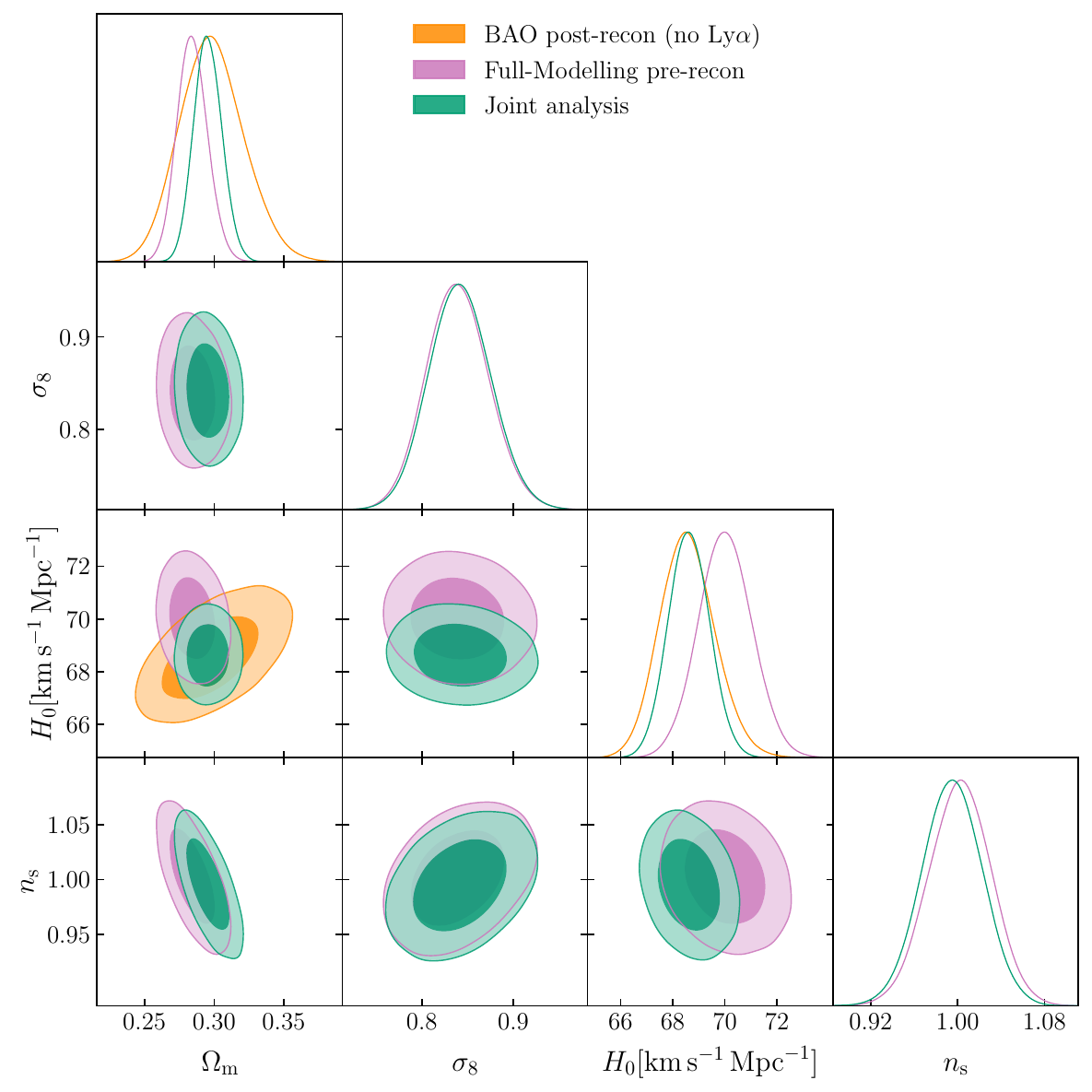}
    \caption{Constraints on $\Lambda$CDM parameters from Full-Modelling pre-reconstruction power spectrum analysis (magenta), BAO post-reconstruction correlation function analysis (orange) and a combined analysis of Full-Modelling and BAO post-recon approach (green) of the combined six galaxy and quasar redshift bins of \cref{tab:Y1data}. The contours display the $68.3\%$ and $95.4\%$ credible intervals. The variables $\sigma_8$ and $S_8$ are derived parameters of the model. The fit uses Gaussian informative priors on $\omega_{\rm b}$ and $n_s$ (see \cref{tab:priors}).}
    \label{fig:LCDM-unblind}
\end{figure}

We now focus on the constraints on the $\Lambda$CDM model based on the Full-Modelling type of fit. We remind the reader that unlike the results presented in \cref{sec:SF-results}, the Full-Modelling scheme assumes $\Lambda$CDM with informative Gaussian priors on the $\Omega_bh^2$ and $n_s$, as listed in \cref{tab:priors}. The results are shown in the triangle plot of \cref{fig:LCDM-unblind} and listed in \cref{tab:FM-results}, where only the variables $\{\Omega_m,\,H_0,\,\omega_b\,({\rm not\, shown}),\,A_s,\,n_s\}$ are independent, and $\sigma_8$ is a derived parameters. The $\omega_b$ parameter has not been shown as it simply follows its BBN prior. We simultaneously fit the uncorrelated six \desidrone\, bins with different nuisance parameters for each, but the same cosmological parameters. The contours are for the Full-Modelling fits to the power spectrum pre-reconstruction multipoles (magenta); fit to the correlation function post-reconstruction multipoles as presented in \cite{DESI2024.III.KP4} (orange), and the joint fit (green). The anisotropic BAO peak position analysis is only sensitive to $H_0$ and $\Omega_m$, and hence the orange contours are only displayed in that subpanel. 
 Additionally, we include the numerical results in \cref{tab:FM-results}, where in this case we also specify the results for the individual bins for completeness. 

\begin{figure}
    \centering
    \includegraphics[width=.49\textwidth]{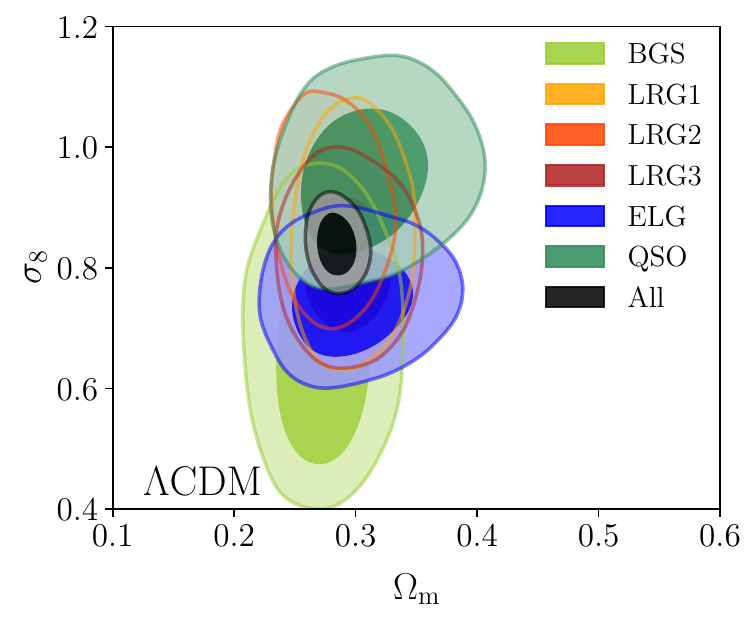}
    \includegraphics[width=.49\textwidth]{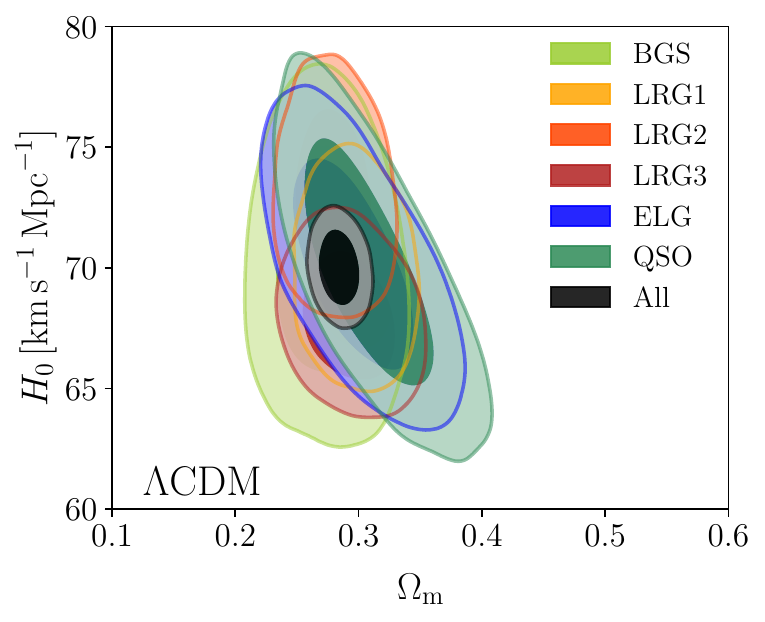}
    \caption{
    Posteriors from individual tracers, and their combination, for the Full-Modelling analysis (without BAO post-reconstruction information), assuming the $\Lambda$CDM cosmological model.
    The contours display the $68.3\%$ and $95.4\%$ credible intervals. As for \cref{fig:LCDM-unblind}, we employ informative Gaussian priors on $n_s$ and $\omega_b$ as those listed in \cref{tab:priors}. The left panel displays $\Omega_m-\sigma_8$, and the right panel $\Omega_m-H_0$.}
    \label{fig:Omegam_s8_H0}
\end{figure}

\begin{table}
    \centering
        \small
    \resizebox{\columnwidth}{!}{
    \begin{tabular}{|c|c|c|c|c|c|c|}
    \hline
    Dataset & $\Omega_m$ & $H_0$ & $\sigma_8$ & $\ln(10^{10} A_\mathrm{s})$ & $n_s$ & $\chi^2/{\rm ndof}$ \\
    \hline\hline
    Full-Modelling alone & & & & & & \\
    \bgs    & $0.272\pm 0.027$          & $70.2^{+3.0}_{-3.3}$  & $0.664^{+0.098}_{-0.14}$  & $2.70^{+0.40}_{-0.45}$ & $0.962\pm 0.040$ & 52 / (72-14) \\
    \lrgo   & $0.297\pm 0.021$          & $69.7^{+1.9}_{-2.2}$  & $0.844^{+0.083}_{-0.095}$ & $3.08\pm 0.24$         & $0.968\pm 0.040$ & 61 / (72-14) \\
    \lrgt   & $0.280\pm 0.020$          & $73.1\pm 2.2$         & $0.888^{+0.071}_{-0.084}$ & $3.10^{+0.20}_{-0.24}$ & $0.979\pm 0.039$ & 52 / (72-14) \\
    \lrgth  & $0.294\pm 0.025$          & $68.0\pm 1.8$         & $0.810^{+0.069}_{-0.078}$ & $3.10^{+0.21}_{-0.24}$ & $0.975\pm 0.038$ & 48 / (72-14) \\
    \elgt   & $0.297^{+0.028}_{-0.036}$ & $70.2\pm 2.9$         & $0.749^{+0.055}_{-0.064}$ & $2.83\pm 0.20$         & $0.969\pm 0.039$ & 51 / (72-15) \\
    \qso    & $0.310^{+0.030}_{-0.041}$ & $70.0\pm 3.4$         & $0.950^{+0.070}_{-0.084}$ & $3.25\pm 0.19$         & $0.976\pm 0.038$ & 62 / (72-15) \\
    All     & $0.284^{+0.010}_{-0.011}$ & $70.0\pm 1.0$         & $0.839\pm 0.034$          & $3.10\pm 0.10$         & $1.002\pm 0.029$ & 340 / (432-14)\\
    \hline
    Full-Modelling + BAO & & & & & & \\
    \bgs    & $0.284\pm 0.024$          & $68.3\pm 2.4$        & $0.662^{+0.096}_{-0.13}$   & $2.73\pm 0.40$         & $0.962\pm 0.040$ & 52 / (73-14)\\
    \lrgo   & $0.307^{+0.018}_{-0.020}$ & $68.8^{+1.3}_{-1.5}$ & $0.835\pm 0.087$           & $3.05\pm 0.22$         & $0.964\pm 0.039$ & 67 / (74-14) \\
    \lrgt   & $0.287\pm 0.020$          & $70.9\pm 1.6$        & $0.889^{+0.072}_{-0.082}$  & $3.17^{+0.21}_{-0.24}$ & $0.979\pm 0.038$ & 54 / (74-14) \\
    \lrgth  & $0.304\pm 0.023$          & $66.8\pm 1.2$        & $0.815^{+0.068}_{-0.076}$  & $3.12\pm 0.22$         & $0.972\pm 0.038$ & 49 / (74-14) \\
    \elgt   & $0.310^{+0.027}_{-0.034}$ & $68.5\pm 2.1$        & $0.755^{+0.054}_{-0.064}$  & $2.86^{+0.17}_{-0.19}$ & $0.969\pm 0.039$ & 51 / (74-15) \\
    \qso    & $0.314^{+0.029}_{-0.039}$ & $69.4\pm 3.1$        & $0.950^{+0.066}_{-0.077}$  & $3.26\pm 0.18$         & $0.976\pm 0.038$ & 62 / (73-15 )\\
    All     & $0.296\pm 0.010$          & $68.63\pm 0.79$      & $0.841\pm 0.034$           & $3.117\pm 0.097$       & $0.994\pm 0.028$ & 352 / (442-14) \\
    \hline
    \end{tabular}
    }
    \caption{Results from Full-Modelling fits assuming a $\Lambda$CDM models with informative Gaussian priors on $\omega_b$ and $n_s$ as listed in \cref{tab:priors}. We display the results of individually fitting each of the six \desidrone\, redshift bins (without combining with BAO post-reconstruction information), as well as the results of a joint fit to all of them (All); and when the BAO post-reconstruction information is also included (FM+BAO). In this last case, we consider only the BAO of each redshift bin when referring to the individual fits, and all bins together (but without the Ly-$\alpha$ BAO) when referring to the joint fit (All). 
    The results of the rows labelled as `All' correspond to those also displayed in \cref{fig:LCDM-unblind}, and the results of the individual fits without BAO to those displayed in \cref{fig:Omegam_s8_H0}.}
    \label{tab:FM-results}
\end{table}

We highlight the consistency among tracers by displaying the individual constraints from the Full-Modelling approach on the selected panels $\sigma_8-\Omega_m$ and $\Omega_m-H_0$ in \cref{fig:Omegam_s8_H0}. The left panel displays a very good consistency in the $\Omega_m$ parameter, which comes in part from the uncalibrated BAO peak position, but also from the shape of the power spectrum on large scales. The already observed good agreement in $\alpha_{\rm iso}$, $\alpha_{\rm AP}$ and $m$ in \cref{fig:SF-unblind} for the ShapeFit analysis is highly consistent with the good agreement in the $\Omega_m$ parameter. On the other hand, the $\sigma_8$ results present more scatter: the high-$z$ quasar bin pushes towards higher values of $\sigma_8$, whereas the low-$z$ BGS towards lower values, with the rest of the tracers between these two trends. These results are fully consistent with the effect already observed for $f\sigma_{s8}$ in \cref{fig:SF-unblind} for the ShapeFit scheme, where the values are low and high for the BGS and QSO, respectively. However, we acknowledge that in both cases, these fluctuations are well within the expected statistical scatter and therefore all the redshift bins can be considered consistent, justifying combining them in a joint fit as we do here. 

From the right panel of \cref{fig:Omegam_s8_H0} we additionally see the projection into the $H_0$ parameter. We observe how the contours at different redshifts present slightly different degenerate directions in the $\Omega_m-H_0$ plane: the direction of elongation of the ellipses rotates as we vary the observed redshift bin. By combining all bins we are able to break degeneracies among these two parameters and obtain significant improvements in the precision of both $\Omega_m$ and $H_0$. As before, we report an excellent consistency on the $H_0$ value from all the tracers. 

\begin{figure}
    \centering
    \includegraphics[width=0.32\linewidth]{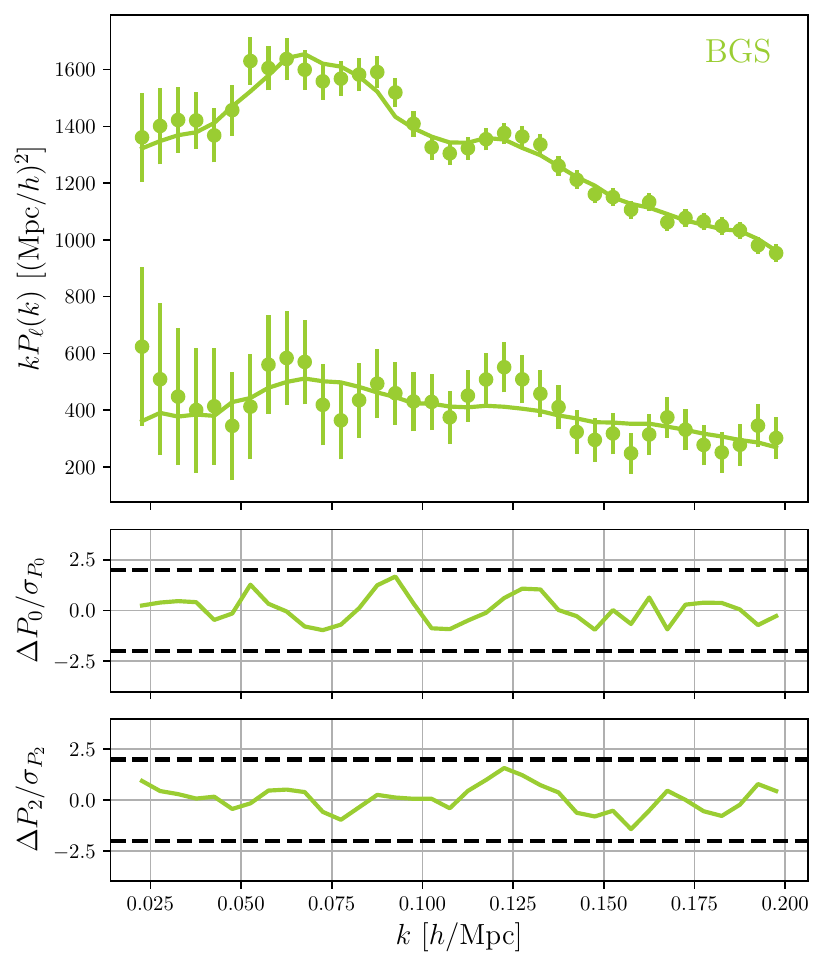}
    \includegraphics[width=0.32\linewidth]{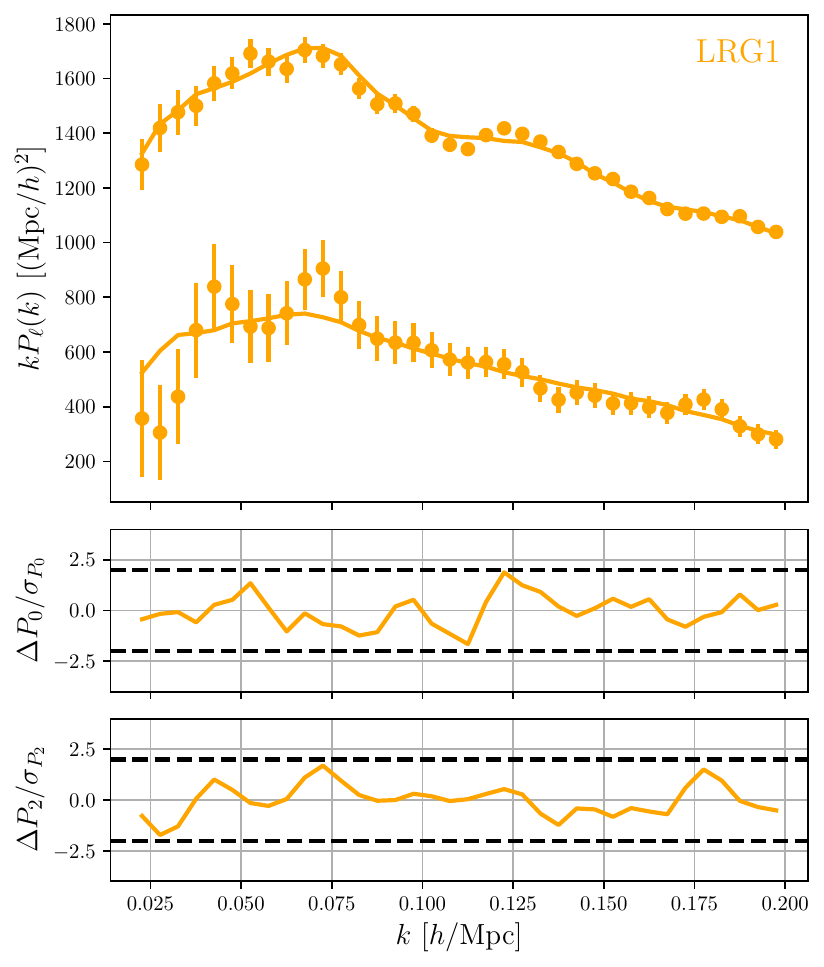}
    \includegraphics[width=0.32\linewidth]{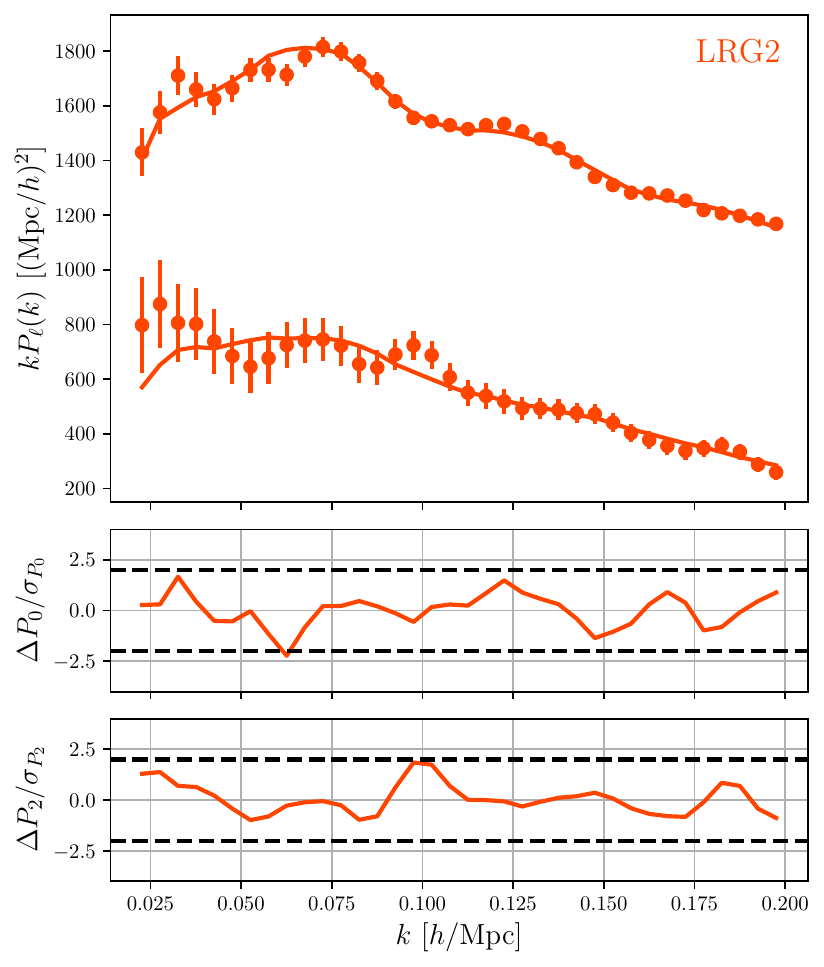}
    
    \includegraphics[width=0.32\linewidth]{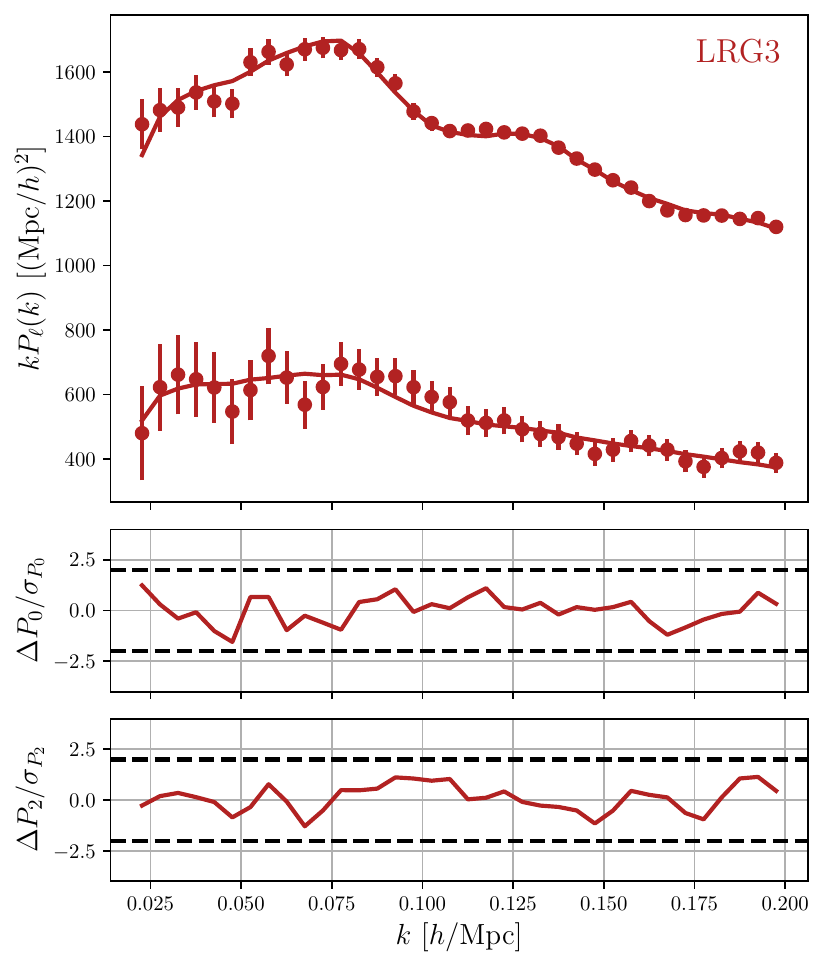}
    \includegraphics[width=0.32\linewidth]{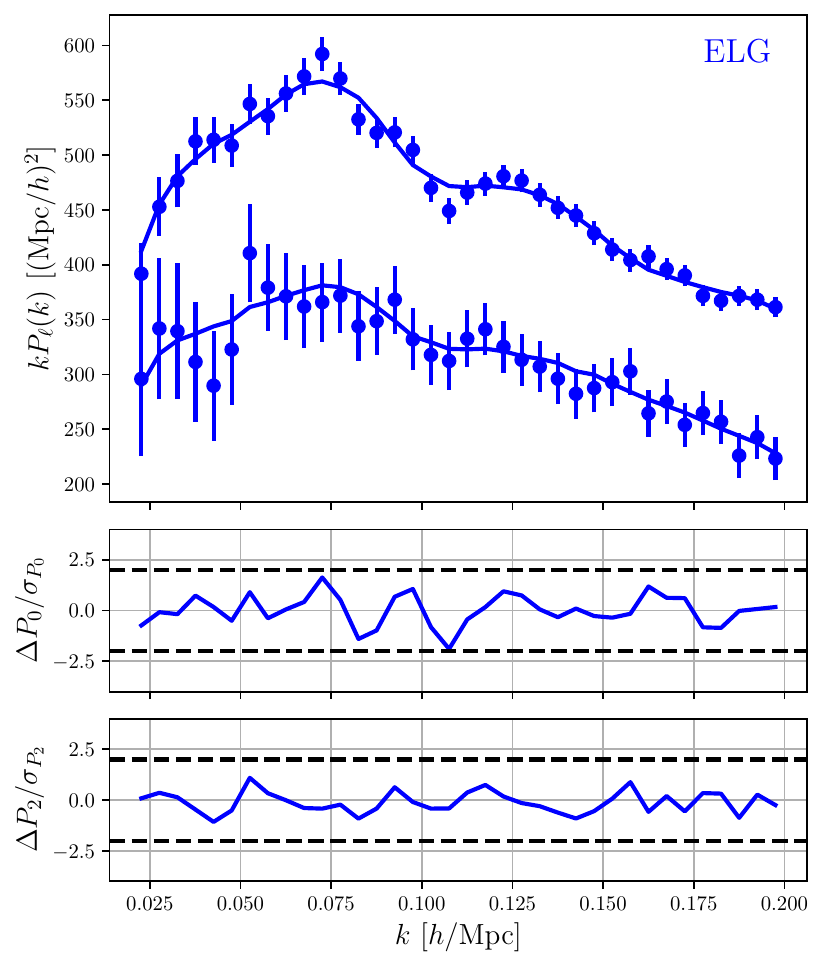}
    \includegraphics[width=0.32\linewidth]{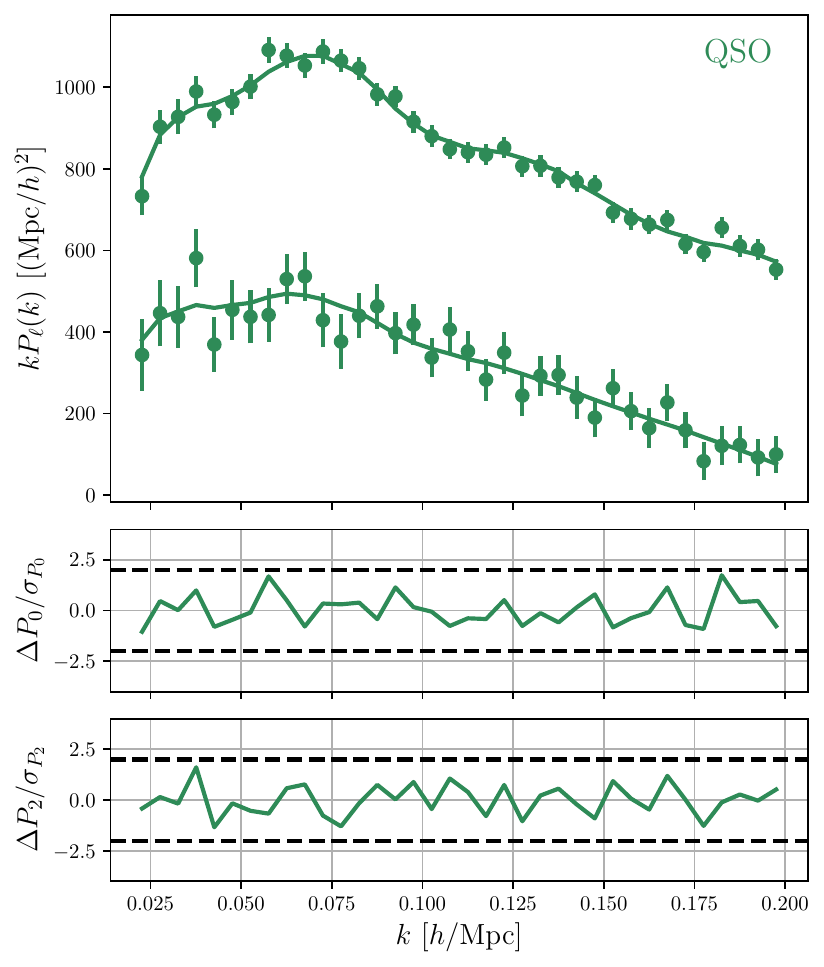}
    \caption{Measured and best-fit monopole and quadrupole from DESI DR1 measurements. Symbols display the measurements and solid lines the best-fitting power spectrum model to each of the individual redshift bins according to the baseline model, as listed in \cref{tab:FM-results} as `Full-Modelling alone'. The errors are $1\sigma$ and include both the statistical and systematic error budget. The panels display the six redshift bins of the tracers presented in Table~\ref{tab:Y1data}, with the same colour scheme employed in previous plots. The bottom subpanels display the residuals between data and model in units of the diagonal error.}
    \label{fig:Pkresults}
\end{figure}

\cref{fig:Pkresults} displays how the best-fitting model for $\Lambda$CDM (solid lines) for the Full-Modelling analysis combined with BAO post-reconstruction (no Ly-$\alpha$), performs with the measured power spectrum multipoles of the six redshift bins of the \desidrone\ data. In general, we confirm that the $\Lambda$CDM model produces a good fit to the data, as shown by the minimum $\chi^2$-per-degree-of-freedom value reported in \cref{tab:FM-results}. 

\subsection{Comparison to previous analyses}\label{sec:comparison}

Prior to \desidrone; the largest galaxy redshift surveys were undertaken as part of the SDSS; these were BOSS \cite{SDSS-DR12-cosmology} and eBOSS \cite{SDSS-DR16-cosmology}.  In order to compare the results from SDSS we have run the DESI baseline Full-Modelling pipeline on publicly available BOSS and eBOSS data \cite{Beutler_and_McDonald21}. Specifically, we used the pre-reconstruction power spectra of the two non-overlapping BOSS LRG redshift bins ($0.2<z<0.5$ and $0.5<z<0.7$) in combination with the eBOSS QSO ($0.8<z<2.2$). For simplicity, we ignore the ELG targets and the rest of the overlapping DR16 LRGs above $z>0.6$. Also, we do not redo the BOSS analysis with the DESI fiducial cosmology (as it would require us to recompute all the power spectra, for SDSS data and mocks), so the fiducial cosmologies used for BOSS and DESI are slightly different. However, our results are independent of the choice of cosmology for the redshift-to-distance conversion, as we have discussed previously. 
We fit the power spectrum from BOSS data with a covariance matrix computed from the \texttt{EZmocks} simulations provided by the BOSS team. For simplicity, the correction factor on the \ezmocks\ covariance from \cref{tab:cov-corr} has not been applied.

\begin{figure}
    \centering
    \includegraphics[width=1.\linewidth]{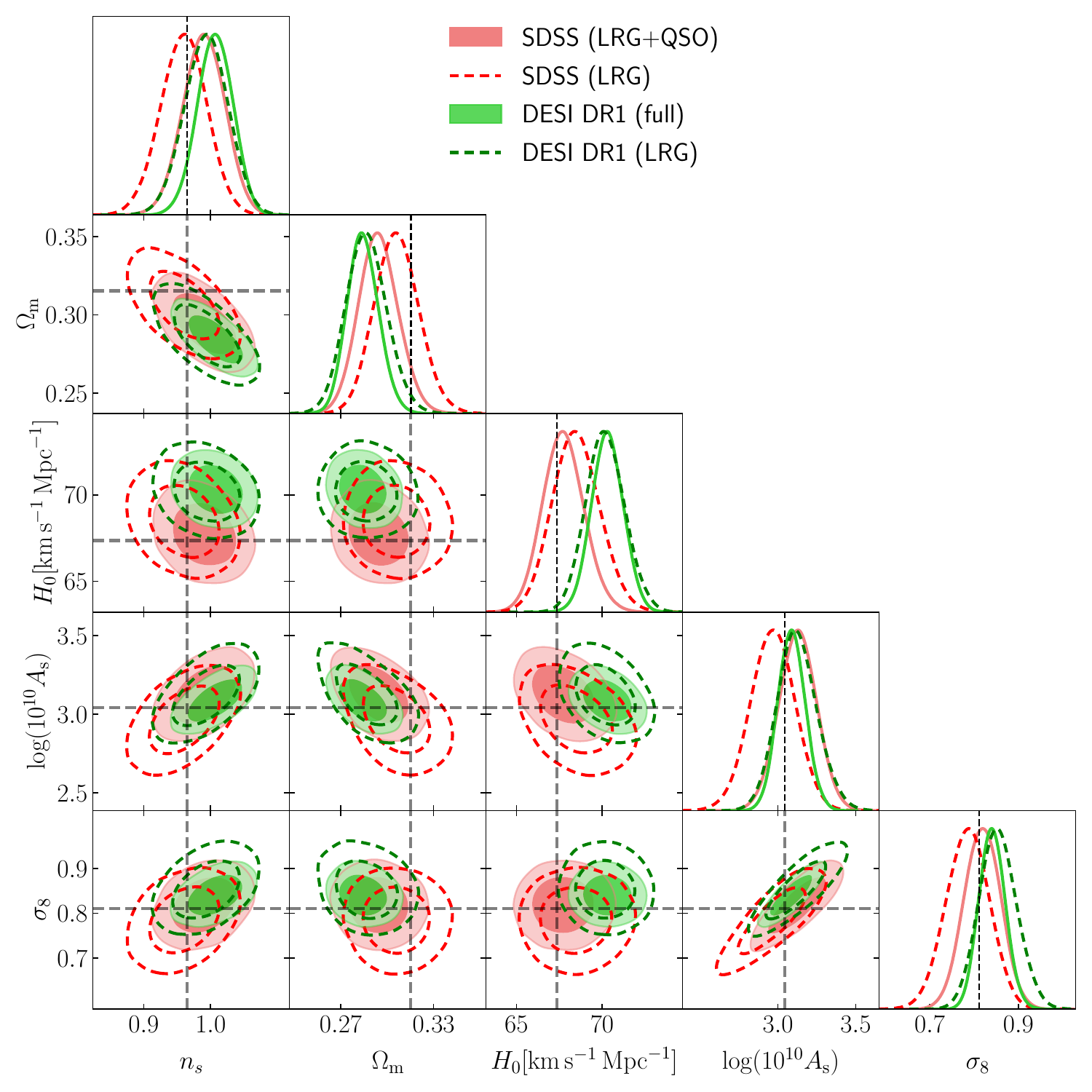}
    \caption{Comparison between SDSS and \desidrone\ Full-Modelling results (without BAO post-reconstruction information) using the same pipeline and Full-Modelling method. The solid red contours show the fit to BOSS LRGs + eBOSS quasar samples and the solid green contours show the fit to the full DESI DR1 footprint (BGS, LRG$\times$3, ELG, QSO). The red and green dashed contours show the results from fitting only non-overlapping BOSS LRGs and DESI DR1 LRGs, respectively, with a $\sim27\%$ of coincident targets. The grey dashed lines show the \emph{Planck} best-fitting cosmology for reference.}
    \label{fig:desi_boss}
\end{figure}

The results are shown in \cref{fig:desi_boss}. Given the difference in sky coverage and redshift range between DESI and SDSS samples, the observed agreement is excellent: we estimate 27\% coincident targets for the LRGs and 15\% for QSO \cite{DESI2024.II.KP3}. If we focus on the SDSS results from LRG+QSO and DESI using all samples (red and green solid contours) we find that the DESI data prefers slightly higher $H_0$ than SDSS data with a $\mathcal{O}(1\sigma)$ shift from \emph{Planck} predictions (displayed as horizontal and vertical dashed lines). We see that the value of $\Omega_m$ is strikingly similar for both surveys with a slight difference if we only look at the LRG sample, where the SDSS LRGs prefer a lower value than the DESI LRGs. 
The higher preference of DESI data for a higher $H_0$ was also observed in the \desidrone\ post-reconstruction BAO analysis and discussed in refs.~\cite{DESI2024.III.KP4,DESI2024.VI.KP7A}. 

Regarding $\sigma_8$, we find both SDSS and \desidrone\ to be consistent with the amplitude of clustering implied by Planck, $\sigma_{8\rm Planck}=0.8111\pm 0.0060$. The official BOSS constraint on $\sigma_8$ was $0.787 \pm 0.047$ \cite{SDSS-DR12-cosmology} with eBOSS quoting $0.850 \pm 0.033$ \cite{SDSS-DR16-cosmology}. However, this was from an older set of models and a different modelling approach and when the BAO postreconstruction analysis was included. Several previous (independent) analyses of BOSS/eBOSS data using an approach closer to ours had found marginal constraints on $\sigma_8$ to be approximately $\mathcal{O}(1\sigma)$ below \emph{Planck} \cite{Beutler2016:1607.03150v1,Satpathy2016:1607.03148v2,Ivanov20,Zhai2022:2203.08999v2,Philcox22,Chen22}. However, the BOSS data are not particularly constraining for such flexible models and $\mathcal{O}(1\sigma)$ shifts up or down in marginal $\sigma_8$ constraints can be caused by projection effects -- as evidenced by the sensitivity to choices in parameterisation -- when the constraining power of the data are not very strong, as pointed out by \cite{Simon23b} when studying the impact of priors and projection effects on $\sigma_8$ and other parameters. The $b_n\sigma_8^n$ bias parameterisation employed in the DESI pipeline results in a $\sigma_8$ constraint of $0.785\pm 0.049$ when fitting BOSS LRG data, while when one reverts to the $(b\sigma_8, b_2, b_s)$ basis the constraints shift down to $0.740\pm 0.051$. The latter constraint is more consistent with the results from previous BOSS analyses that did not scale all of their bias parameters with factors of $\sigma_8$, and such a difference is to be expected when the data are not particularly constraining. This dependence on bias parameterisation and the scaling with $\sigma_8$ is discussed in detail in Appendix B of Ref.~\cite{KP5s2-Maus}. In addition, by comparing the two red posteriors (dashed vs solid) in \cref{fig:desi_boss} we find that the eBOSS quasar sample appears to prefer slightly higher $\sigma_8$ than BOSS LRGs alone. This was also observed in the studies of $f\sigma_8$ of eBOSS QSO \cite{HouDR16QSO,Neveux2020}, where the value of $f\sigma_8$ obtained was higher than the $\Lambda$CDM+\emph{Planck} prediction. We observe a similar trend for \desidrone\ where the quasar sample tends to prefer a higher $\sigma_8$ value, although as we have commented before, this fluctuation remains within the expected statistical range. 

Overall we find the DESI results to be in remarkably good agreement with SDSS.
The slight shift in $H_0$ is expected given the similar variations seen from the \desidrone\ BAO signal and the sensitivity of the $\sigma_8$ constraint to parameterisation can be understood from the structure of the theories being fit.

\section{Conclusion}
\label{sec:conclusion}

We have presented the first measurements and cosmological implications of the galaxy two-point clustering in Fourier space from the DESI galaxies and quasars since the start of data collection in 2021, which forms Data Release 1 \cite{DESI2024.I.DR1}.
We modelled and fitted the entire power spectrum, performing what is referred to as a Full-Shape analysis. By doing so, we extend previous DESI DR1 baryon acoustic oscillation measurements by including additional information from the Alcock-Paczynski effect, redshift-space distortions, and signals from the matter-radiation equality scale.

The clustering is measured using the magnitude-limited Bright Galaxy Survey with $0.1<z<0.4$, Luminous Red Galaxies with $0.4<z<1.1$, Emission Line Galaxies with $0.8<z<1.6$, and quasars with $0.8<z<2.1$, divided into six redshift bins over a $\sim 7,500$ square degree footprint. The large-scale structure catalogues and two-point clustering measurements are described in \cite{DESI2024.II.KP3}.

We explored two ways of analysing the power spectrum data. The first approach consists in fitting a set of compressed model-agnostic variables using the ShapeFit approach \cite{brieden2021}: the isotropic dilation parameter $\alpha_{\mathrm{iso}}$, and the AP parameter $\alpha_{\mathrm{AP}}$, the product of the logarithmic growth rate and the amplitude of matter fluctuations $f\sigma_{8}$, and the slope of the linear power spectrum $m$. This compressed method can be seen as an extension of what was done in previous-generation spectroscopic surveys with the Sloan Digital Sky Survey (SDSS). With only one year of DESI observations, we obtained a combined precision of 4.7\% on $f\sigma_8$, reaching a similar precision with just one year of DESI data than twenty years of observations from the previous generation spectroscopic survey, SDSS. The DESI DR1 galaxy clustering results are in agreement with a $\Lambda$CDM model as predicted by \textit{Planck} based on general relativity.

The second approach consists of directly fitting the cosmological parameters of a given model. We refer to such an approach as Full Modelling.
Within the $\Lambda$CDM model, we combine our Full-Modelling results with the reconstructed DESI DR1 galaxy BAO. Using a Big Bang Nucleosynthesis Gaussian prior on the baryon density parameter, $\omega_b$, and a weak Gaussian prior on the spectral index, $n_s$, we find the matter density is $\Omega_m = 0.296\pm 0.010 $, yielding a 3\% precision measurement that is in agreement with measurements using other cosmological probes. We also constrain the Hubble constant to $H_0 = (68.63 \pm 0.79)[{\rm km\, s^{-1}Mpc^{-1}}]$, which is in agreement with CMB and previous galaxy clustering measurements, but in disagreement with some direct measurements from the local Universe (see \cite{verdeetal24} for a recent review on the Hubble constant measurements). The constraints on the primordial amplitude of the linear power spectrum and the spectral index are $\ln(10^{10} A_s) = (3.117 \pm {0.097})$ and $n_s= 0.994 \pm 0.028$, respectively.
Additionally, we measure the derived parameters related to the amplitude of clustering $\sigma_8 = 0.841 \pm 0.034$. 
This constraint is in excellent agreement with previous galaxy clustering analyses and with CMB results but is slightly higher than the weak gravitational lensing constraints. 
The cosmological interpretation of these results in terms of other models, and in combination with DESI DR1 Ly$\alpha$ and external datasets are presented in the companion paper \cite{DESI2024.VII.KP7B}.

Alongside the state-of-the-art dataset, the \desidrone\ Full-Shape analysis incorporates several significant novel elements. We performed for the first time a Full-Shape blinded analysis in terms of both compressed agnostic variables and in terms of the variables of the $\Lambda$CDM model. This approach is detailed in \cite{KP3s9-Andrade} and was adopted to mitigate confirmation bias. All fiducial choices of the analysis baseline, as well as the size of the systematic errors, are determined on the basis of the tests with mock catalogues and the blinded data catalogues. The true clustering was unveiled only once the blinded catalogues successfully passed a series of unblinding tests (we refer the reader to \cref{sec:unblinding}).

We investigated sources of potential systematics and performed a detailed study of each of them individually. The modelling of the Full-Shape analysis, for both the compressed and full-modelling approaches, relies on the Effective Field Theory (EFT) framework, which allows for a better treatment of small-scale physics thanks to the inclusion of counterterms. We performed a thorough comparison of
four different EFT models and of the compressed and full-modelling approaches, both in Fourier and configuration spaces \cite{KP5s1-Maus,KP5s2-Maus,KP5s3-Noriega,KP5s4-Lai,KP5s5-Ramirez}. In \cite{KP5s7-Findlay}, we studied the impact of the galaxy-halo connection in the Halo Occupation Distribution (HOD) framework, together with the prior-weight effect. The impact of the assumption of the fiducial cosmology, both in terms of ShapeFit and Full-Modelling approaches, is presented in \cite{KP5s8-Gsponer}. The characterisation and mitigation of photometric systematics for Full-Shape analysis, which led to an advanced correction methodology, is described in \cite{KP5s6-Zhao}. Ref.~\cite{KP3s5-Pinon} proposed a method to mitigate the incompleteness due to the fibre-assignment procedure. The characterisation and mitigation of spectroscopic systematics for ELG and for the other DESI tracers are presented in \cite{KP3s4-Yu} and \cite{KP3s3-Krolewski}, respectively. We use a mock-based covariance matrix to perform the Full-Shape analysis on data: a comparison of this estimate with analytic covariance matrix estimate in Fourier space \cite{KP4s8-Alves} and configuration space \cite{KP4s7-Rashkovetskyi} is presented in \cite{KP4s6-Forero-Sanchez}.

The total systematic contribution represents a bit more than $\sfrac{2}{5}$ of the DR1 statistical error. For the first time in a galaxy clustering analysis, we account for the systematic errors at the two-point clustering level directly, which automatically propagates them into the parameters of any model that we fit. The description of this novel methodology is described in \cite{KP5s7-Findlay,KP5s6-Zhao,KP3s5-Pinon}.
 
To quantify the impact of the changes we made in the DESI pipeline, we revisited the SDSS power spectrum measurements using the DESI Full-Shape pipeline. We conclude that the two measurement pipelines yield sufficiently similar results for these datasets to be compatible (see \cref{sec:comparison}), although the DESI DR1 results are more constraining than the earlier clustering measurements from the full SDSS cosmological program. 

We anticipate that the methodology presented in this Full-Shape analysis of the DESI DR1 will pave the way for future analyses of galaxy redshift surveys. The DESI collaboration has already completed the first 3-year of observations and it is currently analyzing this new dataset. Due to the larger volume and improved completeness, this upcoming DESI data release will provide an improvement of about a factor of $\sim2$ in the error bars of cosmological parameters. These will set an even more stringent stress test for the nature of dark matter and dark energy.

\section*{Data Availability}

Data from the plots in this paper will be available on Zenodo as part of DESI's Data Management Plan.
The data used in this analysis will be made public along with Data Release 1 (details in \href{https://data.desi.lbl.gov/doc/releases/}{https://data.desi.lbl.gov/doc/releases/}).

\acknowledgments

This material is based upon work supported by the U.S. Department of Energy (DOE), Office of Science, Office of High-Energy Physics, under Contract No. DE–AC02–05CH11231, and by the National Energy Research Scientific Computing Center, a DOE Office of Science User Facility under the same contract. Additional support for DESI was provided by the U.S. National Science Foundation (NSF), Division of Astronomical Sciences under Contract No. AST-0950945 to the NSF’s National Optical-Infrared Astronomy Research Laboratory; the Science and Technology Facilities Council of the United Kingdom; the Gordon and Betty Moore Foundation; the Heising-Simons Foundation; the French Alternative Energies and Atomic Energy Commission (CEA); the National Council of Humanities, Science and Technology of Mexico (CONAHCYT); the Ministry of Science and Innovation of Spain (MICINN), and by the DESI Member Institutions: \url{https://www.desi.lbl.gov/collaborating-institutions}. Any opinions, findings, and conclusions or recommendations expressed in this material are those of the author(s) and do not necessarily reflect the views of the U. S. National Science Foundation, the U. S. Department of Energy, or any of the listed funding agencies.

The authors are honoured to be permitted to conduct scientific research on Iolkam Du’ag (Kitt Peak), a mountain with particular significance to the Tohono O’odham Nation.

\appendix

\section{Datavectors and covariances for the compressed ShapeFit parameters}
\label{app:SFcov}

Here we report the Gaussian approximation of the ShapeFit covariances for each of the six redshift bins. We start by reporting the results for the ShapeFit fits alone.

For the BGS sample at $z_{\rm eff}=0.30$ we find,
 \begin{equation}
   D_{\rm SF}^{\rm BGS}(0.30)=
 \begin{pmatrix}
D_V/r_{\rm d}(0.30) \\
D_H/D_M(0.30) \\
f\sigma_{s8}(0.30) \\
(m+n) (0.30)
 \end{pmatrix}=
  \begin{pmatrix}
7.788174 \\
3.053800 \\
0.377174 \\
-0.031370
 \end{pmatrix},
 \end{equation}
with the following covariance,
   \begin{equation} 
C_{\rm SF}^{\rm BGS} (0.30) = 10^{-4}
 \begin{pmatrix}
1314.664401 &  -142.669742 &  109.427163 &  -217.509715 \\
  - &  829.170940 &  -158.101737 &  -61.084426 \\
  - &  - &  88.510877 &  -2.272457 \\
  - &  - &  - &  279.702360
 \end{pmatrix}.
\end{equation}
Note that in this case, the parameter $\alpha_{\rm AP}$ is not well constrained, and the information returned on $D_H/D_M$ is highly affected by the flat prior between 0.8 and 1.2. This is due to the low constraining power of the BGS sample, but also that low-$z$ samples are less sensitive to the Alcock-Paczysnki effect\footnote{In the limit of samples at $z=0$, both angular and radial scales are by definition equal and the Alcock-Paczysnki parameter is perfectly unconstrained in the data.}.

For the LRG1 sample at $z_{\rm eff}=0.51$ we find,
 \begin{equation}
   D_{\rm SF}^{\rm LRG1}(0.51)=
 \begin{pmatrix}
D_V/r_{\rm d}(0.51) \\
D_H/D_M(0.51) \\
f\sigma_{s8}(0.51) \\
(m+n) (0.51)
 \end{pmatrix}=
  \begin{pmatrix}
12.514437 \\ 
1.637266 \\ 
0.513635 \\ 
0.027840
 \end{pmatrix},
 \end{equation}
with the following covariance,
   \begin{equation} 
C_{\rm SF}^{\rm LRG1} (0.51) = 10^{-4}
 \begin{pmatrix}
541.309833 &  48.425593 &  -4.652853 &  -37.707751 \\
  - &  97.249820 &  -34.923265 &  -14.418597 \\
  - &  - &  41.295470 &  15.405508 \\
  - &  - &  - &  48.918910 
 \end{pmatrix}.
\end{equation}
For the LRG2 sample at $z_{\rm eff}=0.71$ we find,
 \begin{equation}
   D_{\rm SF}^{\rm LRG2}(0.71)=
 \begin{pmatrix}
D_V/r_{\rm d}(0.71) \\
D_H/D_M(0.71) \\
f\sigma_{s8}(0.71) \\
(m+n) (0.71)
 \end{pmatrix}=
  \begin{pmatrix}
15.675560 \\ 
1.165867 \\ 
0.483623 \\ 
0.046650
 \end{pmatrix},
 \end{equation}
with the following covariance,
   \begin{equation} 
C_{\rm SF}^{\rm LRG2} (0.71) = 10^{-4}
 \begin{pmatrix}
762.457717 &  39.781004 &  -1.896006 &  -62.812849 \\
  - &  36.344098 &  -17.341350 &  -8.333900 \\
  - &  - &  28.119682 &  8.865225 \\
  - &  - &  - &  47.624520
 \end{pmatrix}.
\end{equation}
For the LRG3 sample at $z_{\rm eff}=0.92$ we find,
 \begin{equation}
   D_{\rm SF}^{\rm LRG3}(0.92)=
 \begin{pmatrix}
D_V/r_{\rm d}(0.92) \\
D_H/D_M(0.92) \\
f\sigma_{s8}(0.92) \\
(m+n) (0.92)
 \end{pmatrix}=
  \begin{pmatrix}
19.676985 \\
0.844996 \\ 
0.422164 \\ 
-0.024690
 \end{pmatrix},
 \end{equation}
with the following covariance,
   \begin{equation} 
C_{\rm SF}^{\rm LRG3} (0.92) = 10^{-4}
 \begin{pmatrix}
847.499793 &  26.038900 &  3.257324 &  -37.016091 \\
  - &  16.251088 &  -10.044074 &  -3.698790 \\
  - &  - &  22.370314 &  6.467510 \\
  - &  - &  - &  34.883220 
 \end{pmatrix}.
\end{equation}
For the ELG2 sample at $z_{\rm eff}=1.32$ we find,
 \begin{equation}
   D_{\rm SF}^{\rm ELG2}(1.32)=
 \begin{pmatrix}
D_V/r_{\rm d}(1.32) \\
D_H/D_M(1.32) \\
f\sigma_{s8}(1.32) \\
(m+n) (1.32)
 \end{pmatrix}=
  \begin{pmatrix}
23.861806 \\ 
0.470709 \\ 
0.376715 \\
0.059960
 \end{pmatrix},
 \end{equation}
with the following covariance,
   \begin{equation} 
C_{\rm SF}^{\rm ELG2} (1.32) = 10^{-4}
 \begin{pmatrix}
2342.506886 &  26.159601 &  13.521001 &  -95.336060 \\
  - &  10.309303 &  -6.654663 &  -5.183903 \\
  - &  - &  13.997473 &  9.109619 \\
  - &  - &  - &  43.575710 
 \end{pmatrix}.
\end{equation}
For the QSO sample at $z_{\rm eff}=1.49$ we find,
 \begin{equation}
   D_{\rm SF}^{\rm QSO}(1.49)=
 \begin{pmatrix}
D_V/r_{\rm d}(1.49) \\
D_H/D_M(1.49) \\
f\sigma_{s8}(1.49) \\
(m+n) (1.49)
 \end{pmatrix}=
  \begin{pmatrix}
25.708520 \\
0.426508 \\ 
 0.434858 \\ 
0.064550
 \end{pmatrix},
 \end{equation}
with the following covariance,
   \begin{equation} 
C_{\rm SF}^{\rm QSO} (1.49) = 10^{-4}
 \begin{pmatrix}
3013.788566 &  -2.205101 &  36.332110 &  -98.167826 \\
  - &  5.845806 &  -6.747133 &  -1.913326 \\
  - &  - &  19.785658 &  5.357546 \\
  - &  - &  - &  26.266260
 \end{pmatrix}.
\end{equation}

We now report the results for the ShapeFit fits in combination with the BAO reconstructed fits from \cite{DESI2024.III.KP4}.

For the BGS sample at $z_{\rm eff}=0.30$ we find,
 \begin{equation}
   D_{\rm SF+BAO}^{\rm BGS}(0.30)=
 \begin{pmatrix}
D_V/r_{\rm d}(0.30) \\
D_H/D_M(0.30) \\
f\sigma_{s8}(0.30) \\
(m+n) (0.30)
 \end{pmatrix}=
  \begin{pmatrix}
  7.920703 \\ 
  3.014389 \\ 
  0.396326 \\ 
  -0.049030
 \end{pmatrix},
 \end{equation}
with the following covariance,
   \begin{equation} 
C_{\rm SF+BAO}^{\rm BGS} (0.30) = 10^{-4}
 \begin{pmatrix}
239.901232 &  -61.419576 &  23.441205 &  -33.006574 \\
  - &  726.794608 &  -145.787490 &  -17.577288 \\
  - &  - &  80.924891 &  8.802586 \\
  - &  - &  - &  167.972620
 \end{pmatrix}.
\end{equation}

For the LRG1 sample at $z_{\rm eff}=0.51$ we find,
 \begin{equation}
   D_{\rm SF+BAO}^{\rm LRG1}(0.51)=
 \begin{pmatrix}
D_V/r_{\rm d}(0.51) \\
D_H/D_M(0.51) \\
f\sigma_{s8}(0.51) \\
(m+n) (0.51)
 \end{pmatrix}=
  \begin{pmatrix}
12.621670 \\ 
1.574992 \\ 
0.547632 \\ 
0.020350
 \end{pmatrix},
 \end{equation}
with the following covariance,
   \begin{equation} 
C_{\rm SF+BAO}^{\rm LRG1} (0.51) = 10^{-4}
 \begin{pmatrix}
185.794784 &  -2.289289 &  9.575015 &  -13.518142 \\
  - &  30.722651 &  -14.614473 &  -2.167703 \\
  - &  - &  35.284297 &  13.024004 \\
  - &  - &  - &  42.091060
 \end{pmatrix}.
\end{equation}
For the LRG2 sample at $z_{\rm eff}=0.71$ we find,
 \begin{equation}
   D_{\rm SF+BAO}^{\rm LRG2}(0.71)=
 \begin{pmatrix}
D_V/r_{\rm d}(0.71) \\
D_H/D_M(0.71) \\
f\sigma_{s8}(0.71) \\
(m+n) (0.71)
 \end{pmatrix}=
  \begin{pmatrix}
15.924595 \\ 
1.181734 \\ 
0.479541 \\ 
0.022320
 \end{pmatrix},
 \end{equation}
with the following covariance,
   \begin{equation} 
C_{\rm SF+BAO}^{\rm LRG2} (0.71) = 10^{-4}
 \begin{pmatrix}
310.004892 &  6.627447 &  9.878373 &  -28.045518 \\
  - &  15.261852 &  -8.194175 &  -1.747683 \\
  - &  - &  23.061044 &  7.080468 \\
  - &  - &  - &  45.717820 
 \end{pmatrix}.
\end{equation}
For the LRG3 sample at $z_{\rm eff}=0.92$ we find,
 \begin{equation}
   D_{\rm SF+BAO}^{\rm LRG3}(0.92)=
 \begin{pmatrix}
D_V/r_{\rm d}(0.92) \\
D_H/D_M(0.92) \\
f\sigma_{s8}(0.92) \\
(m+n) (0.92)
 \end{pmatrix}=
  \begin{pmatrix}
19.772506 \\
0.825195 \\ 
0.438454 \\
-0.024940
 \end{pmatrix},
 \end{equation}
with the following covariance,
   \begin{equation} 
C_{\rm SF+BAO}^{\rm LRG3} (0.92) = 10^{-4}
 \begin{pmatrix}
300.214114 &  -2.141102 &  8.238180 &  -8.800301 \\
  - &  5.531413 &  -4.054349 &  -0.329386 \\
  - &  - &  17.011310 &  4.097331 \\
  - &  - &  - &  26.960240 
 \end{pmatrix}.
\end{equation}
For the ELG2 sample at $z_{\rm eff}=1.32$ we find,
 \begin{equation}
   D_{\rm SF+BAO}^{\rm ELG2}(1.32)=
 \begin{pmatrix}
D_V/r_{\rm d}(1.32) \\
D_H/D_M(1.32) \\
f\sigma_{s8}(1.32) \\
(m+n) (1.32)
 \end{pmatrix}=
  \begin{pmatrix}
  24.247096 \\ 
  0.486334 \\ 
  0.372661 \\ 
  0.041720
 \end{pmatrix},
 \end{equation}
with the following covariance,
   \begin{equation} 
C_{\rm SF+BAO}^{\rm ELG2} (1.32) = 10^{-4}
 \begin{pmatrix}
1205.501318 &  -15.845487 &  29.543675 &  -41.167094 \\
  - &  4.516231 &  -3.345970 &  -1.151721 \\
  - &  - &  11.811326 &  8.361394 \\
  - &  - &  - &  38.465670 \\
 \end{pmatrix}.
\end{equation}
For the QSO sample at $z_{\rm eff}=1.49$ we find,
 \begin{equation}
   D_{\rm SF+BAO}^{\rm QSO}(1.49)=
 \begin{pmatrix}
D_V/r_{\rm d}(1.49) \\
D_H/D_M(1.49) \\
f\sigma_{s8}(1.49) \\
(m+n) (1.49)
 \end{pmatrix}=
  \begin{pmatrix}
25.768648 \\ 
0.425629 \\
0.436849 \\
0.061950
\end{pmatrix},
 \end{equation}
with the following covariance,
   \begin{equation} 
C_{\rm SF+BAO}^{\rm QSO} (1.49) = 10^{-4}
 \begin{pmatrix}
2601.484225 &  -2.427625 &  40.332929 &  -77.850995 \\
  - &  5.679769 &  -7.100070 &  -2.754462 \\
  - &  - &  21.028703 &  6.989370 \\
  - &  - &  - &  26.733090 \\
 \end{pmatrix}.
\end{equation}


\section{Consistency tests on the full-modelling approach}

In this appendix, we display the battery of tests performed on unblinded data that demonstrate that the results presented in \cref{sec:results} are robust. These tests are complementary to those shown both in  \cref{subsec:methods-baseline} about the baseline selection, and those on \cref{sec:consistency_unblindeddata_SF} about the robustness of unblinded results, and just re-confirm the same conclusions drawn from those sections. 

\begin{figure}
    \centering
    \includegraphics[width=1.0\linewidth]{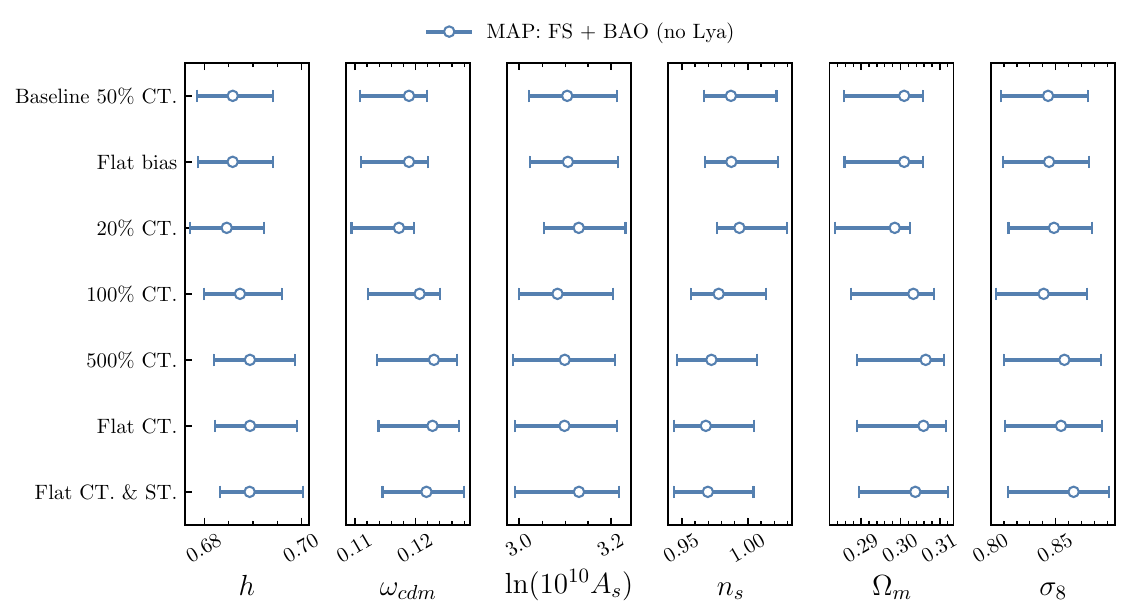}
    \caption{Constraints on $\Lambda$CDM parameters from the joint fit of the six tracers of \desidrone\,data with the BAO reconstruction signal included. We employ the Full-Modelling approach using different choices of priors on counterterms (CT) and stochastic terms (ST), as displayed, and described in \cref{subsec:methods-baseline}. The filled(empty) symbols represent the mean(MAP) value. The rows correspond to
    the same baseline models tested in  \cref{fig:priors}. }    \label{fig:priors_on_data}
\end{figure}

\cref{fig:priors_on_data} displays the performance of different setup options for the priors of counter-terms and stochastic terms, analogously to \cref{fig:priors}, but for the \desidrone\,data instead. We focus only on the Full-Shape combined with BAO data, but the results are very similar to the Full-Shape only datasets. Changing the counter-terms prior can moderately affect some parameters by fractions of the statistical $68\%$ confidence region, similarly to what was displayed already in \cref{fig:priors}. We therefore re-confirm with the actual data that our baseline choice produces robust results.

We have also explored some extra model choices, such as the choice between the Eulerian and Lagrangian  PT modes. The results on the \desidrone\,data are very similar to those already shown in \cref{fig:th-syst} and we do not show them for conciseness. The highest deviation we observe in this choice is on the $\Omega_m$ parameter, with a change of $\Delta\Omega_m\sim0.002$ towards higher values of $\Omega_m$ when the Lagrangian PT is used compared to the Eulerian PT, when using both full-modelling and reconstructed BAO information. Additionally, we have checked that by adding the hexadecapole signal to the \desidrone\,data does not produce any significant changes in the contours, as anticipated by Ref.\cite{KP5s4-Lai} in figure~10. 

\begin{figure}
    \centering
    \includegraphics[width=0.8\linewidth]{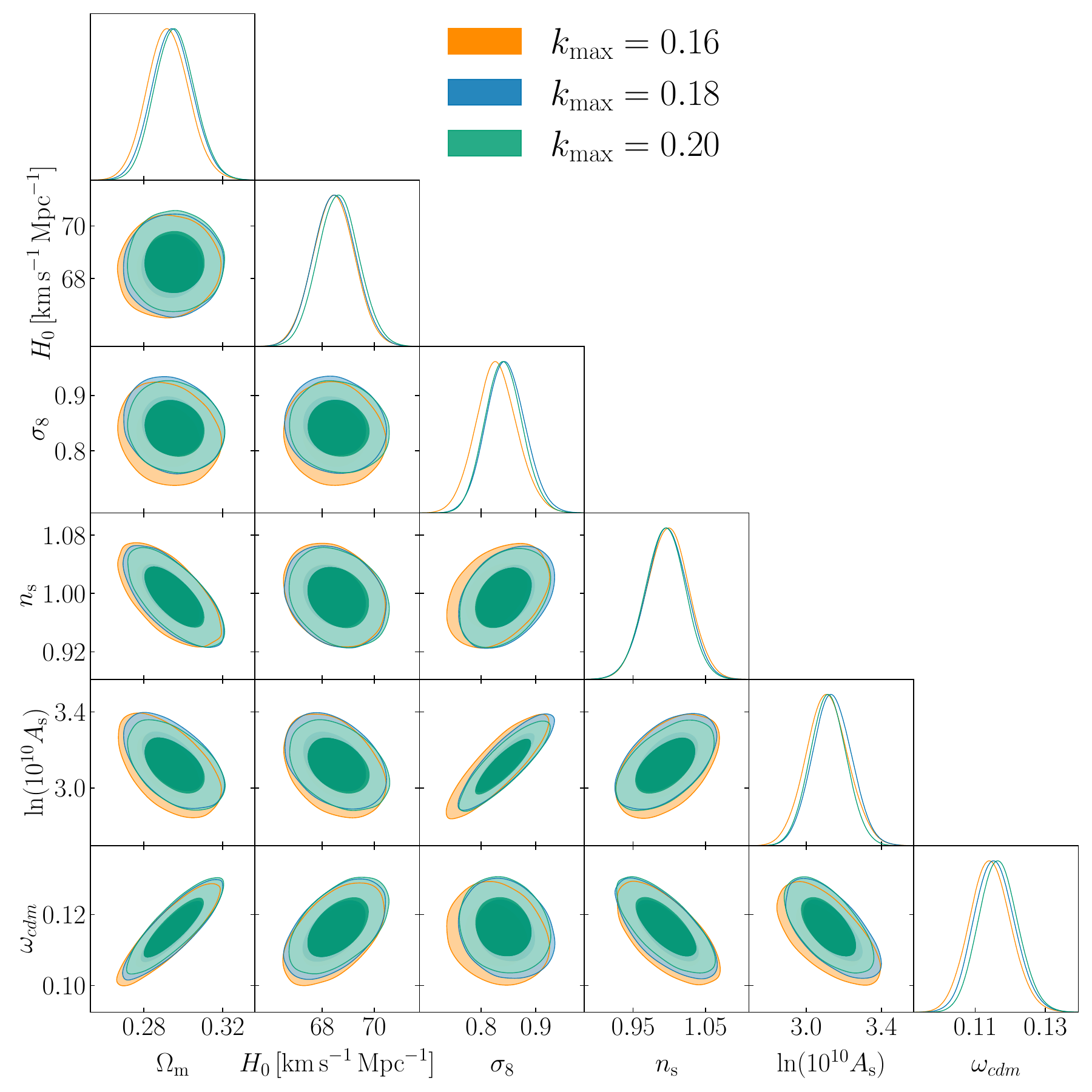}
    \caption{Effect of changing the minimum scale cutoff, $k_{\rm max}$ on the posteriors of the $\Lambda$CDM model for the Full-Modelling type of analysis in combination with the BAO reconstructed data} on \desidrone\, data. We test two larger cutoff scales, $k_{\rm max}=0.16,\,0.18\,h{\rm Mpc}^{-1}$, in addition to the baseline choice of $0.20\,h{\rm Mpc}^{-1}$.
    \label{fig:kmax}
\end{figure}

\begin{figure}
    \centering
    \includegraphics[width=0.8\linewidth]{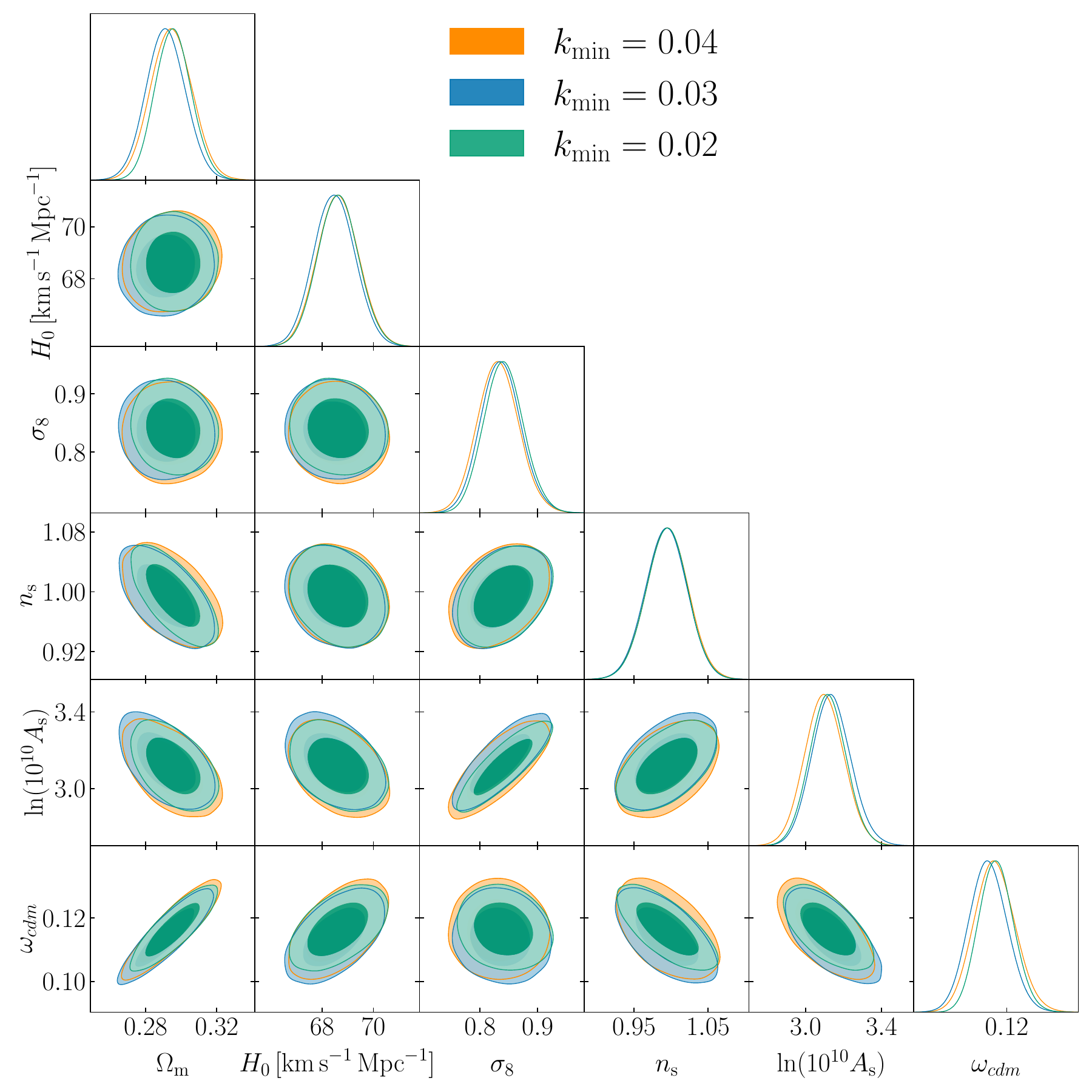}
    \caption{Effect of changing the maximum scale cutoff, $k_{\rm min}$ on the posteriors of the $\Lambda$CDM model for the Full-Modelling type of analysis in combination with the BAO reconstructed data on \desidrone\, data. We test the cutoff scales, $k_{\rm max}=0.03,\,0.04\,h{\rm Mpc}^{-1}$, in addition to the baseline choice of $0.02\,h{\rm Mpc}^{-1}$.}
    \label{fig:kmin}
\end{figure}

In \cref{fig:kmax} we display the impact of different minimum cutoff scales, $k_{\rm max}$ for the power spectrum perturbation theory model on the \desidrone\, data. Refs.\cite{KP5s4-Lai,KP5s2-Maus,KP5s3-Noriega} already validated using mocks that the cutoff scale was around $0.20\,h{\rm Mpc}^{-1}$, which is part of the baseline model. We display two additional conservative cuts, at $k_{\rm max}=0.16,\,0.18\,{h{\rm Mpc}^{-1}}$. We find barely no difference in the position of the maximum of the posteriors nor in their width, validating the robustness of results and statistical errors under changes of the exact $k_{\rm max}$ choice. We have performed a similar analysis changing the $k_{\rm min}$ choice to vary from $0.02\,{h{\rm Mpc}^{-1}}$ to $0.04\,{h{\rm Mpc}^{-1}}$ with no significant changes, as displayed in \cref{fig:kmin}.

\begin{figure}
    \centering
    \includegraphics[width=0.8\linewidth]{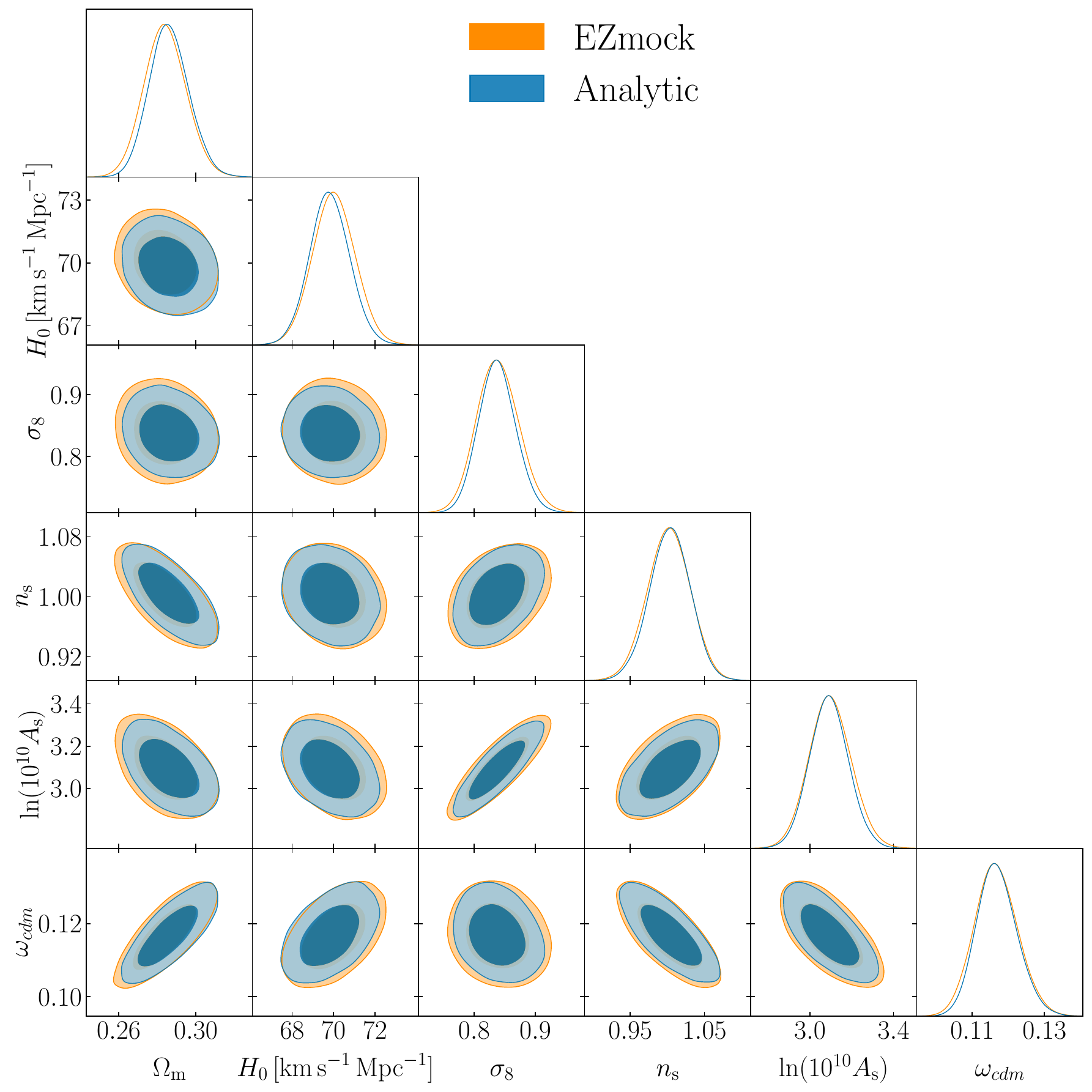}
    \caption{Effect of the choice of the covariance when analyzing the \desidrone\,data. The orange contours display the \texttt{EZmock}-based covariance which is the baseline method employed for the main results of this work. The blue contours display the results for the analytical covariance, as described in \cref{subsec:methods-cov}. }
    \label{fig:cov_on_data}
\end{figure}

We also aim to validate the effect of the mock-based covariance in the final results. We compare the baseline results with those obtained when using an analytical-based covariance as described in \cref{subsec:methods-cov}. As described before, the mock-based covariance is generated from 1000 \texttt{EZmocks}, whereas the analytical approach is based on Gaussian approximation in combination with non-linear clustering methods. 
\cref{fig:cov_on_data} shows very good agreement between the posteriors obtained from the mock-based covariance (the baseline choice) and the alternative analytical covariance choice. We observe that the analytical approach slightly under-estimates the posteriors obtained from the mock-based covariance, likely due to the non-inclusion of all non-linear terms. Thus, we conclude that both methods retrieve very consistent posteriors, and that the baseline method is the most conservative option as it retrieves slightly larger errors. 

\begin{figure}
    \centering
    \includegraphics[width=0.4\linewidth]{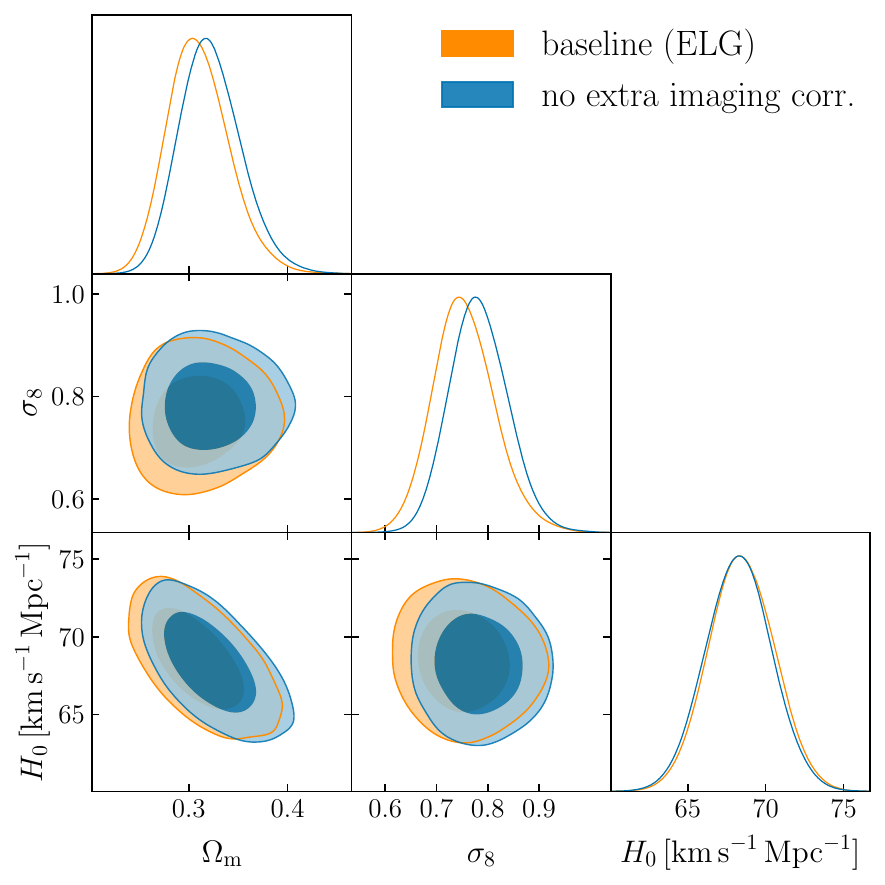}
    \includegraphics[width=0.4\linewidth]{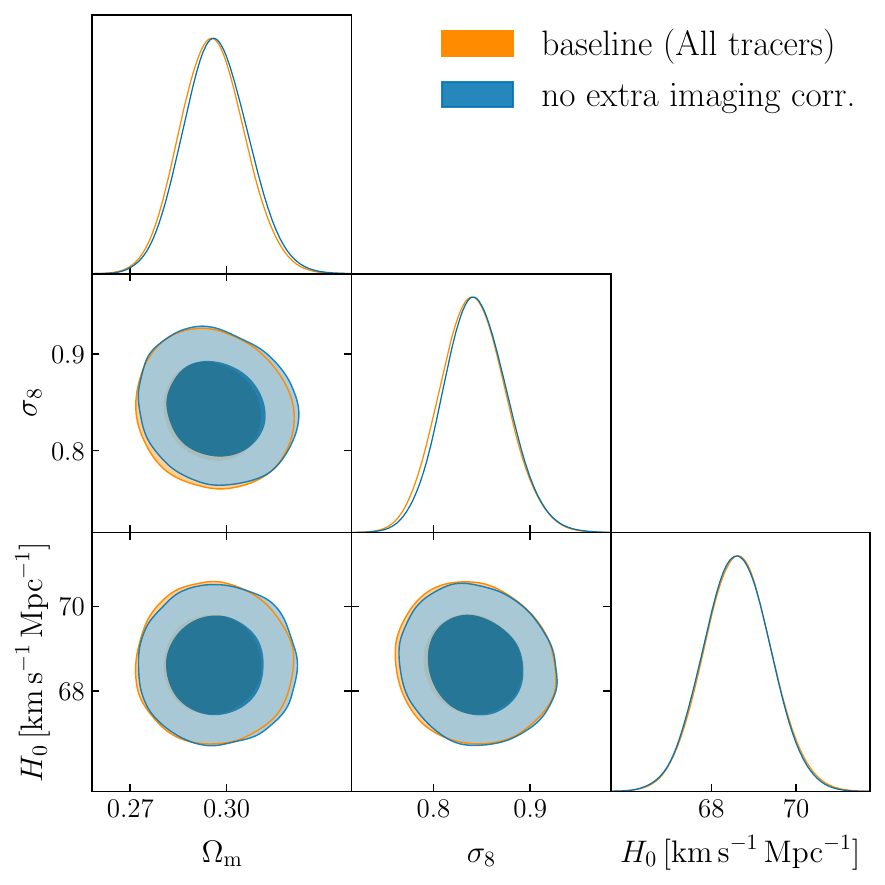}
    \caption{Effect of not applying the imaging systematic correction described in equation~\ref{eq:imaging-corr}, on the ELG sample (left panel) and on all combined galaxy and quasar samples (right panel). Although $n_s$ and $\Omega_b$ are also varied, here we do not display them for conciseness, as the imaging systematic correction impact on those variables is totally negligible. For both panels we display the posteriors obtained when combining the full-shape analysis with the BAO reconstruction information, although only considering the full-shape information does not change the significantly the relative effect.}
    \label{fig:no_imaging_sys}
\end{figure}

For a full description of the effect of the imaging weight correction on the cosmology results for the full-modelling approach we refer the reader to Ref.\cite{KP5s6-Zhao}. In short, we find that dropping the correction terms of equation~\ref{eq:imaging-corr} produces no change for the LRG and BGS samples as expected. Surprisingly, we also find that the effect of these terms is very small for the QSO sample. The largest effect is therefore for the ELG sample (as displayed on the left panel of \cref{fig:no_imaging_sys}), although the overall effect is to modify the position of the posteriors of $\Omega_m$ and $\sigma_8$ by around 1/4 of a sigma at the most. Although not shown, the impact on $\Omega_b h^2$ and $n_s$ is negligible. The panels of~\cref{fig:no_imaging_sys} show the posteriors when the reconstructed BAO information is also added, although the results look very similar if the BAO reconstructed data is not used. However, when all samples are combined (right panel of~\cref{fig:no_imaging_sys}), the overall effect of ignoring these correction terms is again negligible, as the signal is mainly driven by BGS, LRGs and QSO.

\begin{figure}
    \centering
    \includegraphics[width=0.7\linewidth]{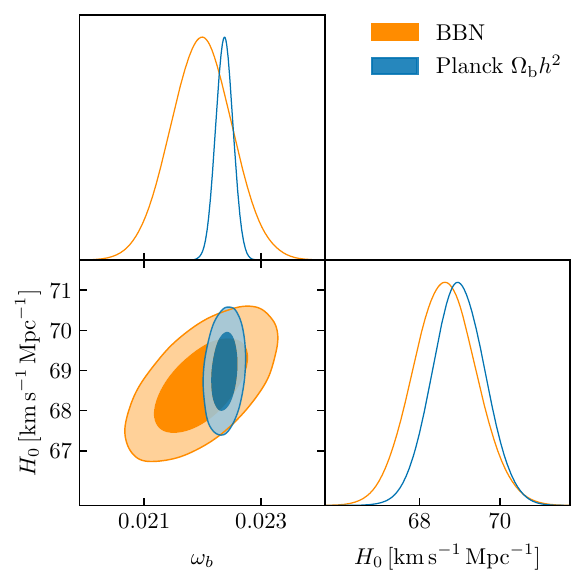}
    \caption{Effect of modifying the physical density baryon prior on the $H_0$ parameter inferred from \desidrone\, data in combination with the BAO reconstructed signal. We choose the BBN-informed prior (as our baseline choice) and the Planck measured value, as labelled. Due to the high consistency between BBN and Planck, the impact on $H_0$ is minor. We do not display the rest of $\Lambda$CDM parameters, as their dependence on $\Omega_bh^2$ (as expected) is negligible.}
    \label{fig:bbn_effect}
\end{figure}

Finally, we also test here the effect of changing the prior on the physical baryon density, $\Omega_bh^2$ from the BBN one to the Planck prior, for the \desidrone\,data in combination with the BAO reconstructed signal. Since this choice mainly affects the determination of $H_0$, we display in \cref{fig:bbn_effect} the contours only for this parameter. We see that the change in prior on $\Omega_bh^2$ does not affect the constraints on $H_0$, as expected given the very good consistency between the physical baryon density inferred from BBN and Planck data.

\section{Fiducial cosmology values for the ShapeFit compressed parameters}
\label{app:fidvalues}

Here we report in \cref{tab:fiducial_values} the fiducial BAO distances and $\sigma_{s8}$ value employed in this paper for the ShapeFit type of analysis. The details of this fiducial cosmology model, based on Planck-$\Lambda$CDM, are reported in the first row of \cref{tab:fid_cosmo}.

\begin{table}[htp]
    \centering
    \begin{tabular}{|c|c|c|c|c|c|c|}
    \hline
        sample (redshift) & $D_M(z)/r_d$ & $D_H(z)/r_d$ & $D_V(z)/r_d$ & $D_H(z)/D_M(z)$ & $\sigma_{s8}(z)$ & $f\sigma_{s8}(z)$   \\
        \hline\hline
        \bgs\ (0.295) & $8.2908$ & $25.8506$ & $8.0663$ & $3.1180$ & $0.6936$ & $0.4723$ \\
        \lrgo\ (0.510) & $13.4928$ & $22.7462$ & $12.8269$ & $1.6858$ & $0.6210$ & $0.4733$ \\
        \lrgt\ (0.706) & $17.6976$ & $20.1727$ & $16.4597$ & $1.1399$ & $0.5638$ & $0.4608$ \\
        \lrgth\ (0.919) & $21.7238$ & $17.7321$ & $19.7356$ & $0.8162$ & $0.5108$ & $0.4398$ \\
        \elgt\ (1.317) & $28.0276$ & $14.0956$ & $24.4318$ & $0.5029$ & $0.4320$ & $0.3944$ \\
        \qso\ (1.491) & $30.3606$ & $12.8359$ & $26.0292$ & $0.4228$ & $0.4042$ & $0.3750$ \\
\hline
          
    \end{tabular}
    \caption{Fiducial values on the angular and radial BAO distances, $D_M$ and $D_H$, in units of the sound horizon scale, for the six DESI DR1 samples. Additionally, we report the isotropic BAO distance, $D_V$ and the Alcock-Paczynski related parameter, $D_H/D_M$. For all redshift bins, we have that the sound horizon scale at drag epoch is $r_d=99.0792\,[\mathrm{Mpc}/h]$. We also report the amplitude of cold+baryon matter fluctuations smoothed in units of $8\,{\rm Mpc}/h$ of the fiducial cosmology, $\sigma_{s8}$.}
    \label{tab:fiducial_values}
\end{table}

\section{Pre-reconstructed two-point clustering of \desidrone}
\label{sec:clustering_blinding}

\begin{figure}
    \centering
    \includegraphics[width=1.\textwidth]{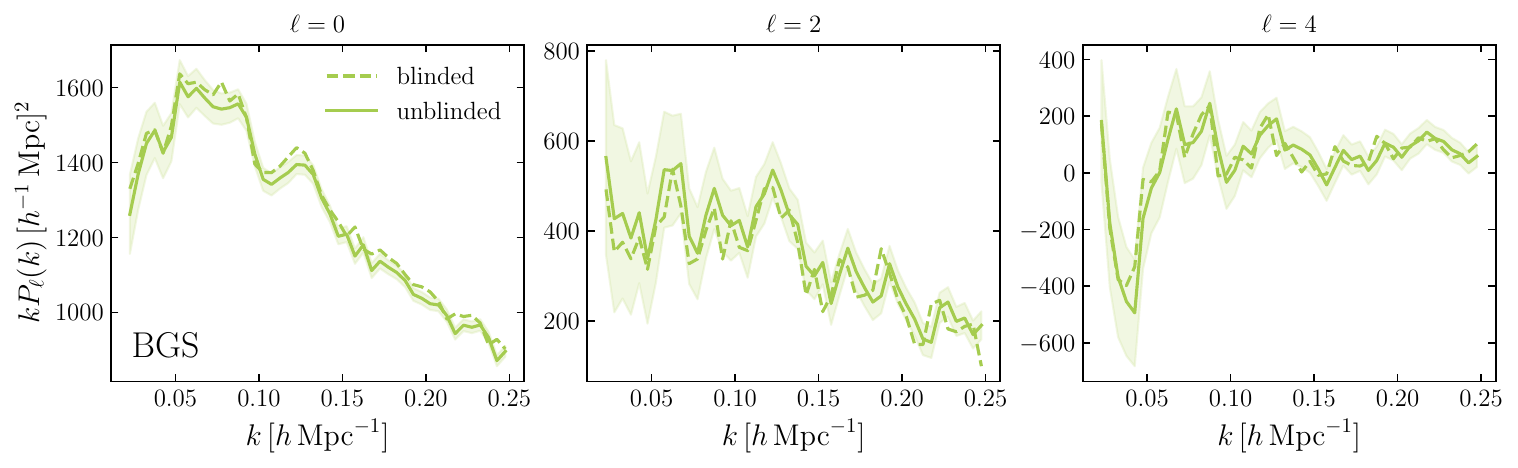} \\
    \includegraphics[width=1.\textwidth]{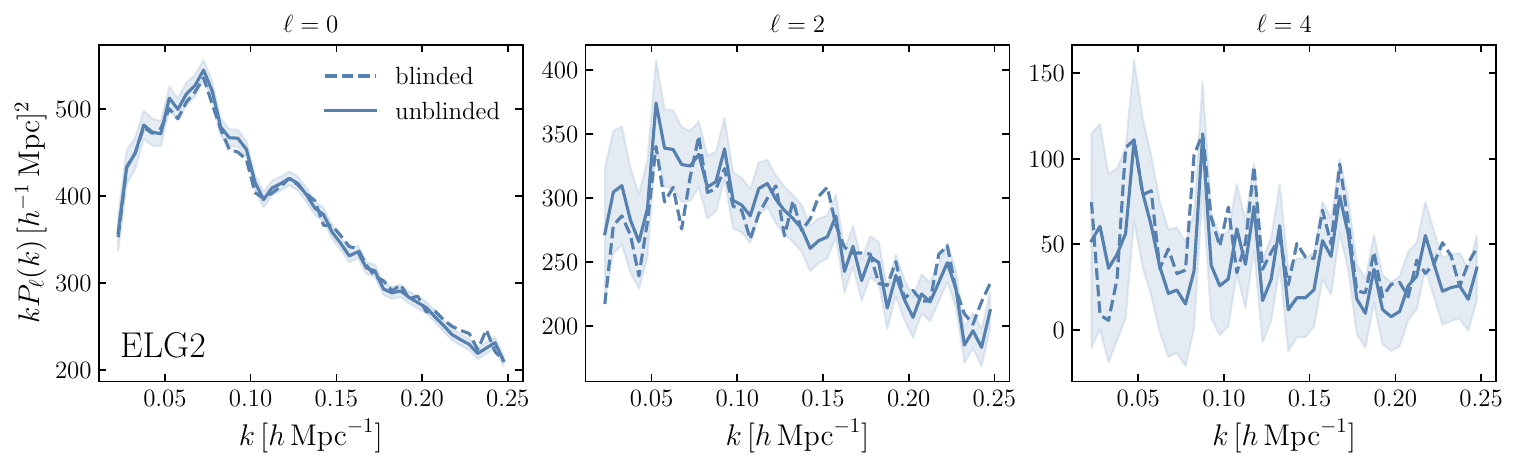} \\
    \includegraphics[width=1.\textwidth]{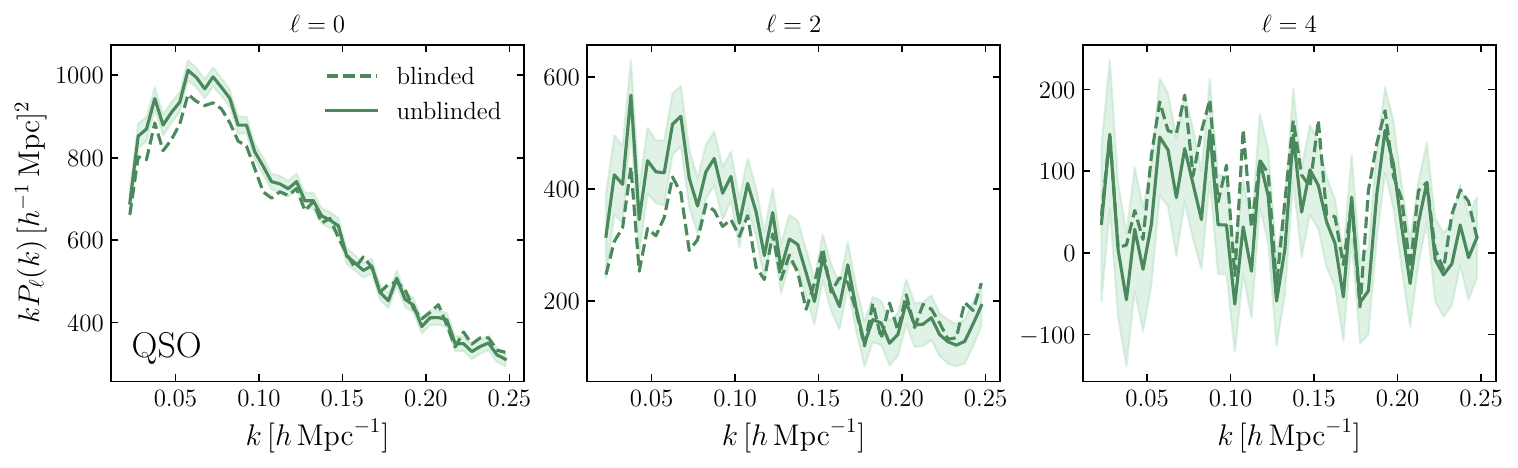} \\
    \caption{Power spectrum multipoles (monopole on the left columns, quadrupole on the middle column, hexadecapole on the right column) corresponding to the BGS (top row), ELG (middle row) and QSO (bottom row)  data samples before (dashed lines) and after unblinding (solid lines). }
    \label{fig:Pk-unblind-1}
\end{figure}

\begin{figure}
    \centering
    \includegraphics[width=1.\textwidth]{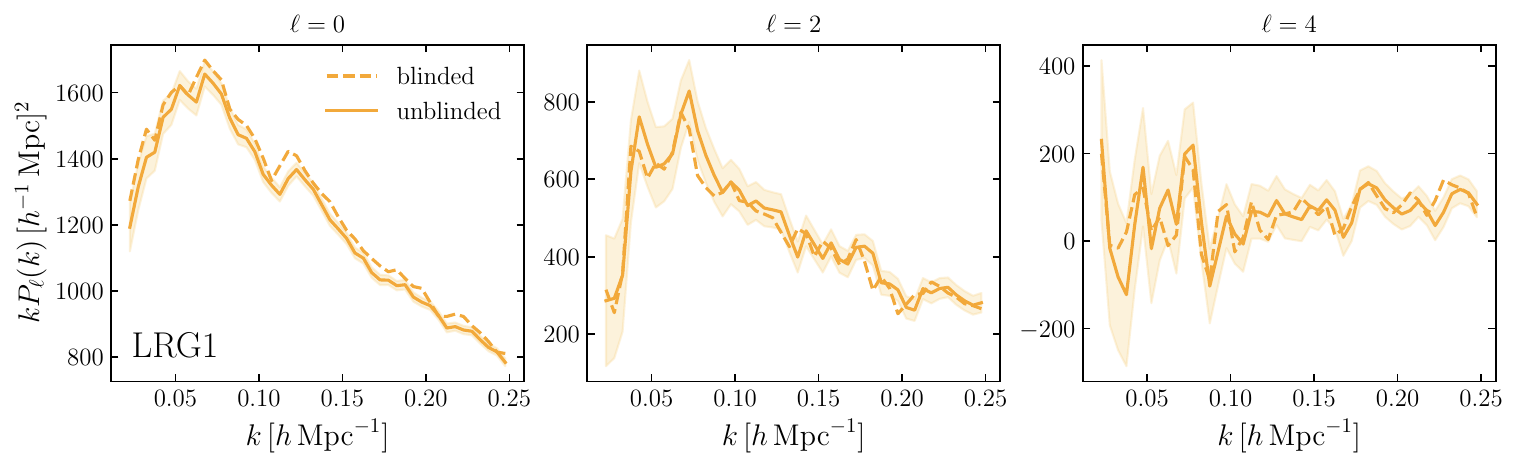} \\
    \includegraphics[width=1.\textwidth]{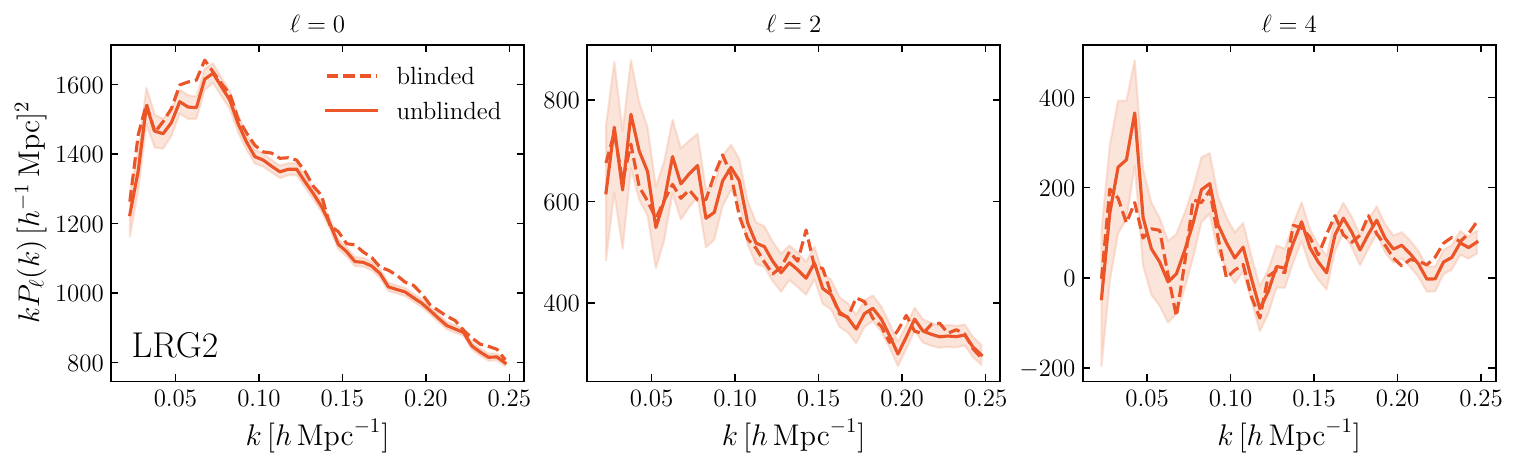} \\
    \includegraphics[width=1.\textwidth]{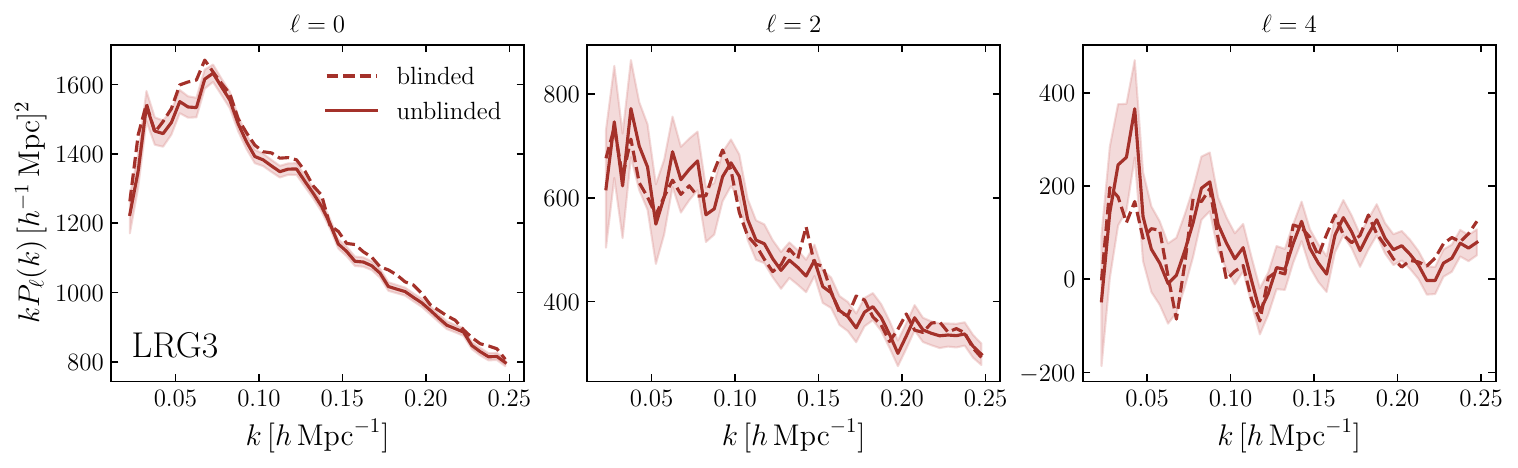} \\
    \caption{Power spectrum multipoles before and after unblinding. Same notation as \cref{fig:Pk-unblind-1} but for the three LRG redshift bins.}
    \label{fig:Pk-unblind-2}
\end{figure}

\cref{fig:Pk-unblind-1} and \cref{fig:Pk-unblind-2} show the three multipoles of the power spectrum of the four DESI samples with the LRG sample being split into three redshift bins for the unblinded (solid) and blinded (dashed) catalogues. The shaded region corresponds to the 1$\sigma$ error from the diagonal terms of the \ezmocks\ covariance matrix. We notice that the blinding procedure successfully modified the BAO position and the amplitude of the quadrupole at large scales. The biggest effect can be observed on the amplitude of the monopole and quadrupole at large scales for QSO. For further validation tests of the blinding procedure for \desidrone\ BAO and Full-Shape galaxy two-point clustering, we refer the reader to our supporting paper \cite{KP3s9-Andrade}.

\section{Gaussian priors on counterterms}\label{app:priors}

Here we explicitly show how we define the priors on counterterms based on their contribution (at $1\sigma$) on the total linear signal.

The galaxy bias term at leading order can be written as,
\begin{equation}
    \delta^{(1)}(k,\mu)=(b+f\mu^2)\delta_L(k)
\end{equation}
where $\delta_L$ is the linear matter\footnote{in fact, only baryons and dark matter, `cb'.} field in real space, $b$ is the (linear) bias, $f$ the logarithmic growth factor and $\mu$ the cosine of the angle between the galaxy pair vector and the line-of-sight.

The counterterm contribution at the $\delta$-level is given by,
\begin{equation}
    \delta_{\rm CT}(k,\mu) = \frac{1}{2}(b\alpha_0 + f\mu^2\alpha_2 + \ldots) k^2 \delta_L(k)
\end{equation}
where $\alpha_i$ are the counterterms, which for this case we only consider up to $\mu^2$ term. 

We then write the linear galaxy signal in redshift space and the first  counter-term  contribution, respectively, as,
\begin{equation}
    P_{\rm lin,\,g}(k,\mu)=\langle \delta^{(1)}(k) {\delta^{(1)}(k)}^*\rangle; \quad  P_{\rm CT,\,g}(k,\mu)=2\langle \delta^{(1)}(k) {\delta_{\rm CT}(k)}^*\rangle
\end{equation}
The Gaussian prior we impose on the counterterms $\alpha_i$ are such that the contribution of $P_{\rm CT,g}$ is a fraction of $X$ of $P_{\rm lin,g}$ at $k=k_{\rm max}$ for $1\sigma$,
\begin{equation}
    X P_{\rm lin}(k_{\rm max},\mu) \geq  |P_{\rm CT}(k_{\rm max},\mu)|
\end{equation}
The $\mu^0$ term implies that $\alpha_0$ must satisfy,
\begin{equation}
    \frac{X}{k_{\rm max}^2}\geq |\alpha_0|
\end{equation}
and similarly for the $\mu^2$ term on $\alpha_2$ (given that we now have fixed the prior on $\alpha_0$),
\begin{equation}
    \frac{X}{k_{\rm max}^2}\geq |\alpha_2|
\end{equation}
For our baseline choice $X=0.5$ (50\% contribution) and $k_{\rm max}=0.20\,h^{-1}{\rm Mpc}$ we find the $1\sigma$ Gaussian constraints used in Table~\ref{tab:priors},
\begin{equation}
   \alpha_{0,2} = \frac{0.5}{0.20^2}\,{\rm Mpc}^2h^{-2}=12.5\,{\rm Mpc}^2h^{-2}  
\end{equation}

\section{Debiasing the inverse of the sum of a sample covariance and an external covariance}\label{app:invcov}

We consider here the case of adding a sample-estimated covariance matrix and some external (e.g. systematic) covariance.

The standard sample covariance matrix estimate for an unknown mean is,
\begin{equation}
    X_{ab} = \frac1{n_S-1} \sum_{i=1}^{n_S} \qty(x_{a,i} - \bar x_a) \qty(x_{b,i} - \bar x_b),
\end{equation}
with $\bar x_a \equiv \frac1{n_S} \sum_{i=1}^{n_S} x_{a,i}$.

The sample covariance estimate is unbiased, meaning that the expectation value is the true covariance: $\ev{\bf X}={\bf C}_s$.
We can write ${\bf X} = {\bf C}_s + {\bf \delta X}$ with $\ev{\bf \delta X}=0$.

The covariance of sample covariance elements estimated in this way are,
\begin{eqnarray} \label{eq:X-covariance}
\cov \qty(X_{ab}, X_{cd}) = \ev{\delta X_{ab} \delta X_{cd}} = \frac{C_{s,ac} C_{s,bd} + C_{s,ad} C_{s,bc}}{n_S-1},
\end{eqnarray}
and the total covariance is,
\begin{equation}
    {\bf C} = {\bf X} + {\bf C}_e,
\end{equation}
and we assume that ${\bf C}_e$ is exact.
Naturally, $\ev{\bf C} = {\bf C}_s + {\bf C}_e$.

Let now ${\bf C}^{-1}$ be a biased estimate due to the noise in ${\bf X}$ and the non-linearity of the inversion operation.
Hence, we can write $\ev{{\bf C}^{-1}} \ne \qty({\bf C}_s + {\bf C}_e)^{-1} \equiv {\bf \Psi}_0$, where the term in the RHS of the non-equality is what we aim to compute.

Let us now expand the inverse of the total covariance,
\begin{align}
    {\bf C}^{-1} & = \qty({\bf C}_s + {\bf C}_e + {\bf \delta X})^{-1} = \qty[\mathbb{I} + \qty({\bf C}_s + {\bf C}_e)^{-1} {\bf \delta X}]^{-1} \qty({\bf C}_s + {\bf C}_e)^{-1} \\
    & = \qty(\mathbb{I} + {\bf \Psi}_0 {\bf \delta X})^{-1} {\bf \Psi}_0 = \qty(\mathbb{I} - {\bf \Psi}_0 {\bf \delta X} + {\bf \Psi}_0 {\bf \delta X} {\bf \Psi}_0 {\bf \delta X} - \dots) {\bf \Psi}_0.
\end{align}
Then let us take the ensemble average (its expectation value),
\begin{equation}
    \ev{{\bf C}^{-1}} \approx \qty(\mathbb{I} - \cancel{{\bf \Psi}_0 \ev{\bf \delta X}} + {\bf \Psi}_0 \ev{{\bf \delta X} {\bf \Psi}_0 {\bf \delta X}}) {\bf \Psi}_0.
\end{equation}
The linear term drops out because $\ev{\bf \delta X}=0$.

If we  designate ${\bf D} \equiv {\bf \Psi}_0 \ev{{\bf \delta X} {\bf \Psi}_0 {\bf \delta X}}$ we then can write,
\begin{align}
    \ev{{\bf C}^{-1}} & \approx \qty(\mathbb{I} + {\bf D}) {\bf \Psi}_0, \\
    {\bf \Psi}_0 & \approx \qty(\mathbb{I} + {\bf D})^{-1} \ev{{\bf C}^{-1}} \approx \qty(\mathbb{I} - {\bf D}) \ev{{\bf C}^{-1}}.
\end{align}
Hence, left-multiplication by $\qty(\mathbb{I} - {\bf D})$ unbiases the estimate of the inverse of the total covariance to the next-to-leading nonvanishing order.

Now, to compute the ${\bf D}$ matrix we first consider its $ij$ element,
\begin{equation}
    D_{ij} = \Psi_{0,ik} \Psi_{0,lm} \ev{\delta X_{kl} \delta X_{mj}} = \frac1{n_S-1} \Psi_{0,ik} \Psi_{0,lm} \qty(C_{s,km} C_{s,lj} + C_{s,kj} C_{s,lm}).
\end{equation}
We can then cast this back to matrix operations:
\begin{equation}
    {\bf D} = \frac1{n_S - 1} {\bf \Psi}_0 {\bf C}_s \qty[\tr({\bf \Psi}_0 {\bf C}_s) \cdot \mathbb{I} + {\bf \Psi}_0 {\bf C}_s].
\end{equation}
In the special case of ${\bf C}_e=0$, ${\bf \Psi}_0 = {\bf C}_s^{-1}$ and we obtain the Hartlap correction ${\bf D} = \frac{N_{\rm bins} + 1}{n_S-1} \cdot \mathbb{I}$.

In the general case, we should use approximations instead of unknown ${\bf \Psi}_0$ and ${\bf C}_s$ in the expression for $\bf D$.
It is natural to take ${\bf C}_s \approx {\bf X}$ and ${\bf \Psi}_0 \approx \qty({\bf X} + {\bf C}_e)^{-1}$.
Thus the practical estimates would be,
\begin{align}
    {\bf \hat D} & = \frac1{n_S - 1} \qty({\bf X} + {\bf C}_e)^{-1} {\bf X} \qty{\tr[\qty({\bf X} + {\bf C}_e)^{-1} {\bf X}] \cdot \mathbb{I} + \qty({\bf X} + {\bf C}_e)^{-1} {\bf X}}, \\
    {\bf \hat \Psi} & = \qty(\mathbb{I} - {\bf \hat D}) \qty({\bf X} + {\bf C}_e)^{-1}.
\end{align}
This unbiased (to next-to-leading order) precision matrix estimate is symmetric (theoretically, numerical errors may lead to small violations).

\bibliographystyle{JHEP}
\bibliography{DESIKP5}


\section{Author Affiliations}
\label{sec:affiliations}

\noindent \hangindent=.5cm $^{1}${Instituto de F\'{\i}sica Te\'{o}rica (IFT) UAM/CSIC, Universidad Aut\'{o}noma de Madrid, Cantoblanco, E-28049, Madrid, Spain}

\noindent \hangindent=.5cm $^{2}${Lawrence Berkeley National Laboratory, 1 Cyclotron Road, Berkeley, CA 94720, USA}

\noindent \hangindent=.5cm $^{3}${Physics Dept., Boston University, 590 Commonwealth Avenue, Boston, MA 02215, USA}

\noindent \hangindent=.5cm $^{4}${Tata Institute of Fundamental Research, Homi Bhabha Road, Mumbai 400005, India}

\noindent \hangindent=.5cm $^{5}${Centre for Extragalactic Astronomy, Department of Physics, Durham University, South Road, Durham, DH1 3LE, UK}

\noindent \hangindent=.5cm $^{6}${Institute for Computational Cosmology, Department of Physics, Durham University, South Road, Durham DH1 3LE, UK}

\noindent \hangindent=.5cm $^{7}${Department of Physics, University of Michigan, Ann Arbor, MI 48109, USA}

\noindent \hangindent=.5cm $^{8}${Leinweber Center for Theoretical Physics, University of Michigan, 450 Church Street, Ann Arbor, Michigan 48109-1040, USA}

\noindent \hangindent=.5cm $^{9}${IRFU, CEA, Universit\'{e} Paris-Saclay, F-91191 Gif-sur-Yvette, France}

\noindent \hangindent=.5cm $^{10}${Institut de F\'{i}sica d’Altes Energies (IFAE), The Barcelona Institute of Science and Technology, Campus UAB, 08193 Bellaterra Barcelona, Spain}

\noindent \hangindent=.5cm $^{11}${Instituto de Ciencias F\'{\i}sicas, Universidad Aut\'onoma de M\'exico, Cuernavaca, Morelos, 62210, (M\'exico)}

\noindent \hangindent=.5cm $^{12}${Instituto Avanzado de Cosmolog\'{\i}a A.~C., San Marcos 11 - Atenas 202. Magdalena Contreras, 10720. Ciudad de M\'{e}xico, M\'{e}xico}

\noindent \hangindent=.5cm $^{13}${Physics Department, Yale University, P.O. Box 208120, New Haven, CT 06511, USA}

\noindent \hangindent=.5cm $^{14}${Department of Physics and Astronomy, University of California, Irvine, 92697, USA}

\noindent \hangindent=.5cm $^{15}${Department of Physics, Kansas State University, 116 Cardwell Hall, Manhattan, KS 66506, USA}

\noindent \hangindent=.5cm $^{16}${Department of Physics \& Astronomy, University of Rochester, 206 Bausch and Lomb Hall, P.O. Box 270171, Rochester, NY 14627-0171, USA}

\noindent \hangindent=.5cm $^{17}${Institute for Astronomy, University of Edinburgh, Royal Observatory, Blackford Hill, Edinburgh EH9 3HJ, UK}

\noindent \hangindent=.5cm $^{18}${Dipartimento di Fisica ``Aldo Pontremoli'', Universit\`a degli Studi di Milano, Via Celoria 16, I-20133 Milano, Italy}

\noindent \hangindent=.5cm $^{19}${Centre for Astrophysics \& Supercomputing, Swinburne University of Technology, P.O. Box 218, Hawthorn, VIC 3122, Australia}

\noindent \hangindent=.5cm $^{20}${NSF NOIRLab, 950 N. Cherry Ave., Tucson, AZ 85719, USA}

\noindent \hangindent=.5cm $^{21}${Department of Physics \& Astronomy, University College London, Gower Street, London, WC1E 6BT, UK}

\noindent \hangindent=.5cm $^{22}${Department of Astronomy and Astrophysics, University of Chicago, 5640 South Ellis Avenue, Chicago, IL 60637, USA}

\noindent \hangindent=.5cm $^{23}${Fermi National Accelerator Laboratory, PO Box 500, Batavia, IL 60510, USA}

\noindent \hangindent=.5cm $^{24}${Korea Astronomy and Space Science Institute, 776, Daedeokdae-ro, Yuseong-gu, Daejeon 34055, Republic of Korea}

\noindent \hangindent=.5cm $^{25}${Institute of Cosmology and Gravitation, University of Portsmouth, Dennis Sciama Building, Portsmouth, PO1 3FX, UK}

\noindent \hangindent=.5cm $^{26}${Departamento de Astrof\'{\i}sica, Universidad de La Laguna (ULL), E-38206, La Laguna, Tenerife, Spain}

\noindent \hangindent=.5cm $^{27}${Instituto de Astrof\'{\i}sica de Canarias, C/ V\'{\i}a L\'{a}ctea, s/n, E-38205 La Laguna, Tenerife, Spain}

\noindent \hangindent=.5cm $^{28}${Department of Physics and Astronomy, University of Sussex, Brighton BN1 9QH, U.K}

\noindent \hangindent=.5cm $^{29}${Departamento de F\'{i}sica, Instituto Nacional de Investigaciones Nucleares, Carreterra M\'{e}xico-Toluca S/N, La Marquesa,  Ocoyoacac, Edo. de M\'{e}xico C.P. 52750,  M\'{e}xico}

\noindent \hangindent=.5cm $^{30}${Institute for Advanced Study, 1 Einstein Drive, Princeton, NJ 08540, USA}

\noindent \hangindent=.5cm $^{31}${Center for Cosmology and AstroParticle Physics, The Ohio State University, 191 West Woodruff Avenue, Columbus, OH 43210, USA}

\noindent \hangindent=.5cm $^{32}${NASA Einstein Fellow}

\noindent \hangindent=.5cm $^{33}${School of Mathematics and Physics, University of Queensland, 4072, Australia}

\noindent \hangindent=.5cm $^{34}${Department of Physics and Astronomy, The University of Utah, 115 South 1400 East, Salt Lake City, UT 84112, USA}

\noindent \hangindent=.5cm $^{35}${Instituto de F\'{\i}sica, Universidad Nacional Aut\'{o}noma de M\'{e}xico,  Cd. de M\'{e}xico  C.P. 04510,  M\'{e}xico}

\noindent \hangindent=.5cm $^{36}${CIEMAT, Avenida Complutense 40, E-28040 Madrid, Spain}

\noindent \hangindent=.5cm $^{37}${Department of Physics \& Astronomy and Pittsburgh Particle Physics, Astrophysics, and Cosmology Center (PITT PACC), University of Pittsburgh, 3941 O'Hara Street, Pittsburgh, PA 15260, USA}

\noindent \hangindent=.5cm $^{38}${Department of Astronomy, School of Physics and Astronomy, Shanghai Jiao Tong University, Shanghai 200240, China}

\noindent \hangindent=.5cm $^{39}${Space Sciences Laboratory, University of California, Berkeley, 7 Gauss Way, Berkeley, CA  94720, USA}

\noindent \hangindent=.5cm $^{40}${University of California, Berkeley, 110 Sproul Hall \#5800 Berkeley, CA 94720, USA}

\noindent \hangindent=.5cm $^{41}${Universities Space Research Association, NASA Ames Research Centre}

\noindent \hangindent=.5cm $^{42}${Center for Astrophysics $|$ Harvard \& Smithsonian, 60 Garden Street, Cambridge, MA 02138, USA}

\noindent \hangindent=.5cm $^{43}${Department of Physics, The Ohio State University, 191 West Woodruff Avenue, Columbus, OH 43210, USA}

\noindent \hangindent=.5cm $^{44}${The Ohio State University, Columbus, 43210 OH, USA}

\noindent \hangindent=.5cm $^{45}${Kavli Institute for Particle Astrophysics and Cosmology, Stanford University, Menlo Park, CA 94305, USA}

\noindent \hangindent=.5cm $^{46}${SLAC National Accelerator Laboratory, Menlo Park, CA 94305, USA}

\noindent \hangindent=.5cm $^{47}${Instituto de Astrof\'{i}sica de Andaluc\'{i}a (CSIC), Glorieta de la Astronom\'{i}a, s/n, E-18008 Granada, Spain}

\noindent \hangindent=.5cm $^{48}${Institute of Physics, Laboratory of Astrophysics, \'{E}cole Polytechnique F\'{e}d\'{e}rale de Lausanne (EPFL), Observatoire de Sauverny, CH-1290 Versoix, Switzerland}

\noindent \hangindent=.5cm $^{49}${Departamento de F\'isica, Universidad de los Andes, Cra. 1 No. 18A-10, Edificio Ip, CP 111711, Bogot\'a, Colombia}

\noindent \hangindent=.5cm $^{50}${Observatorio Astron\'omico, Universidad de los Andes, Cra. 1 No. 18A-10, Edificio H, CP 111711 Bogot\'a, Colombia}

\noindent \hangindent=.5cm $^{51}${Department of Physics, The University of Texas at Dallas, Richardson, TX 75080, USA}

\noindent \hangindent=.5cm $^{52}${Center for Computational Astrophysics, Flatiron Institute, 162 5\textsuperscript{th} Avenue, New York, NY 10010, USA}

\noindent \hangindent=.5cm $^{53}${Scientific Computing Core, Flatiron Institute, 162 5\textsuperscript{th} Avenue, New York, NY 10010, USA}

\noindent \hangindent=.5cm $^{54}${Institut d'Estudis Espacials de Catalunya (IEEC), 08034 Barcelona, Spain}

\noindent \hangindent=.5cm $^{55}${Institute of Space Sciences, ICE-CSIC, Campus UAB, Carrer de Can Magrans s/n, 08913 Bellaterra, Barcelona, Spain}

\noindent \hangindent=.5cm $^{56}${Departament de F\'{\i}sica Qu\`{a}ntica i Astrof\'{\i}sica, Universitat de Barcelona, Mart\'{\i} i Franqu\`{e}s 1, E08028 Barcelona, Spain}

\noindent \hangindent=.5cm $^{57}${Institut de Ci\`encies del Cosmos (ICCUB), Universitat de Barcelona (UB), c. Mart\'i i Franqu\`es, 1, 08028 Barcelona, Spain.}

\noindent \hangindent=.5cm $^{58}${Consejo Nacional de Ciencia y Tecnolog\'{\i}a, Av. Insurgentes Sur 1582. Colonia Cr\'{e}dito Constructor, Del. Benito Ju\'{a}rez C.P. 03940, M\'{e}xico D.F. M\'{e}xico}

\noindent \hangindent=.5cm $^{59}${Departamento de F\'{i}sica, Universidad de Guanajuato - DCI, C.P. 37150, Leon, Guanajuato, M\'{e}xico}

\noindent \hangindent=.5cm $^{60}${Centro de Investigaci\'{o}n Avanzada en F\'{\i}sica Fundamental (CIAFF), Facultad de Ciencias, Universidad Aut\'{o}noma de Madrid, ES-28049 Madrid, Spain}

\noindent \hangindent=.5cm $^{61}${Excellence Cluster ORIGINS, Boltzmannstrasse 2, D-85748 Garching, Germany}

\noindent \hangindent=.5cm $^{62}${University Observatory, Faculty of Physics, Ludwig-Maximilians-Universit\"{a}t, Scheinerstr. 1, 81677 M\"{u}nchen, Germany}

\noindent \hangindent=.5cm $^{63}${Department of Astrophysical Sciences, Princeton University, Princeton NJ 08544, USA}

\noindent \hangindent=.5cm $^{64}${Institut d'Astrophysique de Paris. 98 bis boulevard Arago. 75014 Paris, France}

\noindent \hangindent=.5cm $^{65}${Institute for Fundamental Physics of the Universe, via Beirut 2, 34151 Trieste, Italy}

\noindent \hangindent=.5cm $^{66}${International School for Advanced Studies, Via Bonomea 265, 34136 Trieste, Italy}

\noindent \hangindent=.5cm $^{67}${Kavli Institute for Cosmology, University of Cambridge, Madingley Road, Cambridge CB3 0HA, UK}

\noindent \hangindent=.5cm $^{68}${Department of Astronomy, The Ohio State University, 4055 McPherson Laboratory, 140 W 18th Avenue, Columbus, OH 43210, USA}

\noindent \hangindent=.5cm $^{69}${Department of Physics, Southern Methodist University, 3215 Daniel Avenue, Dallas, TX 75275, USA}

\noindent \hangindent=.5cm $^{70}${The Ohio State University, Columbus, 43210 OH, USA"}

\noindent \hangindent=.5cm $^{71}${Institute of Astronomy, University of Cambridge, Madingley Road, Cambridge CB3 0HA, UK}

\noindent \hangindent=.5cm $^{72}${Department of Physics and Astronomy, University of Waterloo, 200 University Ave W, Waterloo, ON N2L 3G1, Canada}

\noindent \hangindent=.5cm $^{73}${Perimeter Institute for Theoretical Physics, 31 Caroline St. North, Waterloo, ON N2L 2Y5, Canada}

\noindent \hangindent=.5cm $^{74}${Waterloo Centre for Astrophysics, University of Waterloo, 200 University Ave W, Waterloo, ON N2L 3G1, Canada}

\noindent \hangindent=.5cm $^{75}${Graduate Institute of Astrophysics and Department of Physics, National Taiwan University, No. 1, Sec. 4, Roosevelt Rd., Taipei 10617, Taiwan}

\noindent \hangindent=.5cm $^{76}${Schmidt Sciences, 155 W 23rd St, New York, NY 10011, USA}

\noindent \hangindent=.5cm $^{77}${Sorbonne Universit\'{e}, CNRS/IN2P3, Laboratoire de Physique Nucl\'{e}aire et de Hautes Energies (LPNHE), FR-75005 Paris, France}

\noindent \hangindent=.5cm $^{78}${Department of Astronomy and Astrophysics, UCO/Lick Observatory, University of California, 1156 High Street, Santa Cruz, CA 95064, USA}

\noindent \hangindent=.5cm $^{79}${Department of Astronomy and Astrophysics, University of California, Santa Cruz, 1156 High Street, Santa Cruz, CA 95065, USA}

\noindent \hangindent=.5cm $^{80}${Department of Astronomy \& Astrophysics, University of Toronto, Toronto, ON M5S 3H4, Canada}

\noindent \hangindent=.5cm $^{81}${University of Science and Technology, 217 Gajeong-ro, Yuseong-gu, Daejeon 34113, Republic of Korea}

\noindent \hangindent=.5cm $^{82}${Departament de F\'{i}sica, Serra H\'{u}nter, Universitat Aut\`{o}noma de Barcelona, 08193 Bellaterra (Barcelona), Spain}

\noindent \hangindent=.5cm $^{83}${Laboratoire de Physique Subatomique et de Cosmologie, 53 Avenue des Martyrs, 38000 Grenoble, France}

\noindent \hangindent=.5cm $^{84}${Instituci\'{o} Catalana de Recerca i Estudis Avan\c{c}ats, Passeig de Llu\'{\i}s Companys, 23, 08010 Barcelona, Spain}

\noindent \hangindent=.5cm $^{85}${Max Planck Institute for Extraterrestrial Physics, Gie\ss enbachstra\ss e 1, 85748 Garching, Germany}

\noindent \hangindent=.5cm $^{86}${Department of Physics and Astronomy, Siena College, 515 Loudon Road, Loudonville, NY 12211, USA}

\noindent \hangindent=.5cm $^{87}${Department of Physics \& Astronomy, University  of Wyoming, 1000 E. University, Dept.~3905, Laramie, WY 82071, USA}

\noindent \hangindent=.5cm $^{88}${National Astronomical Observatories, Chinese Academy of Sciences, A20 Datun Rd., Chaoyang District, Beijing, 100012, P.R. China}

\noindent \hangindent=.5cm $^{89}${Steward Observatory, University of Arizona, 933 N, Cherry Ave, Tucson, AZ 85721, USA}

\noindent \hangindent=.5cm $^{90}${Aix Marseille Univ, CNRS, CNES, LAM, Marseille, France}

\noindent \hangindent=.5cm $^{91}${Departament de F\'isica, EEBE, Universitat Polit\`ecnica de Catalunya, c/Eduard Maristany 10, 08930 Barcelona, Spain}

\noindent \hangindent=.5cm $^{92}${Aix Marseille Univ, CNRS/IN2P3, CPPM, Marseille, France}

\noindent \hangindent=.5cm $^{93}${University of California Observatories, 1156 High Street, Sana Cruz, CA 95065, USA}

\noindent \hangindent=.5cm $^{94}${Department of Physics \& Astronomy, Ohio University, Athens, OH 45701, USA}

\noindent \hangindent=.5cm $^{95}${Department of Physics and Astronomy, Sejong University, Seoul, 143-747, Korea}

\noindent \hangindent=.5cm $^{96}${Abastumani Astrophysical Observatory, Tbilisi, GE-0179, Georgia}

\noindent \hangindent=.5cm $^{97}${Faculty of Natural Sciences and Medicine, Ilia State University, 0194 Tbilisi, Georgia}

\noindent \hangindent=.5cm $^{98}${Space Telescope Science Institute, 3700 San Martin Drive, Baltimore, MD 21218, USA}

\noindent \hangindent=.5cm $^{99}${Centre for Advanced Instrumentation, Department of Physics, Durham University, South Road, Durham DH1 3LE, UK}

\noindent \hangindent=.5cm $^{100}${Physics Department, Brookhaven National Laboratory, Upton, NY 11973, USA}

\noindent \hangindent=.5cm $^{101}${Beihang University, Beijing 100191, China}

\noindent \hangindent=.5cm $^{102}${Department of Astronomy, Tsinghua University, 30 Shuangqing Road, Haidian District, Beijing, China, 100190}

\noindent \hangindent=.5cm $^{103}${Physics Department, Stanford University, Stanford, CA 93405, USA}

\noindent \hangindent=.5cm $^{104}${Department of Physics, University of California, Berkeley, 366 LeConte Hall MC 7300, Berkeley, CA 94720-7300, USA}

\end{document}